\documentclass[a4paper, 12pt]{article}
\usepackage[utf8]{inputenc}
\usepackage[T1]{fontenc}
\usepackage{babel}
\usepackage{float}
\usepackage[width=170mm,height=257mm]{geometry}
\usepackage{amsmath, amsfonts, amssymb}
\usepackage[pdftex]{graphicx}
\usepackage{lmodern, indentfirst, cite}
\usepackage[position=top, labelfont={color=blue,bf,sf}]{subfig}
\usepackage[textfont={footnotesize}, labelfont={color=blue,bf,sf},
labelsep=endash]{caption}
\usepackage[colorlinks=true, linkcolor=blue, citecolor=blue, urlcolor=blue]{hyperref}

\usepackage[affil-sl]{authblk}
\title{\textbf{\Large Resonant transport and line-type resonances in tilted Dirac cone
double-barrier structures}}
\author[a]{M. Raggui}
\author[a]{O. Habti}
\author[a,b]{A. Kamal}
\author[a]{E.B. Choubabi\thanks{\href{mailto:choubabi.e@ucd.ac.ma}{choubabi.e@ucd.ac.ma}}}
\affil[a]{LPMC Laboratory, Theoretical Physics Group, Faculty of Sciences, Chouaïb
Doukkali University, 24000 El Jadida, Morocco}
\affil[b]{Hassan II University of Casablanca, National Higher School of Arts and Crafts
(ENSAM of Casablanca), LISIME Laboratory, 20670 Casablanca, Morocco}
\providecommand{\pacs}[1]{\noindent \textbf{PACS numbers:} #1\\}
\providecommand{\keywords}[1]{\noindent \textbf{Keywords:} #1}
\begin{document}
%========================================================
\begin{titlepage}
    \maketitle
    \thispagestyle{empty}
    \vspace{2cm}
\begin{abstract}
    We study the transport properties of Dirac fermions in a graphene-based double-barrier
structure composed of two tilted-cone regions separated by a central pristine graphene
region. Using the transfer matrix method, we systematically analyze how different cone
tilts affect Dirac fermion transmission. In reciprocal space, at fixed energy, the Dirac
cones of distinct regions generate isoenergetic conical surfaces (Fermi surfaces). When
these surfaces overlap, their intersections define ``active surfaces'' that enable fermion
transmission. In the symmetric double-barrier configuration, coupling between the barriers
and the central well gives rise to multiple resonance peaks, including line-type
resonances, even within nominally forbidden energy zones. The number and positions of
these resonances depend sensitively on the system parameters. These findings provide new
insights into the role of Dirac cone tilt in complex junctions and may guide the design of
nanoelectronic devices based on two-dimensional tilted-cone materials such as
$\alpha$-(BEDT-TTF)$_2$I$_3$ and borophene.
\end{abstract}
\vspace{1cm}
\end{titlepage}
%========================================================
\section{Introduction}
%========================================================
Since the isolation of graphene from graphite in 2004, condensed matter physics has found
a fertile ground for exploring electronic transport in two-dimensional (2D) systems.
Graphene, a single layer of carbon atoms arranged in a hexagonal lattice, has attracted
considerable attention due to its exceptional properties and wide range of potential
applications in electronics, sensing, and nanoelectronics \cite{1}. One of its most
remarkable features is that the low-energy spectrum is governed by a pseudo-relativistic
Dirac equation, leading to Dirac fermion dynamics analogous to those of massless
particles. In the first Brillouin zone, the energy dispersion is linear around two
inequivalent points, known as the Dirac points \(K\) and \(K'\) \cite{CastroNeto2009}.

Unlike conventional semiconductors, where Dirac fermions are partially reflected at
potential barriers, graphene exhibits the Klein paradox: under certain conditions, Dirac
fermions can transmit through thick electrostatic barriers with unit probability at normal
incidence \cite{3}. This counter intuitive effect, arising from the relativistic nature of
Dirac fermions, opens new possibilities for innovative electronic devices.

In reciprocal space, the energy–wave vector relationship near the Dirac points is
represented by Dirac cones. In pristine graphene, these cones are symmetric, with
identical slopes along the \(k_x\) and \(k_y\) directions. However, in a broader class of
Dirac materials, the cones may be tilted, strongly modifying the transport properties.
Tilted Dirac cones appear in several exotic systems, including the organic conductor
$\alpha$-(BEDT-TTF)$_2$I$_3$~\cite{4,5,Goerbig2008}, 8-Pmmn borophene~\cite{7,8}, and
certain topological insulators~\cite{9,10}. In such systems, the dispersion becomes
anisotropic, with slopes depending on the propagation direction, giving rise to
distinctive transport phenomena such as collimation and directional conductivity
\cite{01,02}. In graphene, tilted Dirac cones can also be engineered through quinoid
lattice deformation, which breaks hexagonal symmetry and produces asymmetric dispersion
near the Dirac points.

Our motivation for studying tilted Dirac cones in a double-barrier configuration stems
from our earlier work on Dirac fermion collimation in tilted-cone heterostructures with a
single barrier \cite{Choubabi2024}. That study revealed intriguing orientation-dependent
effects, which naturally lead to the exploration of more complex geometries. In this
paper, we extend the analysis to a symmetric double-barrier structure composed of
alternating graphene and tilted-cone regions: graphene–tilted cone–graphene–tilted
cone–graphene. The tilt direction of the cones is taken along the \(k_y\)-axis, with both
barriers oriented in the same direction, allowing a clear comparison with our previous
results and isolating the specific role of double-barrier resonances.

\textit{Beyond standard Fabry–Pérot oscillations, a central focus of this work is the
emergence of \emph{line-type resonances}.} These are sharp transmission ridges that
persist over extended energy–momentum (or angular) domains and can even intrude into
nominally forbidden zones. Physically, they originate from quasi-bound states formed in
the central well that hybridize with evanescent modes under the barriers, the latter
acting as effective mirrors. The tilt parameter \(\tau\) breaks the \(k_y\!\to\!-k_y\)
symmetry and provides a powerful knob to shift, multiply, or suppress these line-type
resonances, thereby enabling directional and energy-selective transport control
\cite{Barbier2010,04,Sun2011,RamezaniMasir2010,Baltateanu2019,Alhaidari2012}.

The paper is organized as follows. Section 2 introduces the theoretical model, including
the eigenspinors, energy spectrum, and transfer matrix at the graphene–tilted-cone
interfaces, from which transmission and reflection probabilities are obtained. Section 3
presents the numerical results, highlighting Fabry–Pérot and line-type resonances and the
influence of key system parameters on transport. Finally, Section 4 summarizes our main
findings, their implications for transport in tilted Dirac materials, and possible future
directions.
%========================================================
\section{Theoretical Model}
%========================================================
In this section, we develop the theoretical framework describing the transport of Dirac
fermions through a graphene-based double-barrier structure composed of pristine graphene
and tilted-cone materials, with particular emphasis on the emergence of line-type
resonances.
%========================================================
\subsection{Double-Barrier System and Potential Profile}
%========================================================
\begin{figure}[h!]\centering
    \includegraphics[width=15cm]{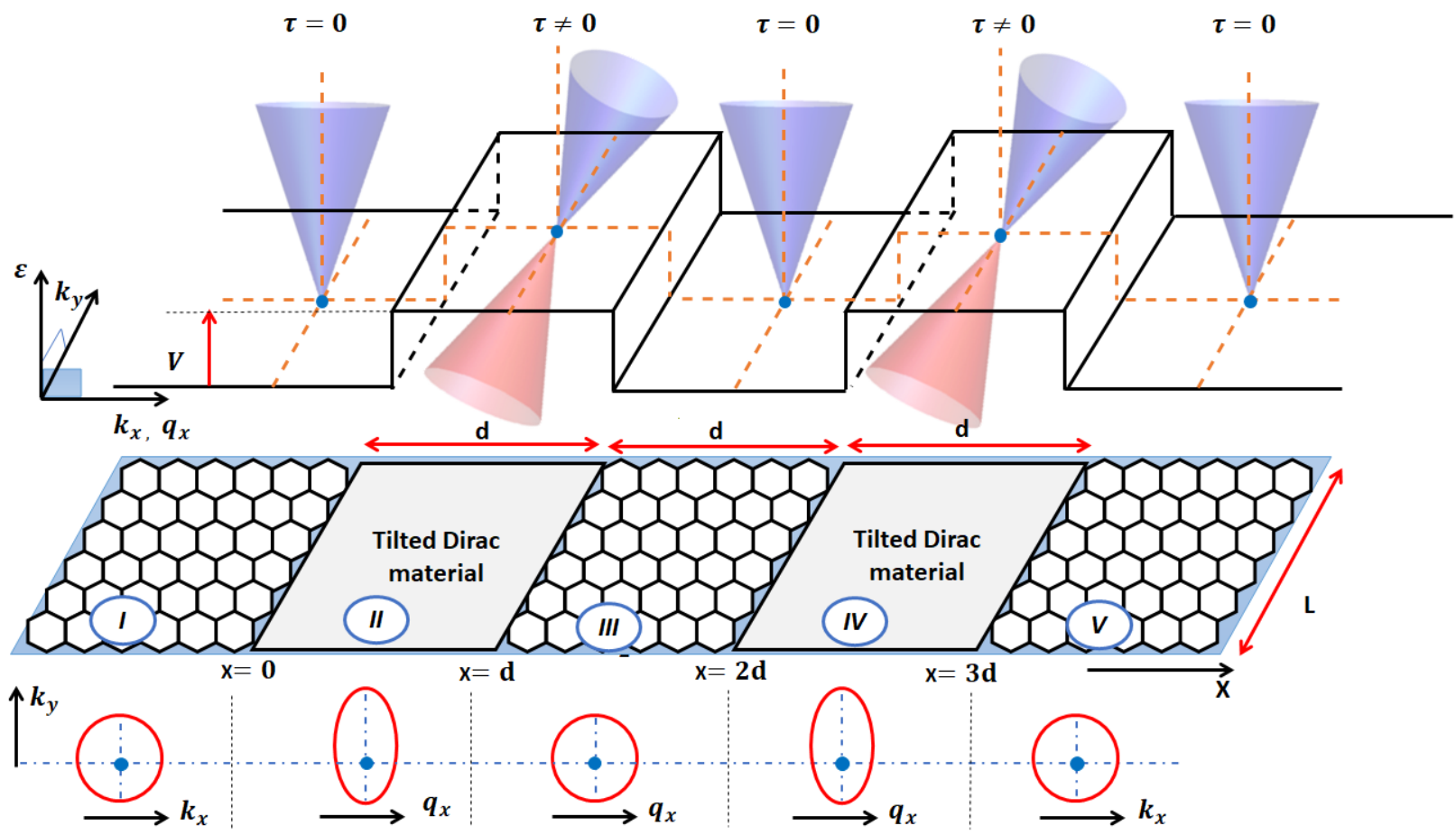}
    \caption{(Color online) Diagram illustrating the five-region system with the potential
profile applied to the different regions. It also shows the Fermi surfaces associated with
the Dirac cones in these different regions.}
    \label{fig0}
\end{figure}

The double-barrier system considered in this work consists of five distinct regions,
labeled $j = 1, 2, 3, 4, 5$. Region $1$, located at the far left, corresponds to a
reference medium such as pristine graphene with a constant zero potential. Adjacent to it
is region $2$, which forms the first barrier of the system. This barrier is composed of a
tilted-cone material characterized by a potential $V$, different from that of region $1$,
thereby modifying the electronic properties relative to its neighbor. Region $3$, situated
at the center, acts as the well or central zone between the two barriers. It can be either
pristine graphene or another tilted-cone material, but with zero potential. The potential
configuration of this region is crucial for determining the transport properties of Dirac
fermions. To the right of region $3$ lies region $4$, which constitutes the second
barrier. It is typically similar in composition to region $2$, with the same potential
$U$, but distinct from the central region. Finally, region $5$, located at the far right,
is generally identical to region $1$, also with a constant zero potential. This final
region serves as the reference for evaluating the transmission and reflection of Dirac
fermions across the entire structure.

Fig.\ref{fig0} illustrates these five regions and the potential profile that varies along
the spatial $ x $-axis , thus creating a double barrier whose transitions influence the
transport properties of Dirac fermions through the system. It is important to note that
regions $2$, $3$ and $4$, which form the double barrier, have the same length $ x = d $.
Additionally, the system has a width $ y = L $.  In this way, we obtain a symmetric system
where the length of the double barrier is $x=3d$.

The reduced potential profile $V(x)$, associated with the five distinct regions of the
system, can be expressed in a dimensionless form after applying the unit reduction
procedure, as given in Eq.\eqref{miral:1}.
%========================================================
\subsection{Dimensionless Formulation and Parameters}
%========================================================
To simplify the notation and facilitate the analytical treatment, we express all physical
quantities in dimensionless units by introducing the reduced energy $\varepsilon$ and the
potential $V$ \cite{RamezaniMasir2010,Choubabi2020}.

The parameter $\kappa_{y}$ denotes the transverse wave vector, corresponding to the
component of momentum perpendicular to the main propagation direction. In contrast,
$\kappa_{x}$ and $\mathbf{q}_{x}$ represent the longitudinal wave vectors, associated with
pristine graphene and the tilted Dirac cone material, respectively. These longitudinal
components govern the propagation of quasiparticles along the $x$-axis, and are strongly
influenced by the geometry and anisotropy of the underlying band structure.

\begin{equation}\label{eq:3}
  \varepsilon=\frac{Ed}{\hbar v_{F}},\quad
  V(x)=\frac{U(x)d}{\hbar v_{F}},\quad
  k_{x}=\kappa_{x}d,\quad
  k_{y}=\kappa_{y}d,\quad
  q_{x}=\mathbf{q}_{x}d
\end{equation}

The dimensionless formulation not only simplifies the notation but also highlights the
scaling laws governing transport in tilted Dirac materials, enabling direct comparison
between systems of different characteristic lengths and velocities.

After rescaling to dimensionless units, the potential profile becomes
\begin{equation}\label{miral:1}
V(x)=V_{j}=\left\{
  \begin{array}{ll}
  V,\quad  \text{if} \quad j=2, 4 \\
   0,\quad  \text{otherwise}
  \end{array}
\right.
\end{equation}
%========================================================
\subsection{Tilted Dirac Hamiltonian}
%========================================================
The effective Hamiltonian describing massless Dirac fermions in the $K$-valley for a
material with isotropic tilted Dirac cones (where \( v_x = v_y = v_F \)) and subjected to
a one-dimensional potential \( V(x) \) along the \( x \)-axis can be formulated as
follows~\cite{Goerbig2008,Betancur2022,Tan2022,Mojarro2022,Nguyen2018}:

\begin{equation}\label{eq001}
H = v_F \begin{pmatrix}
0 & \pi^\dagger \\
\pi & 0
\end{pmatrix}
+ v_t p_y \begin{pmatrix}
1 & 0 \\
0 & 1
\end{pmatrix}
+ V(x) \begin{pmatrix}
1 & 0 \\
0 & 1
\end{pmatrix},
\end{equation}

where  $ \pi = p_x + i p_y $ is the momentum operator, with $ p_x = -i \hbar \partial_x $
and $ p_y = -i \hbar \partial_y $ representing the momentum operators in the $ x $ and $ y
$ directions, respectively. $ \pi^\dagger $ is the Hermitian conjugate of $ \pi $. The
first term, $ v_F \begin{pmatrix} 0 & \pi^\dagger \\ \pi & 0 \end{pmatrix} $, describes
the kinetic energy of Dirac fermions, where $ v_F $ is the Fermi velocity.

The second term, $ v_t p_y \begin{pmatrix} 1 & 0 \\ 0 & 1 \end{pmatrix} $, introduces an
additional velocity component $ v_t $, originating from the tilting of the Dirac cones
along the $ k_y $ direction~\cite{Islam2018,Trescher2015}. This tilt modifies the dispersion
relation and impacts the transport properties of the quasiparticles by breaking the
isotropy of the cones.

The last term, $ V(x) \begin{pmatrix} 1 & 0 \\ 0 & 1 \end{pmatrix} $, represents the
external one-dimensional electrostatic potential, which varies along the $ x $ axis and
shifts the fermionic energy spectrum depending on position. Such models are widely used to
investigate electronic transport in tilted Dirac systems and related heterostructures.

It is important to note that, in this work, Planck’s constant $ \hbar $ and the Fermi
velocity $ v_F $ are set to unity ($ \hbar = v_F = 1 $) to simplify the formalism,
yielding dimensionless quantities. The tilt parameter is then defined as $ \tau =
\tfrac{v_t}{v_F} = v_t $.

In the different regions $j$ of the structure, the generalized Hamiltonian can be
expressed in the form of Eq.~\eqref{eq001}, which simplifies to:
\begin{equation}\label{eq2}
H =
\begin{pmatrix}
 \tau  k_y+V (x) & -i \partial_{x}-i k_y  \\
 -i \partial_{x}+i k_y & \tau  k_y+V (x)
\end{pmatrix}.
\end{equation}

In this formulation, we assume that the commutator $[H, p_y] = 0$ holds, which implies
that the wave function in region $j$ can be written as a separable form
$\psi_j(x,y) = \varphi_j(x) e^{i k_y y}$,
where $k_y$ is the conserved transverse momentum.

When $\tau = 0$ and $V(x) = 0$, the Hamiltonian reduces to that of pristine graphene in
regions $1$, $3$ and $5$~\cite{Goerbig2008,Choubabi2024,Kamal2018}.
%========================================================
\subsection{Eigenvalue Problem and Spinor Solutions}
%========================================================
The solution of the eigenvalue equation for the Hamiltonian $H$ in region $j$ provides
several key results regarding the dynamics of Dirac fermions in tilted-cone
systems~\cite{Choubabi2024,Das2020,Kong2021}.
First, we obtain the eigen-spinor,

\begin{equation}\label{eq1}
\psi_{j}=  \frac{1}{\sqrt{2}}
\begin{pmatrix}
 e^{i k_j x} & e^{-i k_j x} \\
 s_j e^{i \theta _j} e^{i k_j x} & -s_j e^{-i \theta _j} e^{i k_j x}
\end{pmatrix}
\begin{pmatrix}
 \alpha _j \\
 \beta _j
\end{pmatrix}
e^{i k_y y },
\end{equation}

which represents the quantum state of Dirac fermions in region $j$, where $\alpha _j$ and
$\beta _j$ are the amplitudes of forward and backward propagating modes,
respectively~\cite{Das2020,Kamal2018}.

The corresponding energy spectrum is given by

\begin{equation}\label{eq0}
    \varepsilon_j=\tau_j k_{y}+V_j+s_j\sqrt{k^{2}_{j}+k^{2}_{y}},
\end{equation}

which establishes the relation between the particle’s energy and its wave vector. Since
the system is adiabatic and isolated, the energy $\varepsilon_j=\varepsilon$ is conserved.
The longitudinal wave vector $k_j$ then follows as

\begin{equation}\label{eq0k}
k_{j}=\sqrt{(\varepsilon-\tau_j k_{y}-V_j)^{2}-k^{2}_{y}},
\end{equation}

corresponding to the wave vector component along the $x$-axis in region
$j$~\cite{Choubabi2024,Pattrawutthiwong2021,Kong2021}.

The eigen-spinor encodes the orientation of Dirac fermions in pseudospin space, and when
combined with the energy spectrum, provides a complete description of the fermionic state
in each region. The parameter $s_j = \text{sign}(\varepsilon - \tau_j k_y - V_j)$
distinguishes between the conduction ($s_j=+1$) and valence ($s_j=-1$) band solutions.

The propagation angle $\theta_j$ corresponds to the direction of the wave vector
$\vec{k}_j=(k_j, k_y)$ with respect to the $x$-axis, defined as

\begin{equation}
\theta_j = \arctan\left(\frac{k_y}{k_j}\right).
\end{equation}

This angle is crucial for determining the trajectory of Dirac fermions in the $(x,y)$
plane and plays a central role in analyzing transmission through regions with tilted
cones~\cite{Choubabi2024,Pattrawutthiwong2021,Kapri2020}.

Altogether, these results provide the foundation for analyzing how tilted Dirac cones and
potential barriers shape transport phenomena, including resonance effects and line-type
resonances in double-barrier systems.
%========================================================
\subsection{Special Cases: Pristine Graphene and Tilted Regions}
%========================================================
In regions $j=1, 3, 5$, corresponding to pristine graphene, the parameters take the values
$s_j=1$, $\tau_j=0$, $k_j=k_x$, $\theta_j=\theta$, and $V_j=0$. In this limit, the
Hamiltonian reduces to that of massless Dirac fermions near the $K$ point in the Brillouin
zone~\cite{CastroNeto2009,Katsnelson2006}, expressed as

\begin{equation}\label{eq2}
H =
\begin{pmatrix}
 0 & -i \partial_{x}-i k_y  \\
 -i \partial_{x}+i k_y & 0
\end{pmatrix}.
\end{equation}

The eigenvalue problem in these regions ($j=1,3,5$) yields the spinor

\begin{equation}
\psi_{j}=  \frac{1}{\sqrt{2}}
\begin{pmatrix}
 e^{i k_x x} & e^{-i k_x x} \\
 e^{i \theta } e^{i k_x x} & - e^{-i \theta } e^{i k_x x}
\end{pmatrix}
\begin{pmatrix}
 \alpha _j \\
 \beta _j
\end{pmatrix}
e^{i k_y y },
\end{equation}

where $(\alpha_1, \alpha_3, \alpha_5) = (1, a, t)$ and $(\beta_1, \beta_3, \beta_5) = (r,
b, 0)$. Here, $r$ and $t$ are the reflection and transmission amplitudes, while $a$ and
$b$ describe the propagation amplitudes inside the central region.

The dispersion relation is given by
\begin{equation}
\varepsilon = \sqrt{k^{2}_{x}+k^{2}_{y}},
\end{equation}
with the longitudinal wave vector
\begin{equation}
k_{x} = \sqrt{\varepsilon^{2}-k^{2}_{y}}.
\end{equation}

In the tilted regions $j=2,4$, corresponding to the barriers, the cones are inclined by
$\tau$, with parameters $s_j=\pm 1$, $\tau_j=\tau$, $k_j=q_x$, $\theta_j=\phi$, and
$V_j=V$. The associated spinors are

\begin{equation}
\psi_{j}=  \frac{1}{\sqrt{2}}
\begin{pmatrix}
 e^{i q_x x} & e^{-i q_x x} \\
 s_j e^{i \phi} e^{i q_x x} & -s_j e^{-i \phi} e^{i q_x x}
\end{pmatrix}
\begin{pmatrix}
 \alpha _j \\
 \beta _j
\end{pmatrix}
e^{i k_y y },
\end{equation}

with propagation amplitudes $(\alpha_2, \alpha_4) = (a_1, a_2)$ and $(\beta_2, \beta_4) =
(b_1, b_2)$. The dispersion relation in these regions reads

\begin{equation}\label{eq0000}
\varepsilon = \tau k_{y} + V \pm \sqrt{q^{2}_{x}+k^{2}_{y}},
\end{equation}

which highlights the modification of the band structure induced by
tilting~\cite{Goerbig2008,Soluyanov2015,Xu2023,Pattrawutthiwong2021}.
%========================================================
\subsection{Fermi Surfaces and Classification of Tilted Cones}
%========================================================
The Fermi surfaces in region $j$ can be described by the Cartesian
equation~\cite{Choubabi2024,Tan2022}

\begin{equation}\label{eq5}
q_x^2 + \left(1-\tau^2\right)\,k_y^2 +2\,\tau\,k_y\left(\varepsilon-V\right) -
\left(\varepsilon-V\right)^2 =0.
\end{equation}

Here, $\tau = e$ acts as the eccentricity of the conic section, a central parameter
controlling the geometry of the Fermi surface. Its value directly determines the
electronic anisotropy of the material and hence the propagation of Dirac fermions across
the double barrier~\cite{Trescher2015,Sinha2019}.

\begin{figure}[H]\centering
\hspace{-10mm}
\subfloat[Untilted]{
        \includegraphics[scale=0.10]{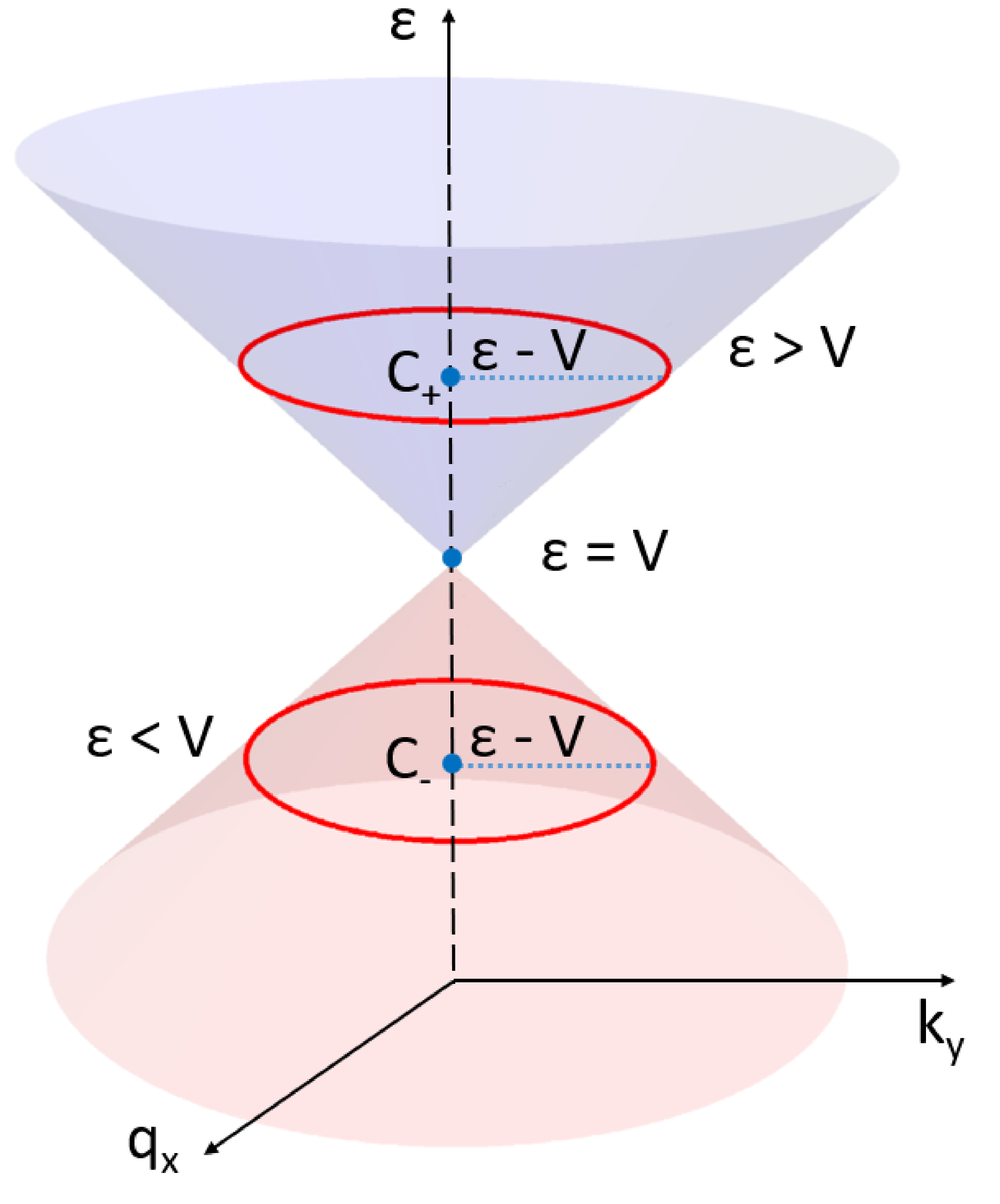}
        \label{fig02:SubFigA}
    }\hspace{-5mm}
	\subfloat[Type $I$]{
        \includegraphics[scale=0.10]{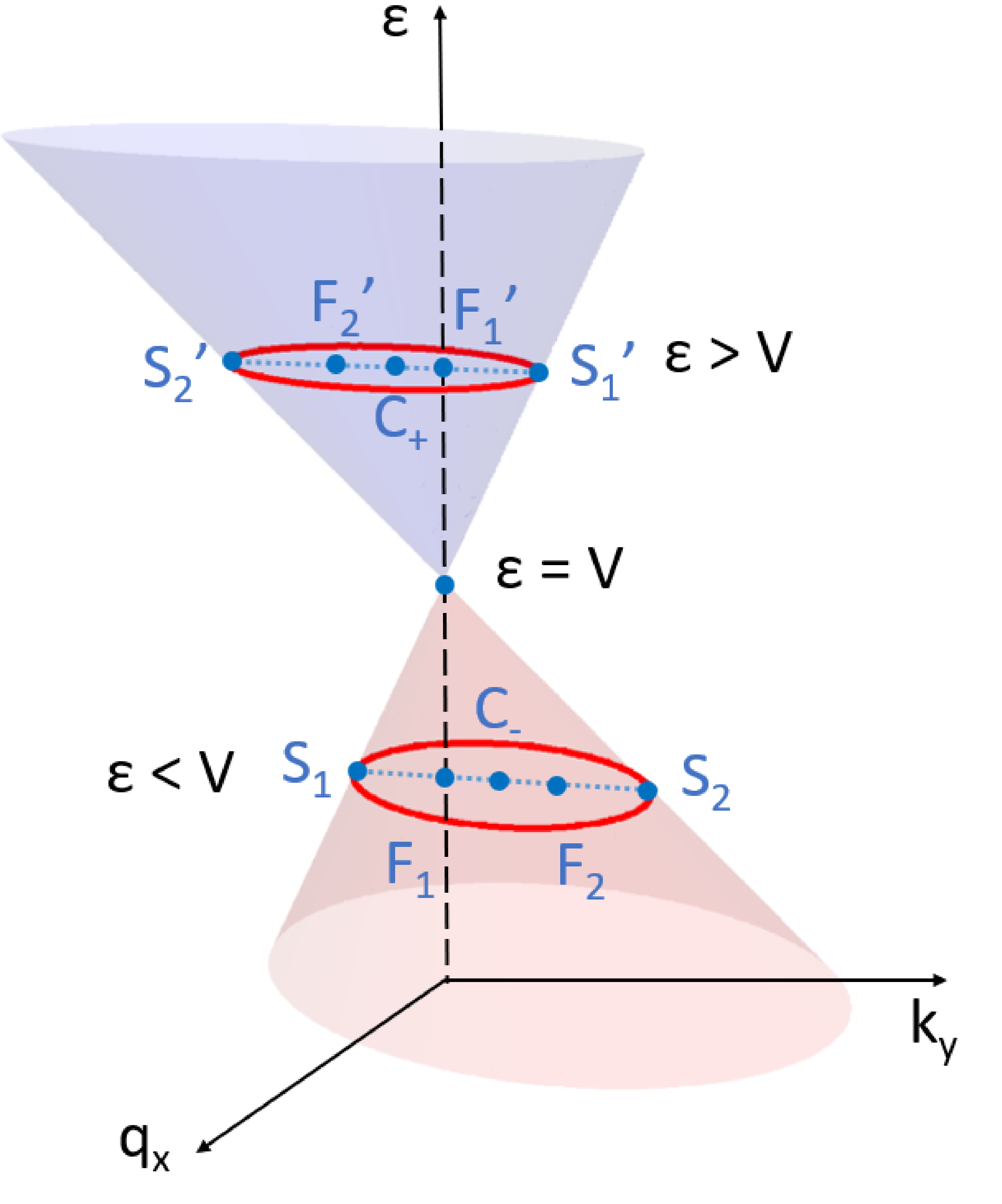}
        \label{fig02:SubFigB}
    }\hspace{-5mm}
	\subfloat[Type $II$]{
        \includegraphics[scale=0.11]{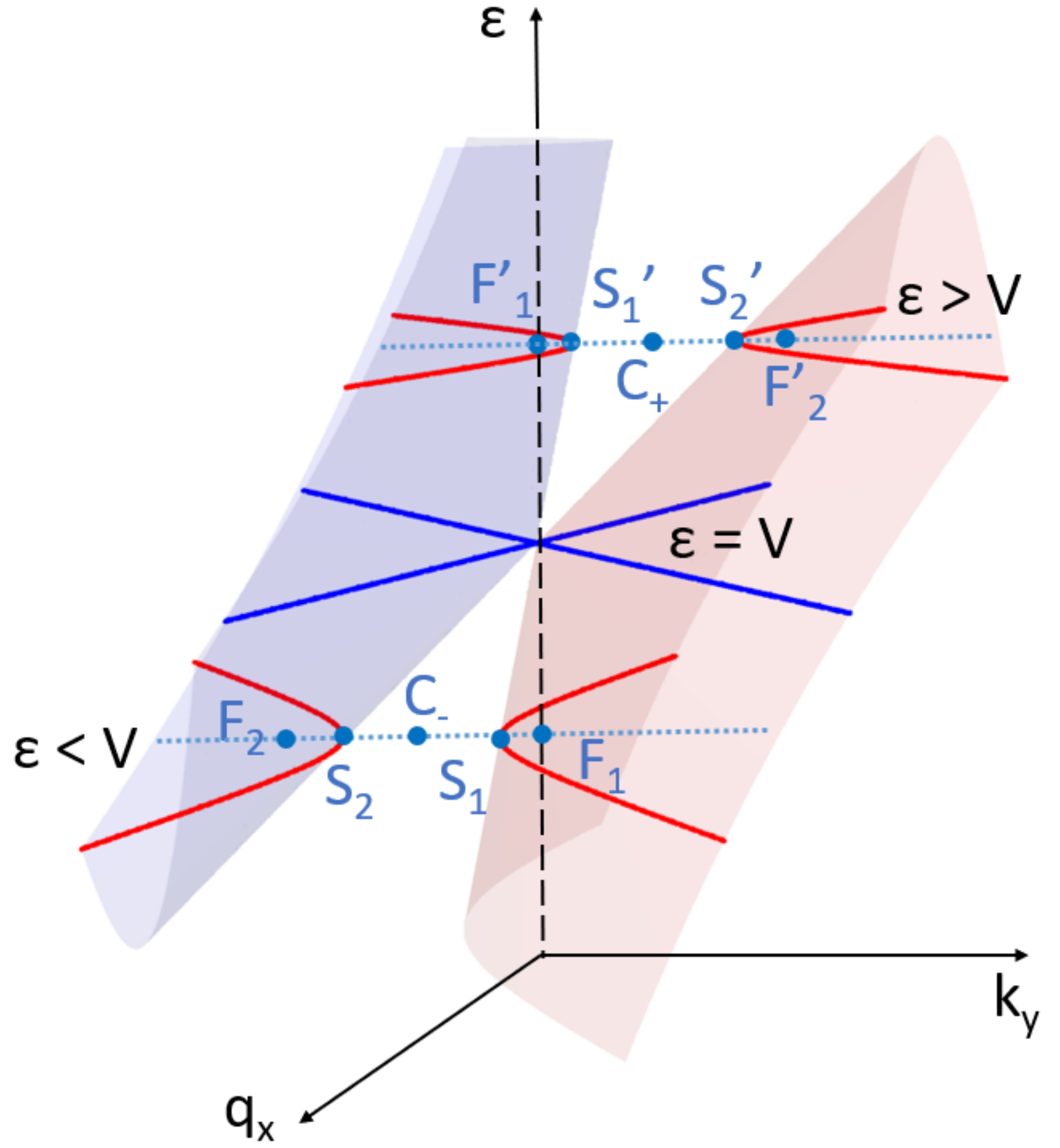}
        \label{fig02:SubFigC}
    }\hspace{-3mm}
	\subfloat[Type $III$ ]{
        \includegraphics[scale=0.10]{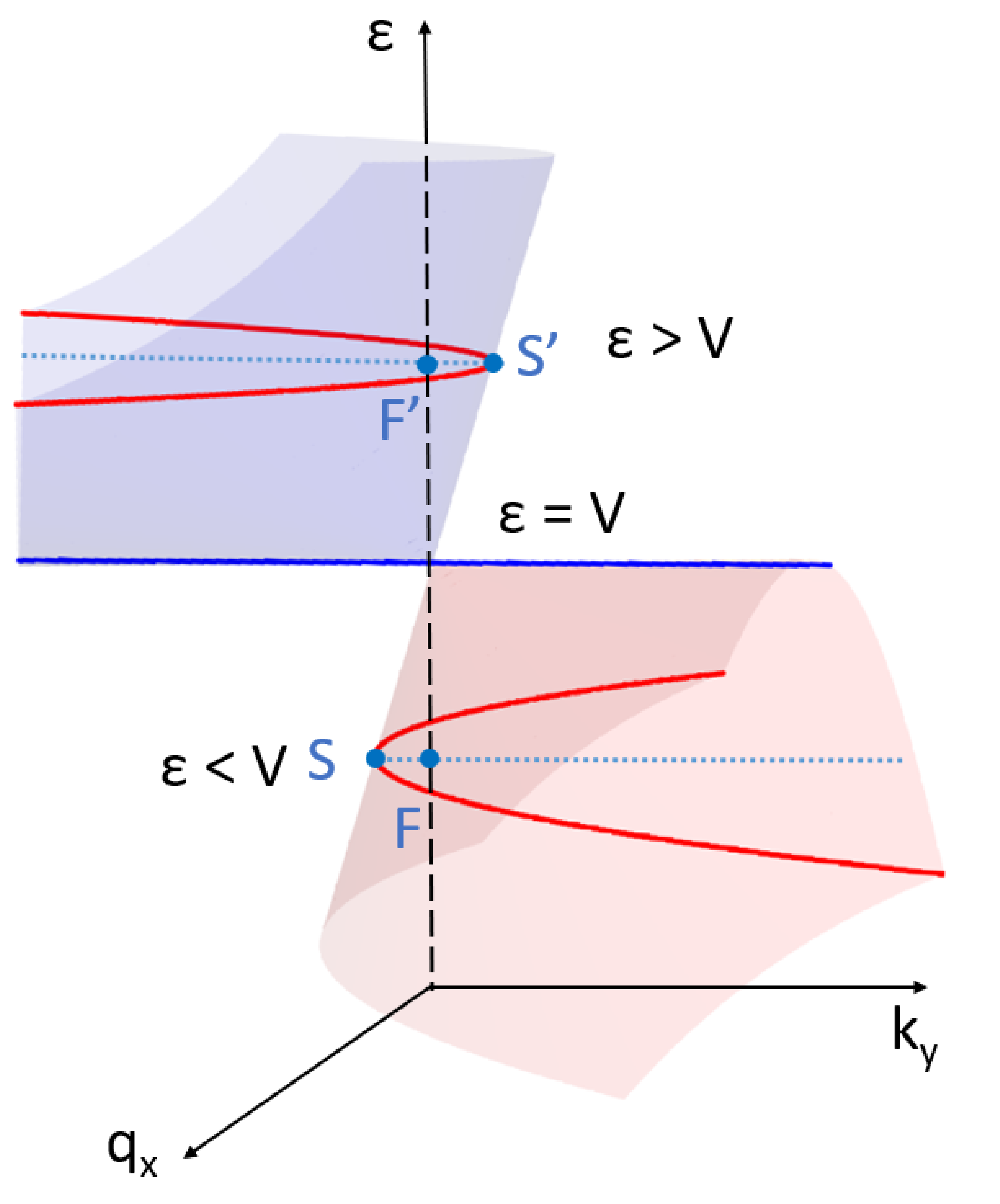}
        \label{fig02:SubFigD}
    }
	\caption{(Color online) Different types of tilted Dirac cones in reciprocal space,
showing Fermi surfaces corresponding to distinct energy levels. These surfaces illustrate
the geometry of the conical curves and the location of Dirac points where conduction and
valence bands meet.}
	\label{fig02}
\end{figure}

Specifically, when $e=0$ (untilted, Fig.~\ref{fig02:SubFigA}), the Fermi surface is
circular, reflecting isotropy. For $0 < e < 1$ (Type I, Fig.~\ref{fig02:SubFigB}), it
becomes elliptical, revealing mild anisotropy. At $e=1$ (Type III,
Fig.~\ref{fig02:SubFigD}), the surface turns parabolic, marking a transition in the
electronic response. Finally, for $e>1$ (Type II, Fig.~\ref{fig02:SubFigC}), the Fermi
surface is hyperbolic, associated with strong anisotropy and exotic transport
signatures~\cite{Choubabi2024,Wild2022,Pattrawutthiwong2021,Tan2022,Canton2011}.

These classifications highlight how the eccentricity and tilt of Dirac cones directly
govern transport behavior, and they provide the theoretical basis for analyzing resonances
in graphene–tilted cone heterostructures~\cite{Choubabi2024}.
%========================================================
\subsection{Active Surfaces and Collimation Effects}
\begin{figure}[H]\centering
\hspace{-10mm}
	\subfloat[$\tau=0$]{\includegraphics[scale=0.19]{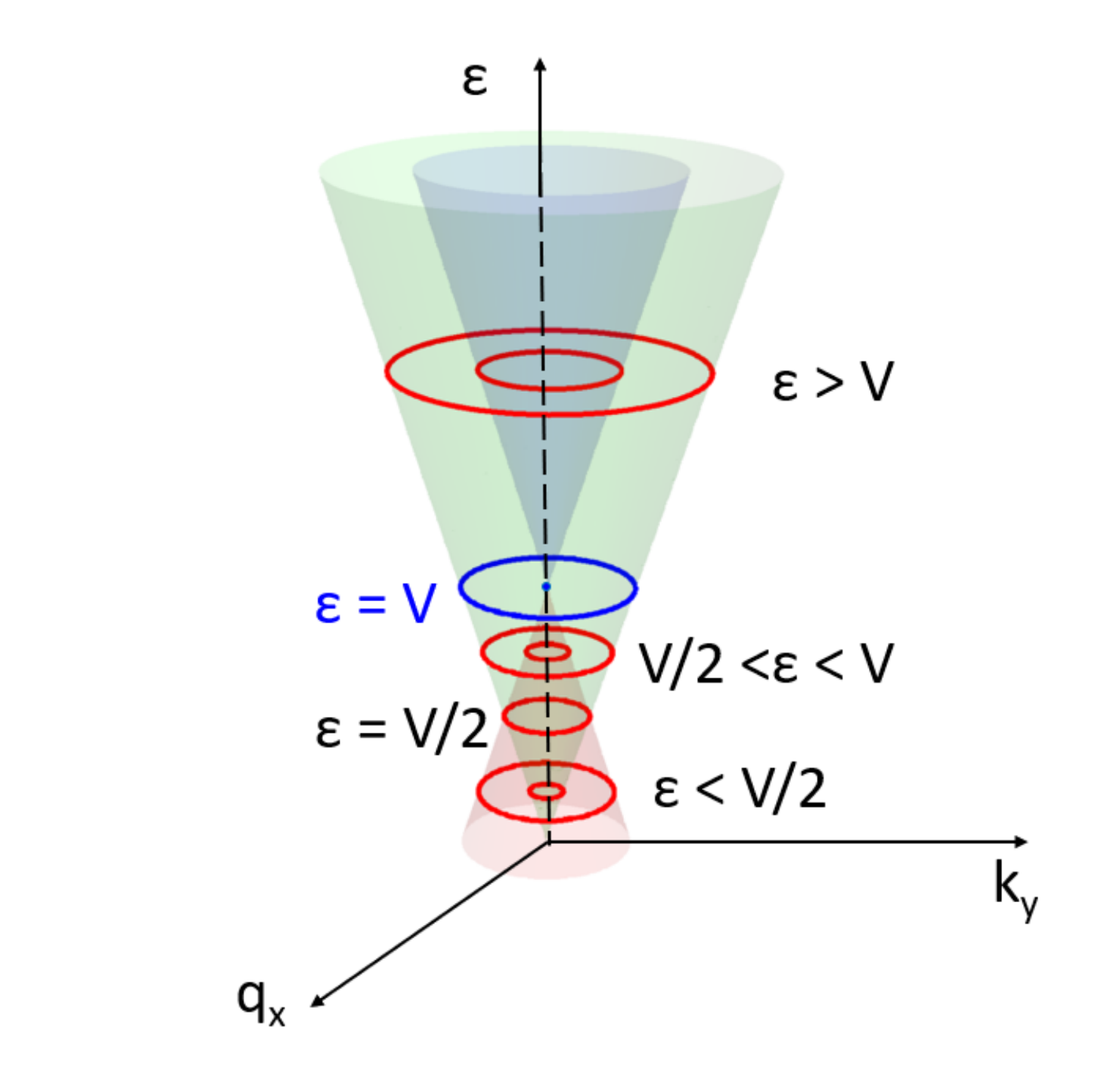}\label{fig03:SubFigA}}
	\hspace{-3mm}
	\subfloat[$0<\tau<1$]{\includegraphics[scale=0.17]{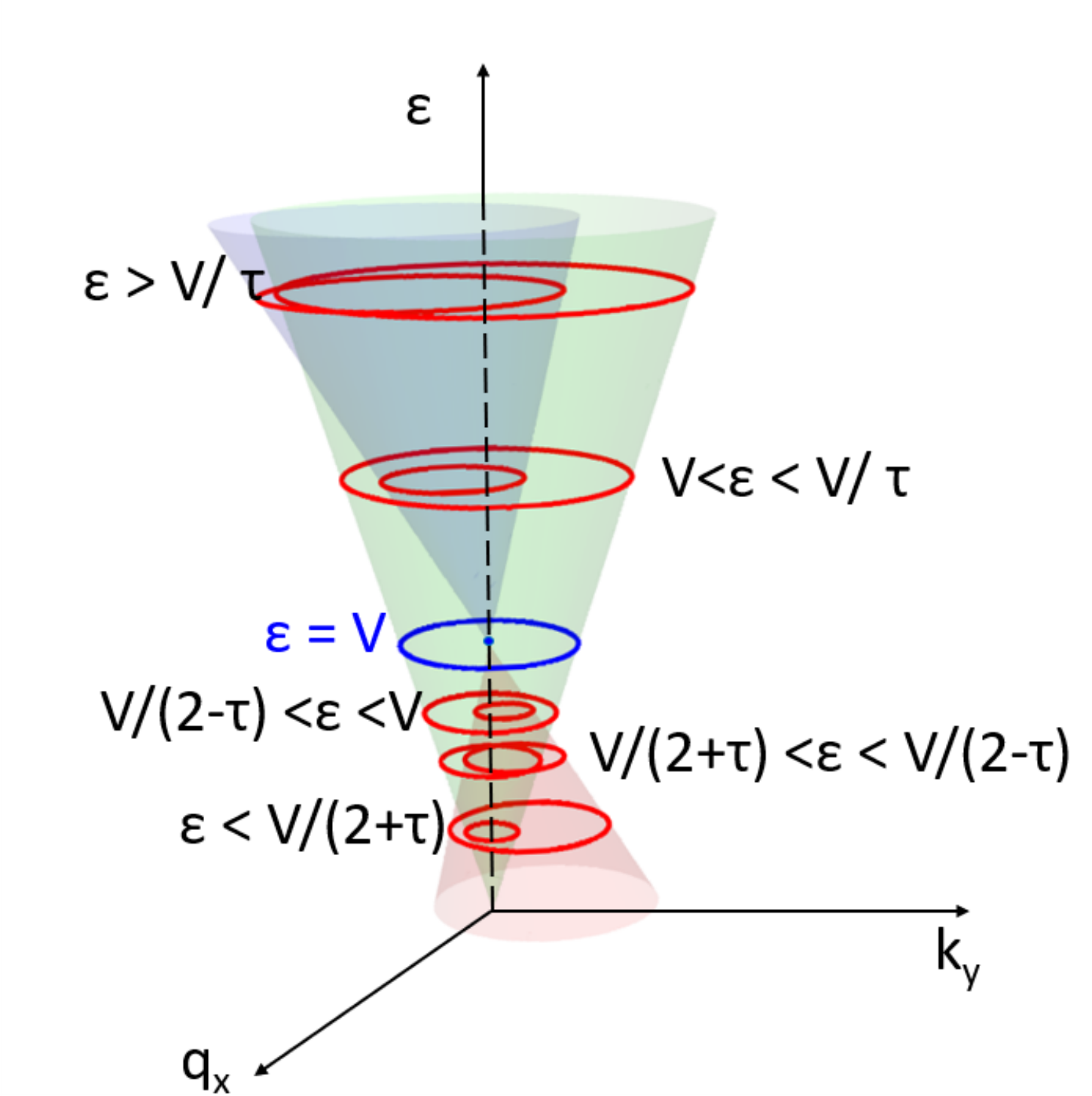}\label{fig03:SubFigB}}
	\hspace{-2mm}
	\subfloat[$\tau = 1$]{\includegraphics[scale=0.11]{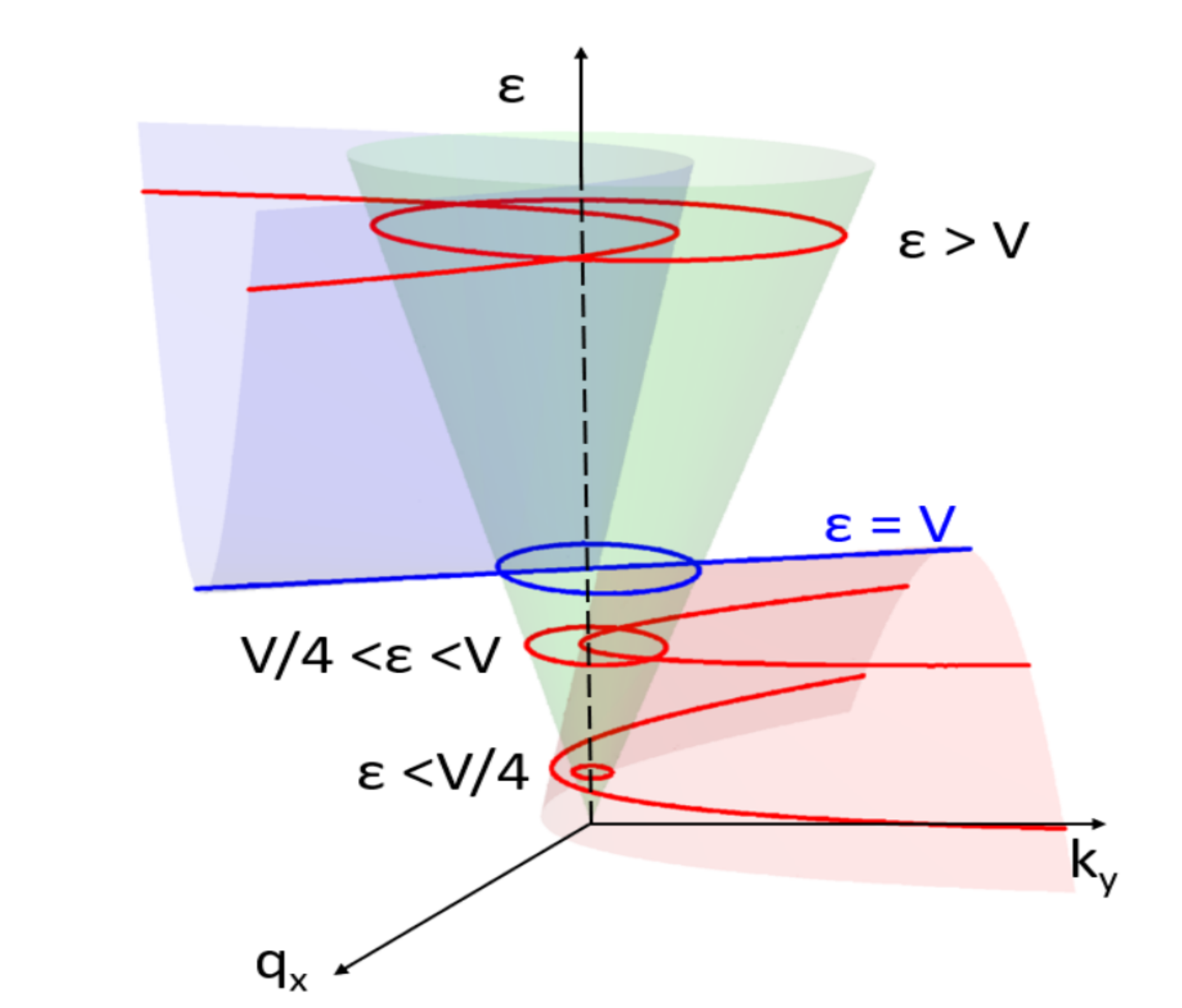}\label{fig03:SubFigC}}
	\hspace{-4mm}
	\subfloat[$\tau > 1$]{\includegraphics[scale=0.095]{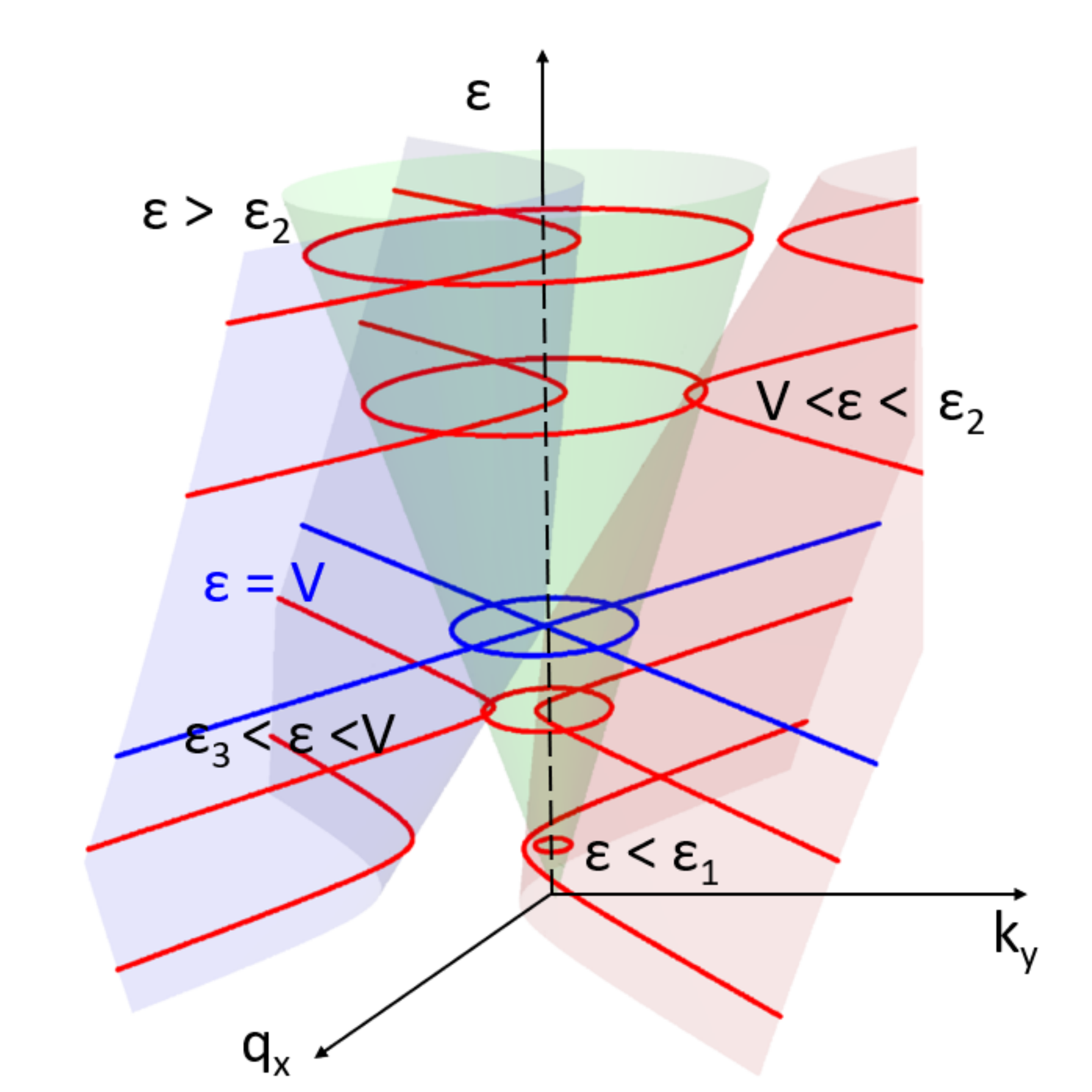}\label{fig03:SubFigD}}
	\caption{(Color online) Evolution of active surfaces, resulting from collimation
during the diffusion of Dirac fermions, as a function of propagation energy $\varepsilon$,
for different values of $\tau$.}
	\label{fig0003}
\end{figure}

%========================================================
The collimation effect arises from the overlap of Dirac cones in different regions of the
heterostructure, producing intersections between their respective Fermi surfaces
(Fig.~\ref{fig0003}). These intersections, referred to as \emph{active surfaces},
correspond to momentum-space zones where electronic states from adjacent regions are
matched, thereby enabling efficient transmission of Dirac fermions across the interfaces.
Such alignment is crucial for maintaining phase coherence and strongly impacts the
transport properties of the system~\cite{Katsnelson2006,Pattrawutthiwong2021}.

When the conic sections of different regions intersect, they define propagation channels
in which fermions experience reduced backscattering and enhanced transparency. In this
regime, the transmission is dominated by trajectories confined within the overlap region,
giving rise to directional transport or beam-like propagation, often described as fermion
collimation~\cite{Yesilyurt2018,Choubabi2020,Choubabi2024}. The extent of this effect depends sensitively on the tilt
parameter $\tau$, the Fermi energy, and the barrier potential.

As shown in Fig.~\ref{fig0003}, the geometry of the active surfaces evolves as a function
of propagation energy $\varepsilon$ and the tilt parameter $\tau$. In the untilted case
($\tau=0$, Fig.~\ref{fig03:SubFigA}), the overlap between Fermi surfaces is isotropic,
leading to uniform transmission channels. For $0<\tau<1$ (Type I,
Fig.~\ref{fig03:SubFigB}), the anisotropy of the cones generates elliptical intersections,
producing partial collimation along preferred directions. At $\tau=1$ (Type III,
Fig.~\ref{fig03:SubFigC}), the overlap becomes parabolic, marking the transition from
closed to open electronic orbits. Finally, when $\tau>1$ (Type II,
Fig.~\ref{fig03:SubFigD}), the hyperbolic intersections induce highly anisotropic
propagation, with fermions guided along narrow angular windows.

These active surfaces thus act as momentum filters, selecting the directions in which
Dirac fermions can propagate coherently across the double barrier. This mechanism is
directly connected to Fabry–Pérot–type resonances and contributes to the appearance of
line-type resonances in tilted-cone heterostructures~\cite{Montambaux2009,Enderlin2010,Jellal2012}. The phenomenon
highlights the fundamental role of collimation in graphene-based and tilted Dirac
materials, opening possibilities for electron optics and directionally selective
nanoelectronic devices~\cite{Rickhaus2015,Xu2023}.

%========================================================
\subsection{Transfer Matrix Formalism}
%========================================================
To compute the transmission coefficient of Dirac fermions across the double–barrier
structure, we employ the transfer matrix method. This technique, widely used in quantum
mechanics and mesoscopic physics, provides a systematic framework for analyzing wave
propagation through multilayer systems and potential barriers~\cite{Ando1991,Walker1992}. In the context of graphene and other Dirac materials, the transfer matrix approach has been
extensively applied to study tunneling, resonant transport, and Klein tunneling
phenomena~\cite{Katsnelson2006,Barbier2010}.

By expressing the boundary conditions of the spinor wave functions at each interface in
terms of transfer matrices $\mathcal{M}_j(x)$, we can track the evolution of the fermionic
states across adjacent regions. The transfer matrix that relates the spinor in the $j$-th
region to that in the $(j+1)$-th region at position $x$ is defined as:

\begin{equation}
	\mathcal{M}_j(x)=
	\begin{pmatrix}
 e^{i\,k_j\,x} &  e^{-i\,k_j\,x}\\
 s_j e^{i\,\theta_j}\,e^{i\,k_j\,x} & -s_j\,e^{-i\,\theta_j}\,e^{-i\,k_j\,x}
	\end{pmatrix}^{-1}
	\begin{pmatrix}
 e^{i\,k_{j+1}\,x} &  e^{-i\,k_{j+1}\,x}\\
 s_{j+1} e^{i\,\theta_{j+1}}\,e^{i\,k_{j+1}\,x} &
-s_{j+1}\,e^{-i\,\theta_{j+1}}\,e^{-i\,k_{j+1}\,x}
	\end{pmatrix}.
\end{equation}

The total transfer matrix of the system, obtained as the ordered product of the interface
matrices, is given by:
\begin{equation}\label{eq6}
    \mathcal{M}_{N/2} = \prod_{j=1}^N \mathcal{M}_j((j-1)d),
\end{equation}
where $N$ is the total number of interfaces in a structure with $N/2$ barriers.

The relation between the amplitudes of the incident, reflected, and transmitted waves is
then expressed as:
\begin{equation}
    \begin{pmatrix}
    1 \\
    r_{N/2}
    \end{pmatrix}
    =
    \mathcal{M}_{N/2}
    \begin{pmatrix}
    t_{N/2} \\
    0
    \end{pmatrix}
    =
    \begin{pmatrix}
    M_{11} & M_{12} \\
    M_{21} & M_{22}
    \end{pmatrix}
    \begin{pmatrix}
    t_{N/2} \\
    0
    \end{pmatrix}
    =
    \begin{pmatrix}
    M_{11}t_{N/2} \\
    M_{21}t_{N/2}
    \end{pmatrix}.
\end{equation}

From this, the transmission and reflection amplitudes are obtained as:
\begin{equation}
t_{N/2} = \frac{1}{M_{11}}, \quad r_{N/2} = \frac{M_{21}}{M_{11}}.
\end{equation}
Since the input and output regions are identical, the transmission and reflection
coefficients are simply:
\begin{equation}
T_{N/2} = |t_{N/2}|^2, \quad R_{N/2} = |r_{N/2}|^2.
\end{equation}

This formalism allows us to analyze in detail the resonance peaks (Fabry–Pérot–type
resonances) as well as the transmission minima arising between them, highlighting the
interplay between barrier geometry, tilt parameter, and incidence angle.
%========================================================
\subsection{Single-Barrier Transmission}
%========================================================
The transmission coefficient \( T_1 \) for a single potential barrier can be expressed as:
\begin{equation}\label{eq020}
T_{1} = \frac{\cos^2\theta\, \cos^2\phi}{\cos^2\theta\, \cos^2\phi\, \cos^2(d q_{x}) +
\sin^2(d q_{x})\, \left(1 - s \sin\theta\, \sin\phi\right)^2}.
\end{equation}

This formula shows that the transmission of Dirac fermions depends on the incidence angle
\( \theta \), the refraction angle \( \phi \), the barrier width \( d \), and the
longitudinal wave vector component \( q_x \). The denominator reflects interference
effects within the barrier: the oscillatory term \( \cos^2(d q_x) \) governs
Fabry–Pérot–like oscillations, while the factor \( \left(1 - s \sin\theta\,
\sin\phi\right)^2 \) encodes the role of chirality in the transmission
process~\cite{Katsnelson2006}.

Perfect transmission (\( T_1 = 1 \)) occurs at resonance conditions when \( \cos(d q_x) =
0 \), i.e., \( q_x d = n\pi \) with \( n \in \mathbb{Z} \). These conditions correspond to
the formation of quasi-bound states inside the barrier, leading to constructive
interference~\cite{Barbier2010,RamezaniMasir2010}. The associated resonance energies are:
\begin{equation}\label{eq012}
  \varepsilon = V + \tau k_{y} \pm \sqrt{\left( \frac{n \pi}{d} \right)^2 + k_{y}^2}.
\end{equation}
Such resonances play a central role in shaping the transmission spectrum as a function of
both energy and incidence angle, and they are often referred to as line-type resonances in
tilted Dirac systems.

In addition to resonance peaks, the formalism also predicts transmission minima. According
to Eq.~\eqref{eq020}, transmission is suppressed when:
\[
\cos(d q_{x}) = 0 \quad \Longleftrightarrow \quad d q_{x} = \frac{(2n + 1)\pi}{2}, \quad n
\in \mathbb{Z}.
\]
In this situation, destructive interference maximizes reflection, and the minimum
transmission reads:
\begin{equation}\label{eq3}
  T_1^{\min} = \frac{\cos^2 \theta\, \cos^2 \phi}{\left(1 - s \sin \theta\, \sin
\phi\right)^2}.
\end{equation}
The corresponding resonance energies are:
\begin{equation}\label{eq013}
\varepsilon = V + \tau k_{y} \pm \sqrt{\left( \frac{(2n+1) \pi}{2d} \right)^2 + k_{y}^2}.
\end{equation}

These expressions emphasize the critical role of chirality and barrier geometry in
governing interference effects, determining whether Dirac fermions undergo constructive
tunneling (perfect transmission) or destructive interference (suppressed transmission).
%========================================================
\subsection{Double-Barrier Transmission}
%========================================================
To further investigate the resonance peaks in the case of a double barrier, we derive the
analytical expression of the transmission coefficient through such a structure, building
on the results obtained for a single barrier~\cite{Katsnelson2006,Barbier2010,Choubabi2024}.
Equation~\ref{eq1}, which defines the spinor solution in a given region, involves the
following matrix:
\[
M_j(x) =
\begin{pmatrix}
e^{i k_j x} & e^{-i k_j x} \\
s_j e^{i \theta_j} e^{i k_j x} & -s_j e^{-i \theta_j} e^{-i k_j x}
\end{pmatrix},
\]
which explicitly depends on the spatial coordinate \(x\). This allows us to introduce the
following translation relation:
\begin{equation}
M_j(x + d) = M_j(x).\, T_j(d),
\end{equation}
with the translation matrix
\[
T_j(d) =
\begin{pmatrix}
e^{i k_j d} & 0 \\
0 & e^{-i k_j d}
\end{pmatrix}.
\]
The inverse relation then reads:
\begin{equation}
M_j^{-1}(x + d) = T_j^{-1}(d).\, M_j^{-1}(x).
\end{equation}

Using this relation and equation~\ref{eq6}, the total transfer matrix of the double
barrier is written as:
\begin{equation}\label{eq7}
\mathcal{M}_{2} = \prod_{j=1}^{4} \mathcal{M}_j((j-1)d) = \mathcal{M}_1.\, T_1^{-1}(d).\,
\mathcal{M}_1.\, T_1(2d),
\end{equation}
where the single–barrier transfer matrix is
\begin{equation}
\mathcal{M}_1 = \prod_{j=1}^{2} \mathcal{M}_j((j-1)d) = M_1(0)^{-1} \cdot M_2(0) \cdot
M_2(d)^{-1} \cdot M_1(d).
\end{equation}

Denoting the components of \(\mathcal{M}_1\) by \( m_{ij} = (\mathcal{M}_1)_{ij} \), the
total double–barrier transfer matrix becomes
\begin{equation}
\mathcal{M}_2 =
\begin{pmatrix}
B_{11} & B_{12} \\
B_{21} & B_{22}
\end{pmatrix}.
\begin{pmatrix}
e^{i k_x d} & 0 \\
0 & e^{-i k_x d}
\end{pmatrix},
\end{equation}
with the coefficients
\begin{equation}
\begin{aligned}
B_{11} &= m_{11}^2 e^{-i k_x d} + m_{12} m_{21} e^{i k_x d}, \\
B_{12} &= m_{11} m_{12} e^{-i k_x d} + m_{12} m_{22} e^{i k_x d}, \\
B_{21} &= m_{21} m_{11} e^{-i k_x d} + m_{22} m_{21} e^{i k_x d}, \\
B_{22} &= m_{21} m_{12} e^{-i k_x d} + m_{22}^2 e^{i k_x d}.
\end{aligned}
\end{equation}

The key element for transmission is the top-left entry of \(\mathcal{M}_2\):
\begin{equation}\label{eq030}
(\mathcal{M}_2)_{11}=M_{11} = \frac{1}{t_2}= m_{11}^2 + m_{12} m_{21} e^{2 i k_x d},
\end{equation}
where
\begin{equation}\label{eq031}
m_{11} = \frac{1}{t_1}=\frac{1}{|t_1|} e^{-i\phi_{t_{1}}}, \quad
m_{12} = \frac{r_1^*}{t_1^*}, \quad
m_{21} = \frac{r_1}{t_1}.
\end{equation}

By substituting these coefficients, the double–barrier transmission probability is
obtained as
\begin{equation}
T_{2} = t_2 t_2^* = \frac{T_1^2}{4R_1 \cos^2\left(2 k_x d+ \phi_{t_{1}}\right) + T_1^2},
\end{equation}
where \(T_1\) and \(R_1\) are the transmission and reflection coefficients of a single
barrier.

This expression highlights the Fabry–Pérot–like interference produced by multiple
reflections inside the central well~\cite{RamezaniMasir2010,Sun2011,Dakhlaoui2021}. Constructive
interference occurs when
\begin{equation}\label{eq011}
2 k_x d + \phi_{t_{1}} = \left(2n+1\right)\frac{\pi}{2}, \quad n \in \mathbb{Z},
\end{equation}
leading to perfect transmission \(T_2=1\). In contrast, destructive interference arises
when
\begin{equation}\label{eq0111}
2 k_x d + \phi_{t_{1}} = n\pi, \quad n \in \mathbb{Z},
\end{equation}
giving the minimum transmission
\begin{equation}
T_2^{\mathrm{min}} = \frac{T_1^2}{T_1^2 + 4 R_1}.
\end{equation}

As $T_1 \to 1$, the minima increase toward unity, reflecting the nearly transparent nature
of the barriers even between resonances. This interplay between barrier transparency,
accumulated phase, and cone tilt produces a spectrum of sharp resonances and
antiresonances, which are highly tunable via barrier width \(d\), potential height \(V\),
tilt parameter \(\tau\), and incidence angle.

This analysis shows how line-type resonances and Fabry–Pérot interference govern tunneling
in graphene–tilted cone heterostructures, providing a theoretical foundation for the
numerical results presented in the next section.
%========================================================
\section{Results and Discussions}
%========================================================
The study of transport properties in graphene-based heterostructures has revealed a broad
range of quantum interference phenomena, including Klein tunneling, Fabry–Pérot–like
oscillations, and resonance-assisted tunneling~\cite{Katsnelson2006,Young2009,Allain2011}.
When tilted Dirac cones are present, additional degrees of freedom arise due to the
anisotropic dispersion, which modifies tunneling conditions and gives rise to novel
resonance features~\cite{Goerbig2008,Montambaux2009}.
Double-barrier configurations are particularly suitable to explore such effects, since the
intermediate region between the barriers acts as a quantum cavity supporting quasi-bound
states that strongly affect transmission~\cite{Alhaidari2012,Sun2011,Baltateanu2019,RamezaniMasir2010}.

A central focus of this work is the characterization of the so-called \textit{line-type
resonances}, which result from the interplay between propagating and evanescent modes in
tilted Dirac materials. These resonances, absent in conventional semiconductors, play a
crucial role in shaping the transmission spectrum and highlight the unique interference
physics of Dirac fermions in graphene and related systems~\cite{Alhaidari2012,Sun2011,Baltateanu2019}.

In the following subsections, we provide a detailed numerical analysis of these transport
features, emphasizing the influence of cone tilt, barrier geometry, and incidence
conditions on the resonance spectrum.
%========================================================
\subsection{General Transmission Features}
%========================================================
Understanding the general features of quantum transport in Dirac materials is essential
for both fundamental physics and device engineering. In graphene and related
two-dimensional systems, potential barriers give rise to a wide variety of tunneling
regimes, from Klein tunneling at normal incidence to Fabry–Pérot–like oscillations under
resonant conditions~\cite{Katsnelson2006,Young2009,Allain2011}.
With tilted Dirac cones, transport becomes even richer due to anisotropic dispersion and
modified interference conditions, which strongly influence the resonance
spectrum~\cite{Goerbig2008,Nguyen2018,AlMarzoog2024}. In this context, \textit{line-type resonances} emerge as a
distinctive signature of double-barrier heterostructures, producing sharp transmission
peaks even in classically forbidden energy regions.

Here, we present numerical results for Dirac fermion transport through a symmetric
double-barrier structure aligned along the $x$-axis. The tilt parameter $\tau$ of the
Dirac cone is assumed to vanish ($\tau=0$) outside the barriers and in the intermediate
region, while it may take finite values ($\tau \neq 0$) inside the potential zones (see
figure).

From a practical perspective, particularly for graphene-based resonant tunneling devices
(RTDs) and nanoscale energy filters, analyzing the transmission spectrum and the resonance
peaks predicted by the analytical model is of prime importance~\cite{Britnell2013,Zhang2023}. The
so-called line-type resonances appear for all values of $\tau$, originating from the
repeated back-and-forth motion of Dirac fermions inside the central well, confined between
two evanescent zones created by the barriers. These resonances correspond to quasi-bound
states whose hybridization with propagating states produces sharp transmission peaks
extending into forbidden zones.

Physically, this mechanism is analogous to a Fabry–Pérot interferometer of a special kind:
rather than propagating waves reflecting between barriers, it is the evanescent modes that
act as effective “mirrors,” confining fermions in the well and creating interference
patterns~\cite{Allen2017,Campos2012}. In this sense, line-type resonances can be seen as a new
class of Fabry–Pérot resonances. The analogy can also be drawn with a vibrating string,
where standing waves are sustained by repeated reflections at the ends of the string.

The position, width, and intensity of these resonances depend strongly on the tilt
parameter $\tau$, as well as on the structural parameters of the device such as barrier
height and thickness. This strong tunability paves the way for applications in resonant
tunneling devices and energy-selective filters, offering precise control of transmission
as a function of energy and incidence angle~\cite{Feenstra2012,Britnell2013}.
%========================================================
\subsection{Transmission in the $(\varepsilon, k_y)$ Plane}
%========================================================
The transmission spectrum of Dirac fermions across a double-barrier structure exhibits a
rich variety of features when represented in the $(\varepsilon,k_y)$ plane. Such maps are
widely used to visualize Fabry–Pérot oscillations, Klein tunneling, and the emergence of
bound states in graphene and related Dirac
materials.
The introduction of tilted Dirac cones further enriches these spectra by breaking the
symmetry between positive and negative transverse momenta, thus modifying resonance
conditions and the distribution of quasi-bound states~\cite{Goerbig2008}.
\begin{figure}[!h]\centering
    \subfloat[$\tau=0$]{
    \hspace{-0.5cm}\includegraphics[scale=0.16]{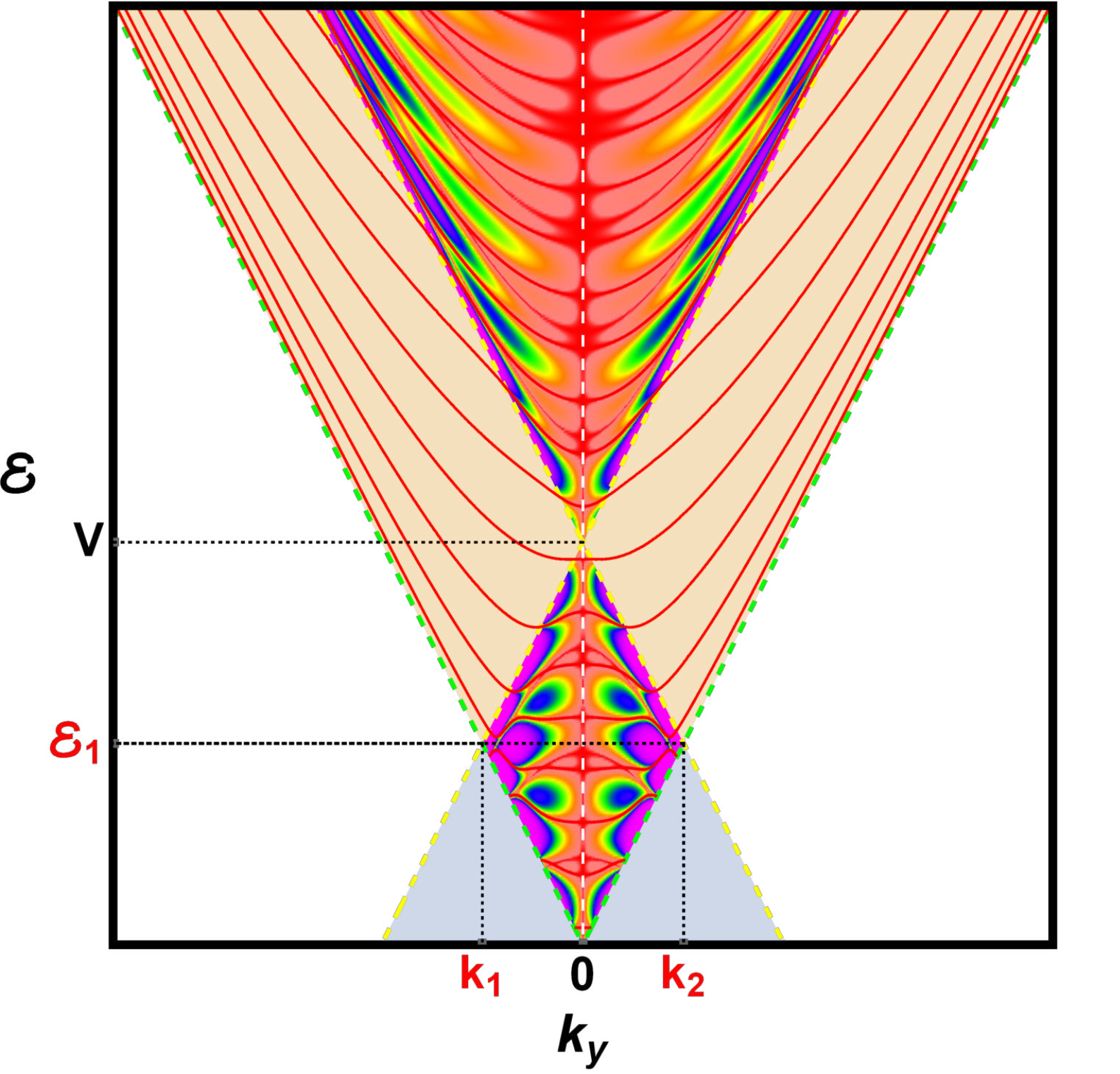}
    \label{fig04:SubFigA}}
    \subfloat[][$\tau=0.5$]{
    \hspace{-0.37cm}\includegraphics[scale=0.16]{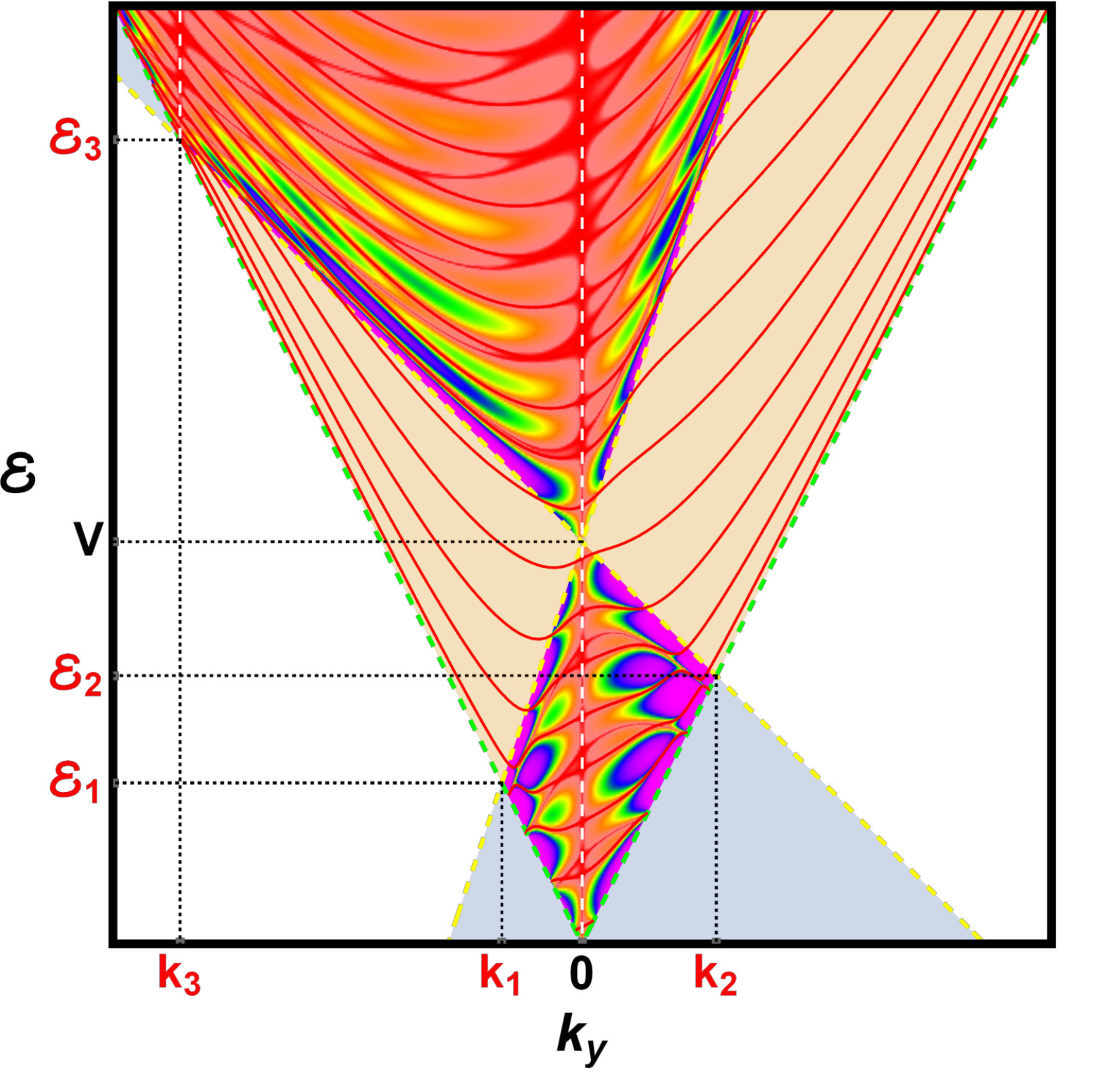}
    \label{fig04:SubFigB}}
    \subfloat[][$\tau=1$]{
    \hspace{-0.4cm}\includegraphics[scale=0.16]{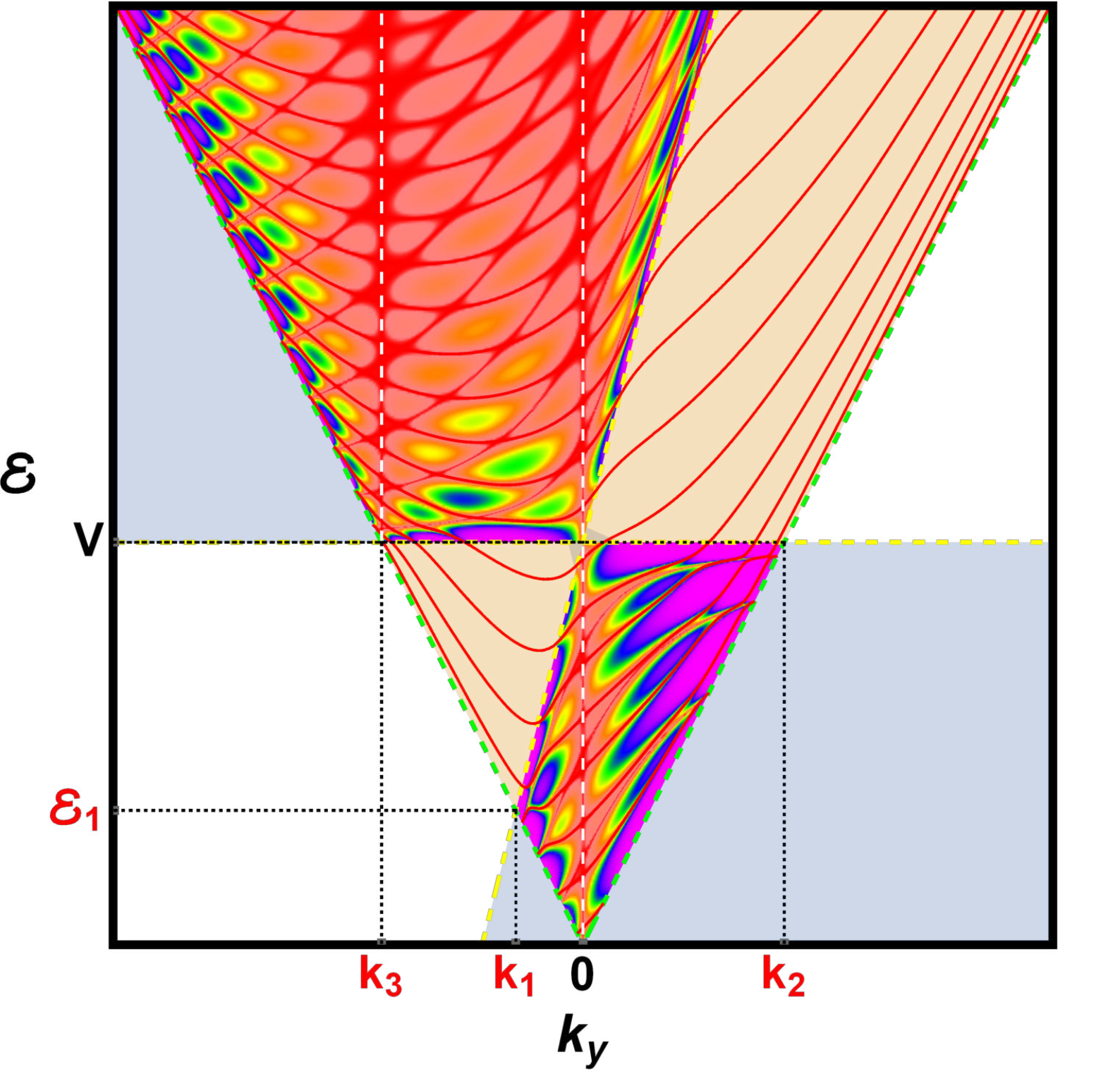}\label{fig04:SubFigC}}
     \subfloat[][$\tau=2$]{
     \hspace{-0.27cm}\includegraphics[scale=0.16]{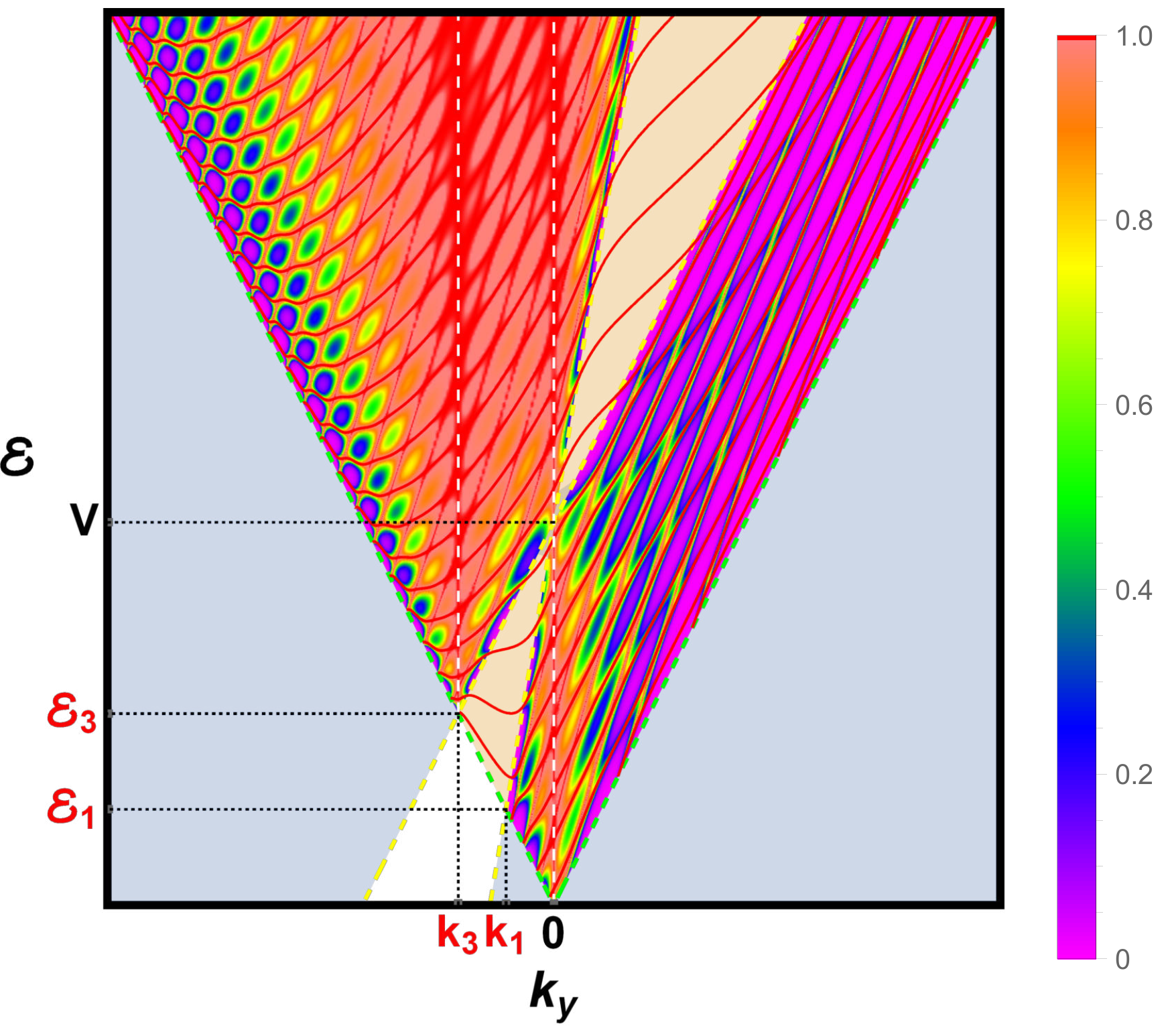}\label{fig04:SubFigD}}\\
\subfloat[$\tau=0$]{
    \hspace{-0.8cm}\includegraphics[width=0.25\linewidth]{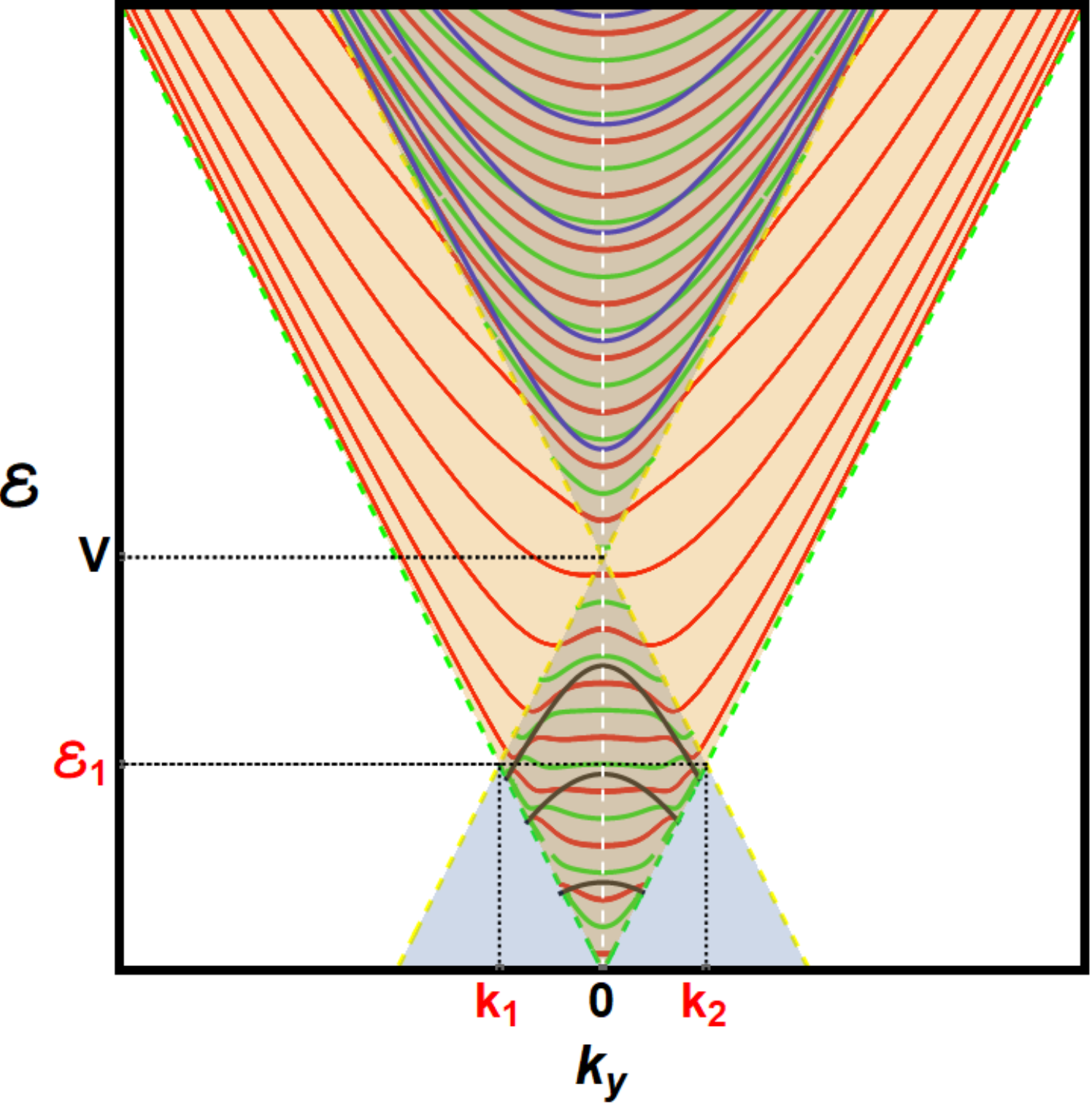}\label{fig04:SubFigE}}
    \subfloat[][$\tau=0.5$]{
   \includegraphics[width=0.25\linewidth]{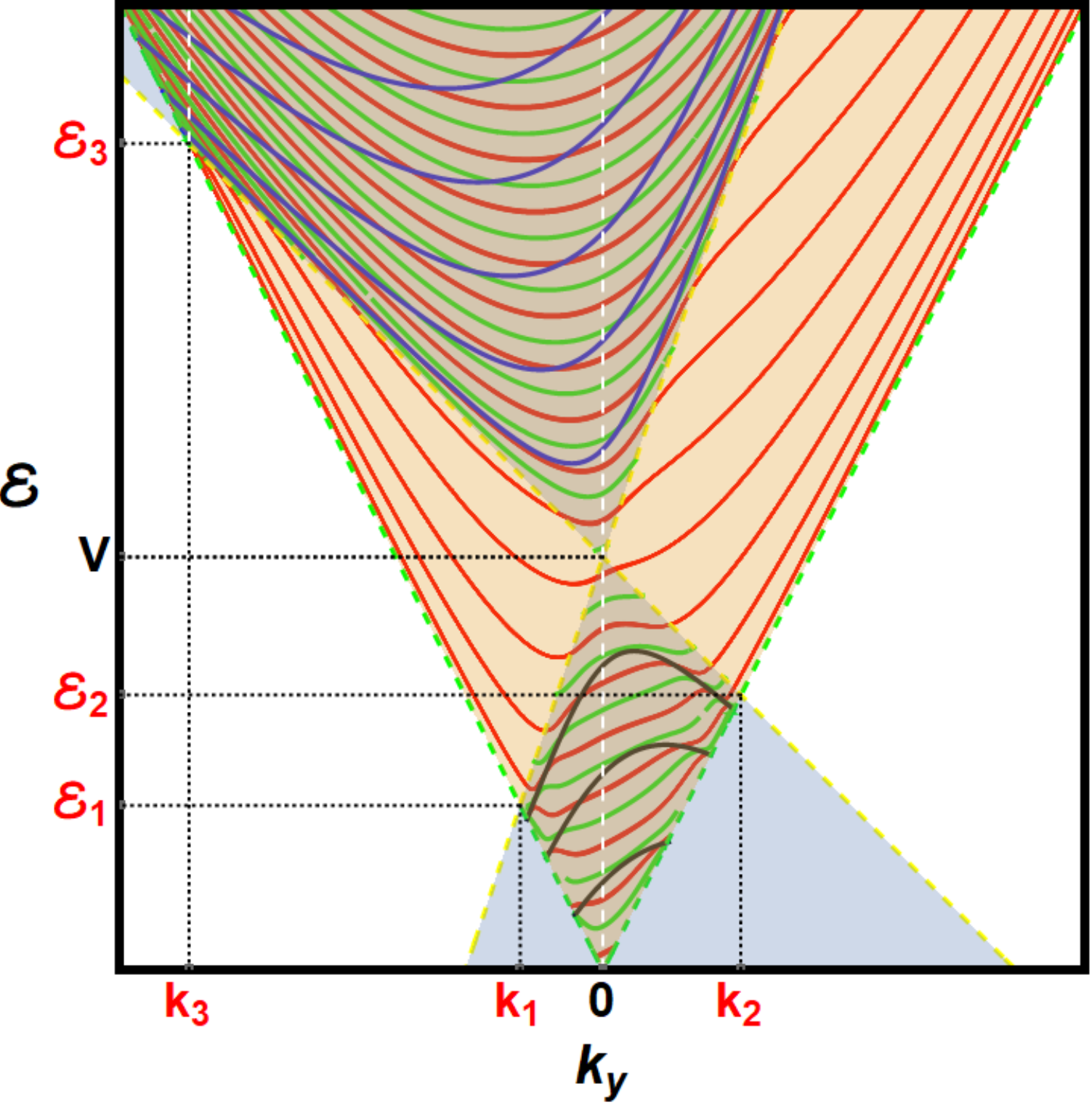}\label{fig04:SubFigF}}
   \subfloat[][$\tau=1$]{
   \includegraphics[width=0.25\linewidth]{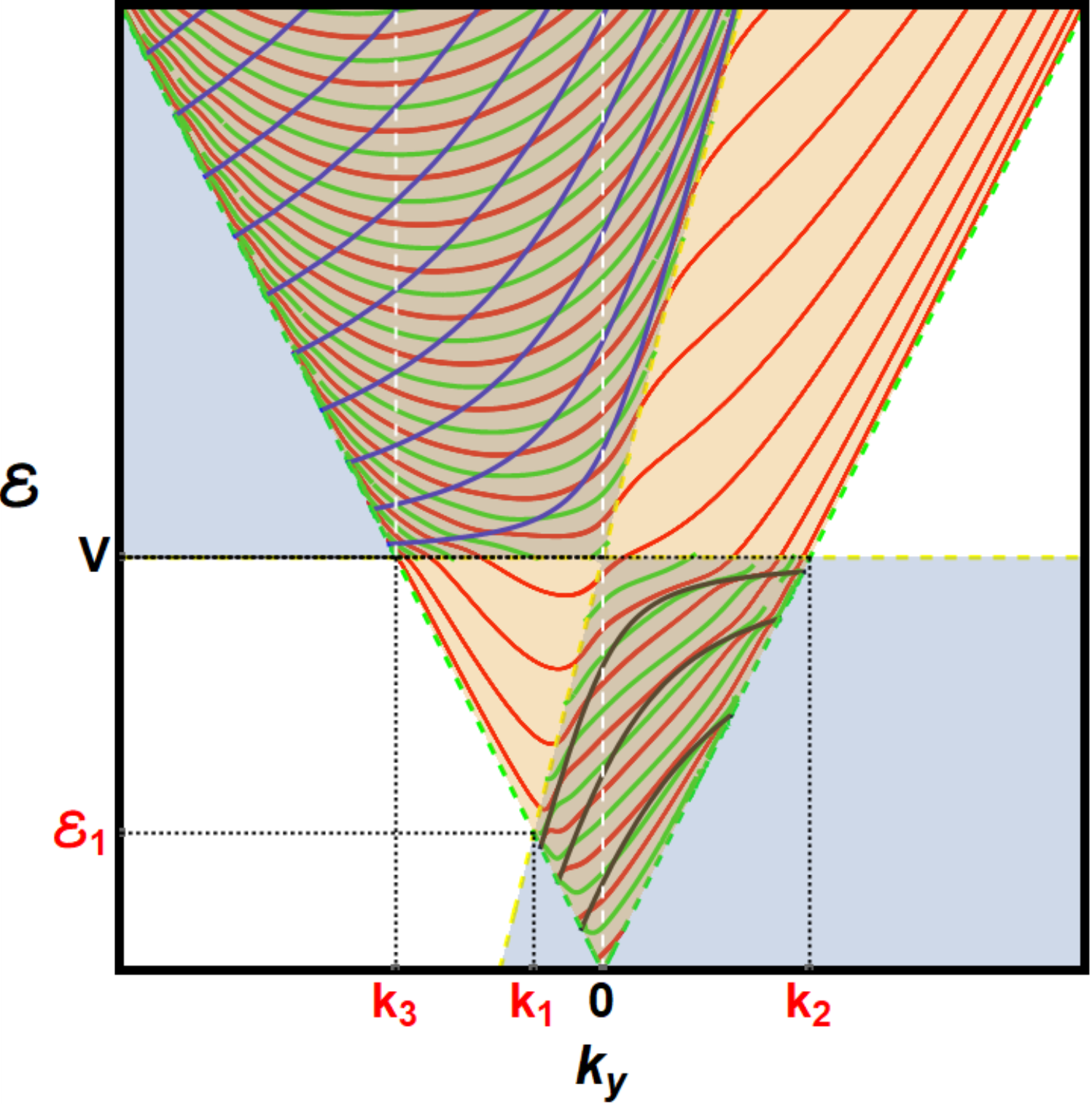}\label{fig04:SubFigG}}
   \subfloat[][$\tau=2$]{
   \includegraphics[width=0.25\linewidth]{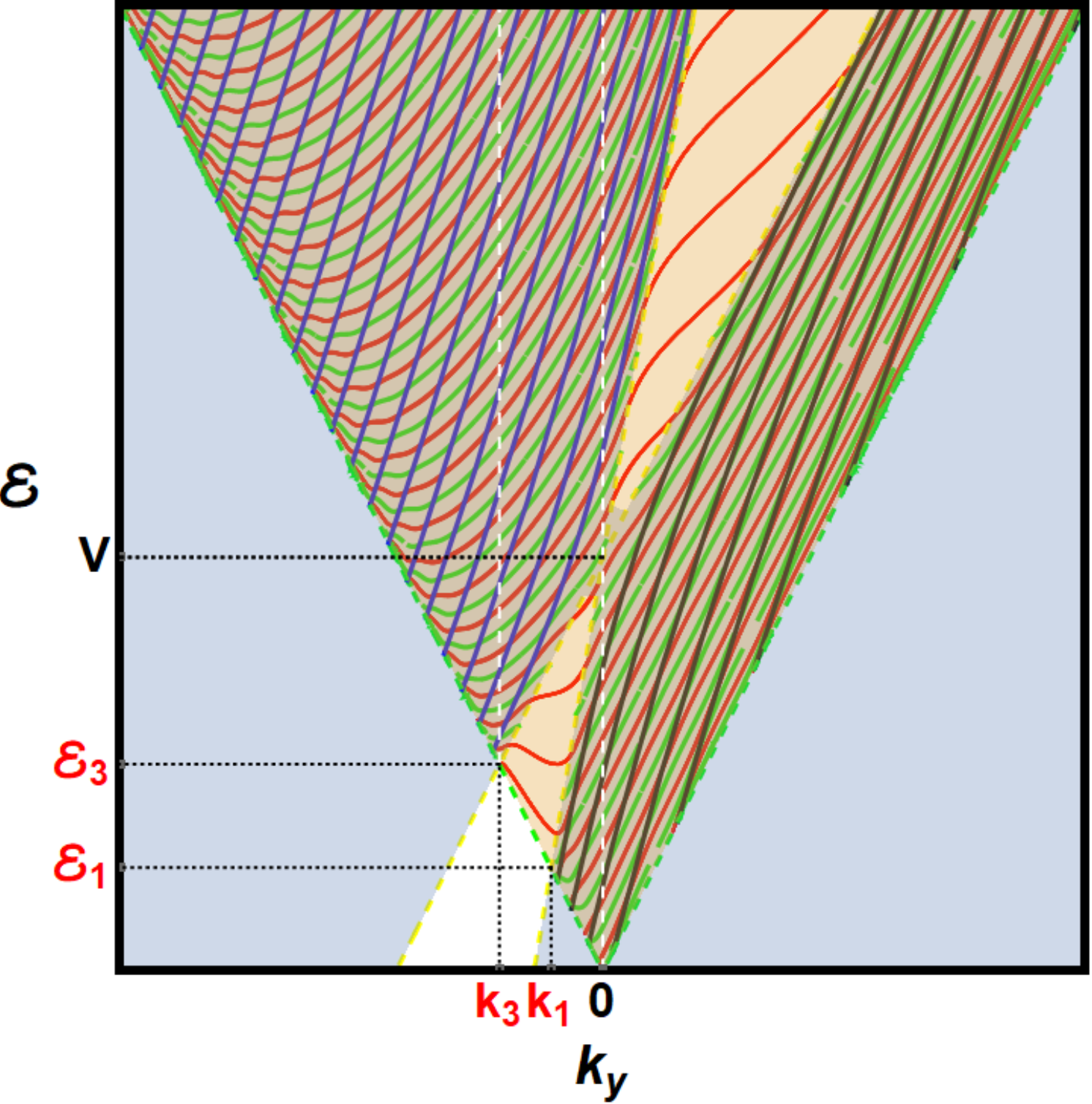}\label{fig04:SubFigH}}
    \caption{(Color online) Density plot of the transmission probability $T$ in the
($\varepsilon$, $k_y$) plane for a double-barrier structure. Both barriers have the same
height $V=3$, the same width $d$, and are separated by a distance $d=4$. The plot is shown
for four values of the tilt parameter $\tau$. The incident energy $\varepsilon$ is shown
on the vertical axis and the transverse momentum $k_y$ on the horizontal axis.}
\label{fig04}
\end{figure}
Figure~\ref{fig04} illustrates the effect of the tilt parameter $\tau$ on the transmission
density. For $\tau=0$, the spectrum is perfectly symmetric with respect to $k_y=0$:
Fabry–Pérot oscillations manifest as regular parabolic patterns in the allowed regions,
while line-type resonances extend as branches into the forbidden (yellow) zones. For
$\tau=0.5$, the line-type resonances penetrate deeper into the forbidden regions, as a
direct consequence of the Fermi surface shift induced by tilt. At $\tau=1$, the critical
energies $\varepsilon_2=\tfrac{V}{2-\tau}$ and $\varepsilon_3=\tfrac{V}{\tau}$ coincide
($\varepsilon_2=\varepsilon_3=V$), producing a straight line that marks the location of
Dirac points. For $\tau=2$, the critical coordinates
$(k_2,\varepsilon_2)=(\tfrac{V}{2-\tau},\tfrac{V}{2-\tau})$ diverge, leading to a spectrum
enriched with peaks in the allowed zones, while the number of bound states in the
forbidden regions is strongly reduced. These features highlight the role of cone tilt in
reshaping transport anisotropy.

The distribution of line-type resonances is also strongly $\tau$-dependent. For $0 < \tau
\leq 1$, in the forbidden zone on the left-hand side of $k_y=0$, these resonances connect
the conduction and valence bands of the tilted cone. In contrast, for $\tau > 1$, such
connections appear in both forbidden zones, reflecting a more intricate interplay between
propagating and evanescent modes. Reversing the sign of the tilt mirrors the entire
transmission profile with respect to $k_y=0$, as expected from symmetry arguments.

Remarkably, the Klein paradox persists across all these configurations: perfect
transmission ($T=1$) is always observed at normal incidence ($k_y=0$), as well as at
$k_y=-V/\tau$, where adjacent regions share the same effective refractive index
($n_1=n_2$), with $n_1=\varepsilon$ and $n_2=\varepsilon-\tau k_y -V$ ~\cite{Choubabi2024}. In these situations, barriers become completely transparent regardless of
their width, height, or the incident energy~\cite{Beenakker2008,Katsnelson2006,Allain2011}.

Beyond this universal transparency, additional families of resonances can be identified.
Barrier resonances appear when the longitudinal wave vector in the barrier, $q_x$,
satisfies $q_x d = n\pi$, yielding constructive interference.
Well resonances arise when the longitudinal wave vector in the well, $k_x$,
satisfies Eq.~\ref{eq011}, leading to standing-wave–like quasi-bound states.

These resonances give rise to perfect transmission peaks distributed in two regimes. For
$\varepsilon < V$, the peaks correspond to tunneling resonances (black parabolic curves in
Figs.~\ref{fig04:SubFigE}--\ref{fig04:SubFigH}), with energies determined by
Eq.~\ref{eq012}. For $\varepsilon > V$, the peaks are associated with propagating
resonances (blue curves), also given by Eq.~\ref{eq012}. Bound states confined between the
barriers yield additional perfect transmission peaks when Eq.~\ref{eq011} is satisfied
(red curves), while minima of the transmission coefficient follow Eq.~\ref{eq0111} (green
curves).

Altogether, these results demonstrate that tilted Dirac cones not only shift and reshape
resonance patterns, but also enrich the transmission spectrum with highly anisotropic
line-type resonances that can be exploited for device applications such as angle-selective
filters and resonant tunneling diodes~\cite{Britnell2013,Feenstra2012}.
%========================================================
\subsection{Energy-Resolved Cuts at Fixed $k_y$}
%========================================================
To gain further insight into the role of the tilt parameter on transmission, it is
instructive to analyze energy-resolved cuts of the density maps presented in
Figure~\ref{fig04}. This so-called \textit{cutting method} has been widely used to
identify tunneling regimes and resonance structures in graphene-based double-barrier
systems~\cite{RamezaniMasir2010,Kundu2010}.

\begin{figure}[!h]\centering
    \subfloat[$\tau=0$,\quad $k_{y}=1$]{
   \hspace{-0.7cm}\includegraphics[scale=0.17]{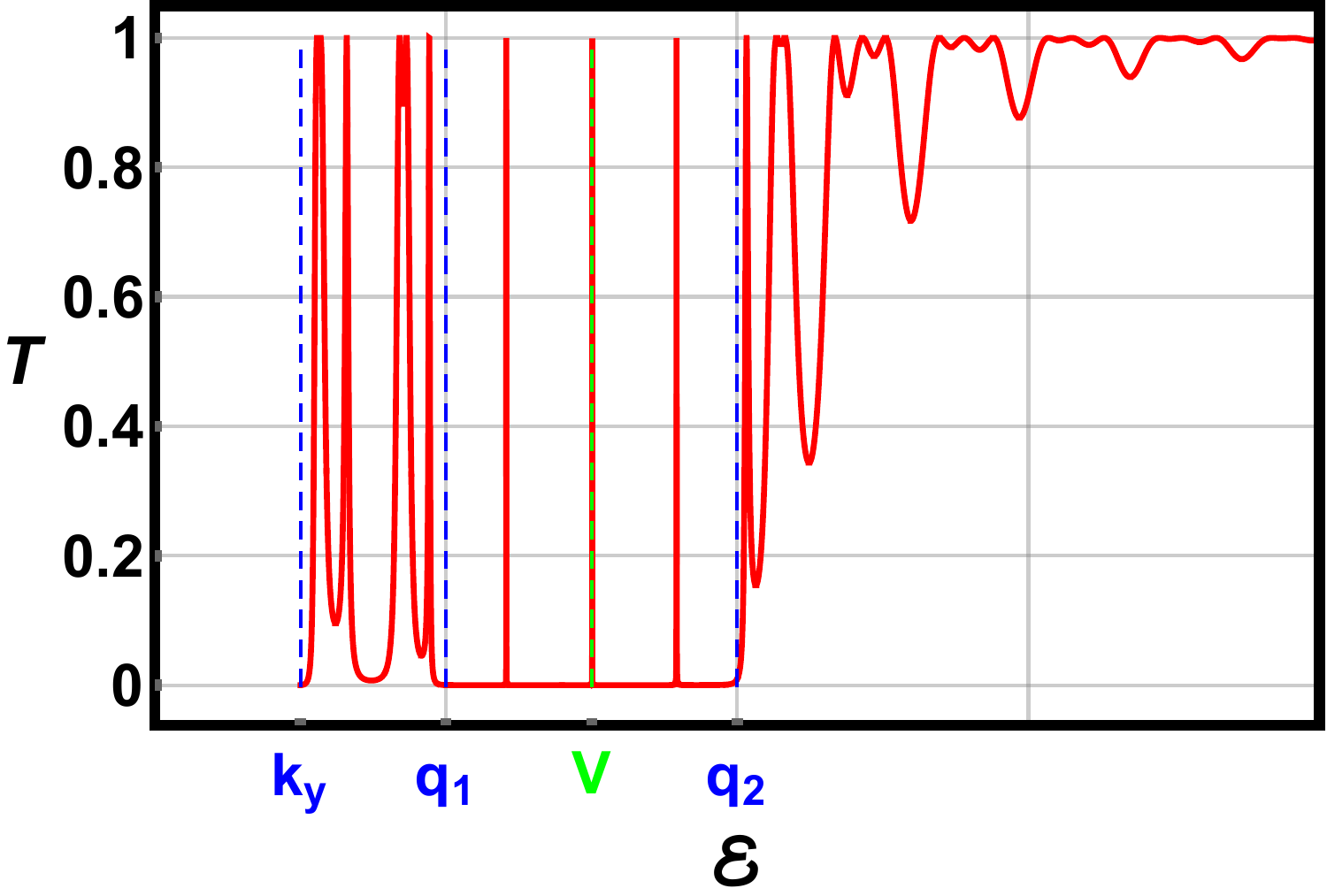}
    \label{fig05:SubFigA}}
    \subfloat[$\tau=0.5$,\quad $k_{y}=1$]{
    \includegraphics[scale=0.17]{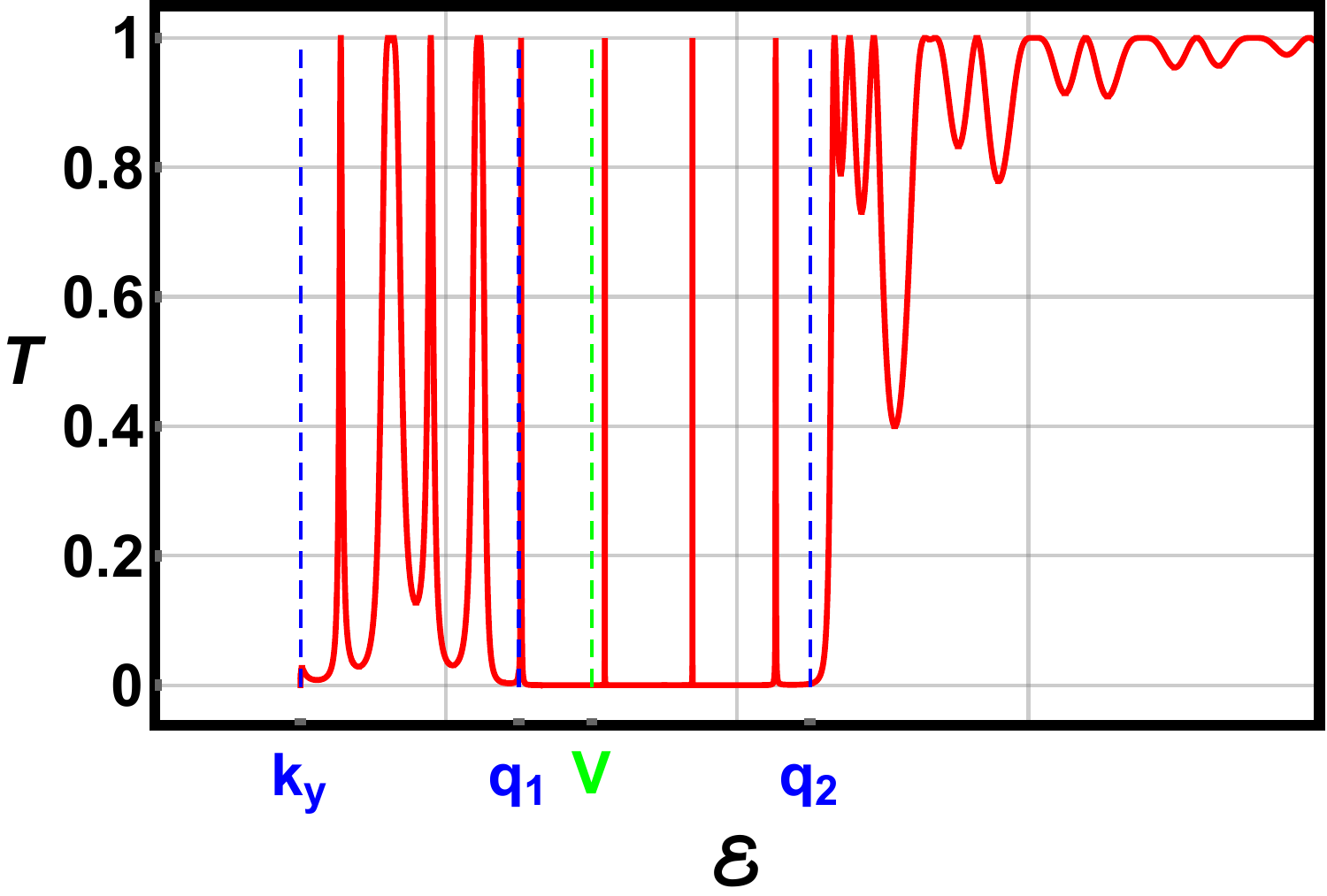}
    \label{fig05:SubFigB}}
    \subfloat[$\tau=1$,\quad $k_{y}=1$]{
    \includegraphics[scale=0.17]{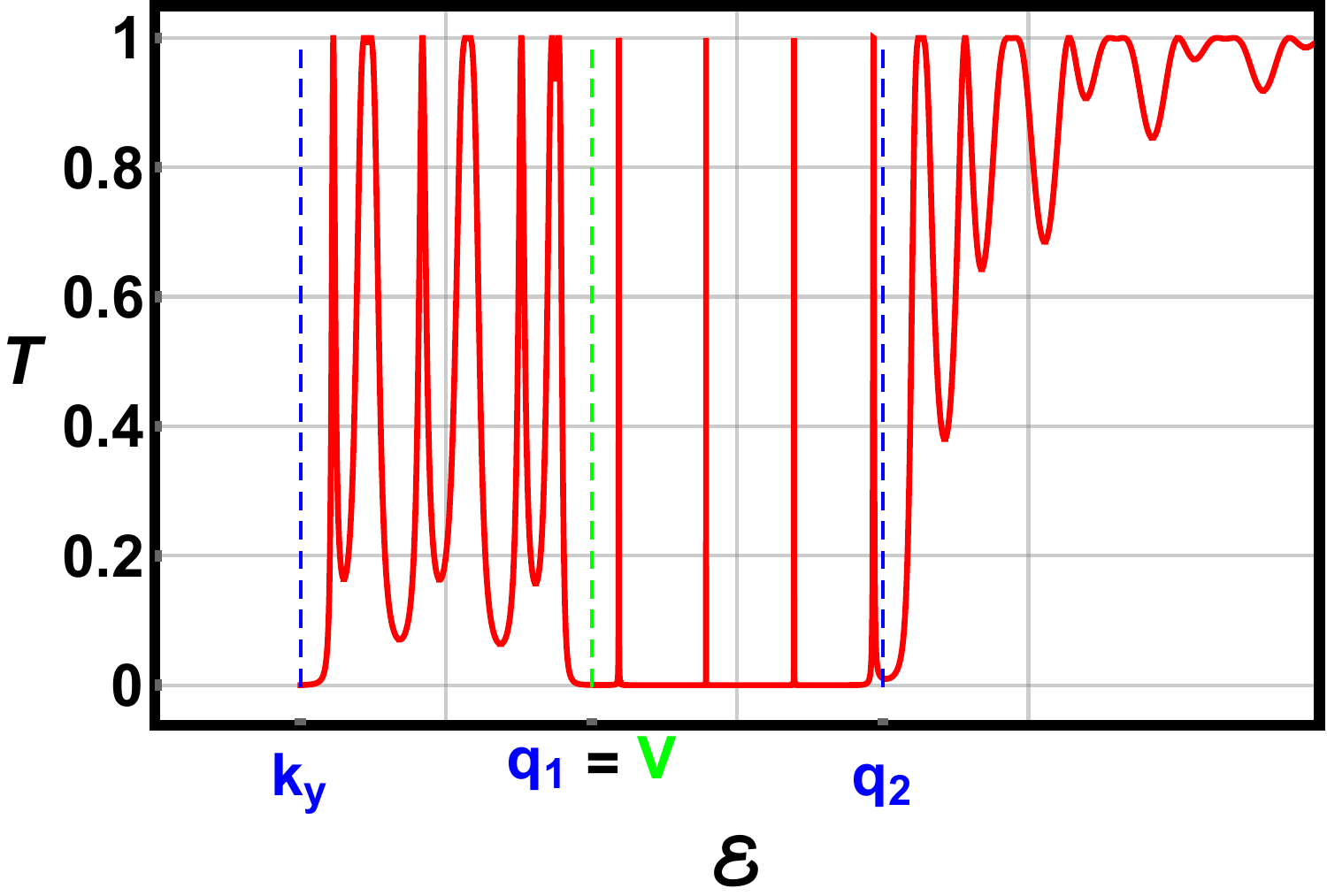}\label{fig05:SubFigC}}
     \subfloat[$\tau=2$,\quad $k_{y}=1$]{
     \includegraphics[scale=0.17]{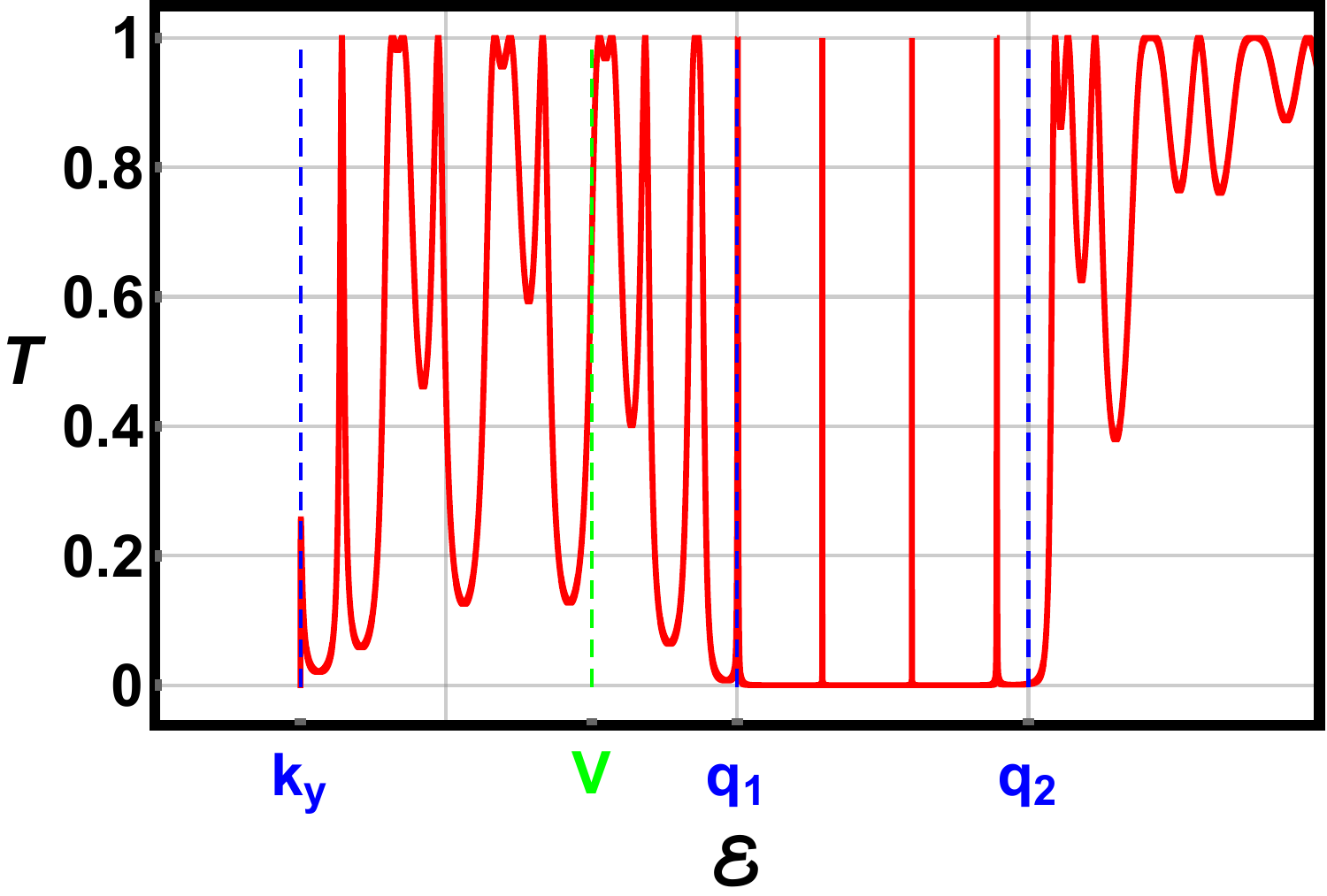}\label{fig05:SubFigD}}\\
      \subfloat[$\tau=0$,\quad $k_{y}=-1$]{
    \hspace{-0.7cm}\includegraphics[scale=0.17]{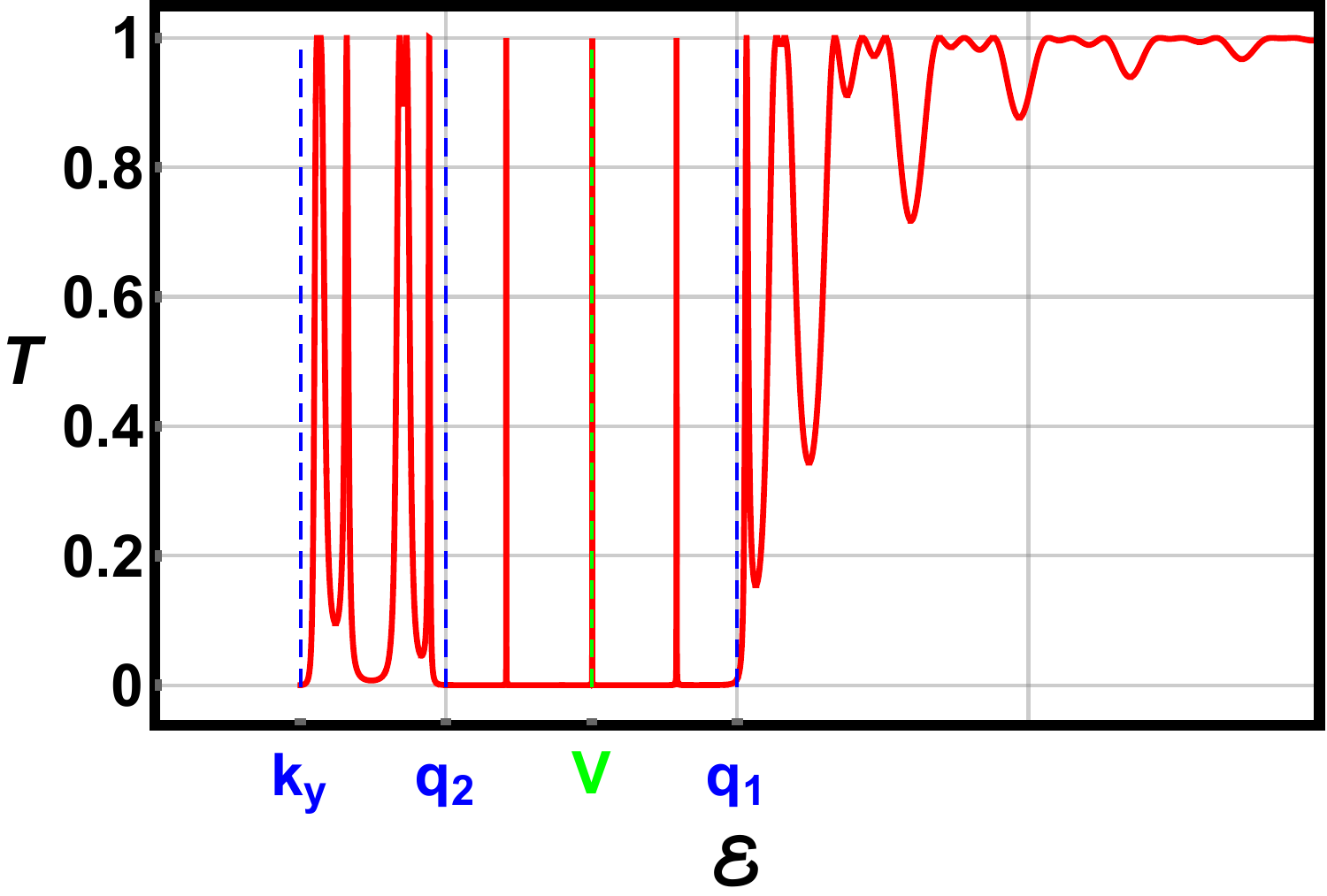}
    \label{fig05:SubFigE}}
    \subfloat[$\tau=0.5$,\quad $k_{y}=-1$]{
    \includegraphics[scale=0.17]{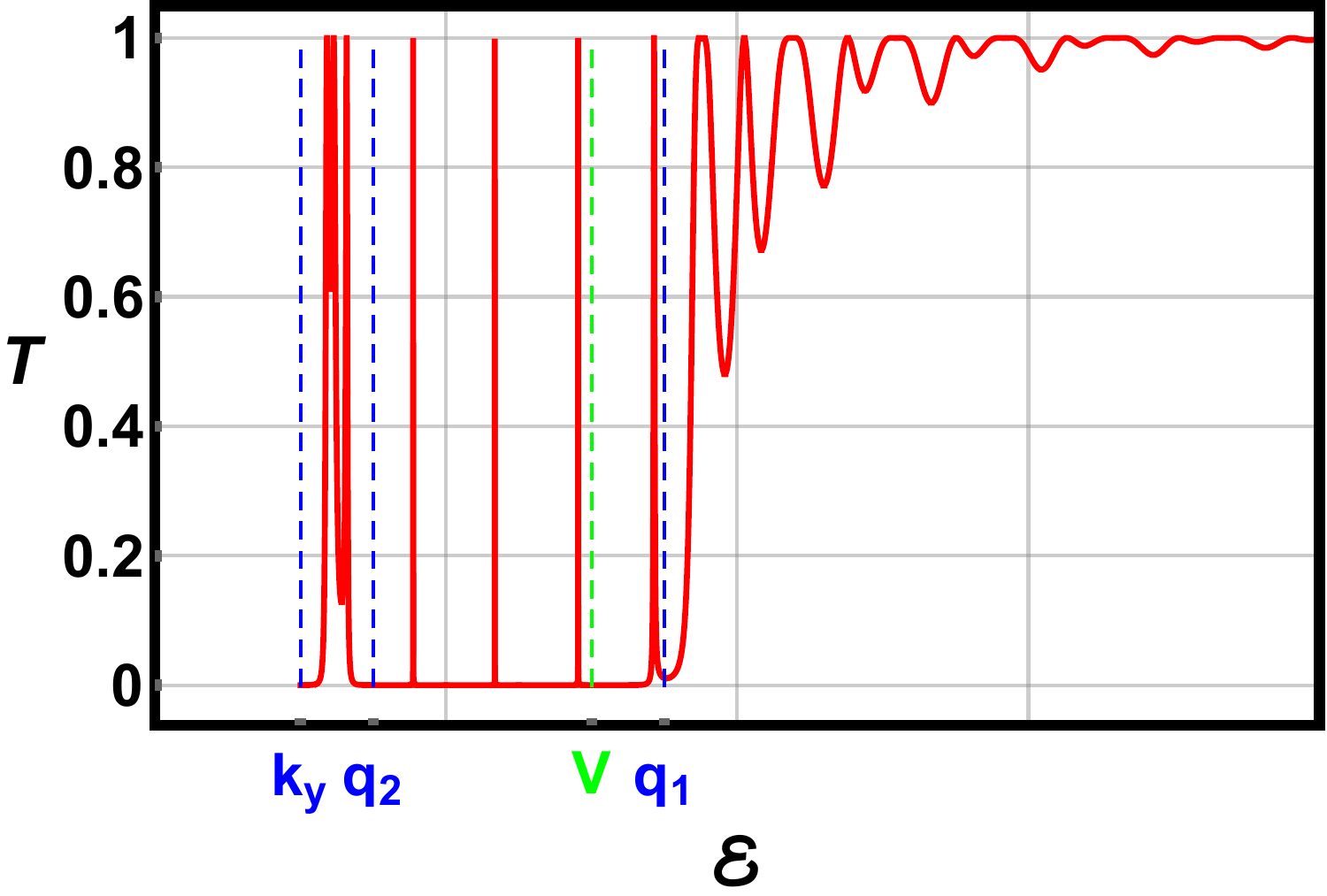}
    \label{fig05:SubFigF}}
    \subfloat[$\tau=1$,\quad $k_{y}=-1$]{
    \includegraphics[scale=0.17]{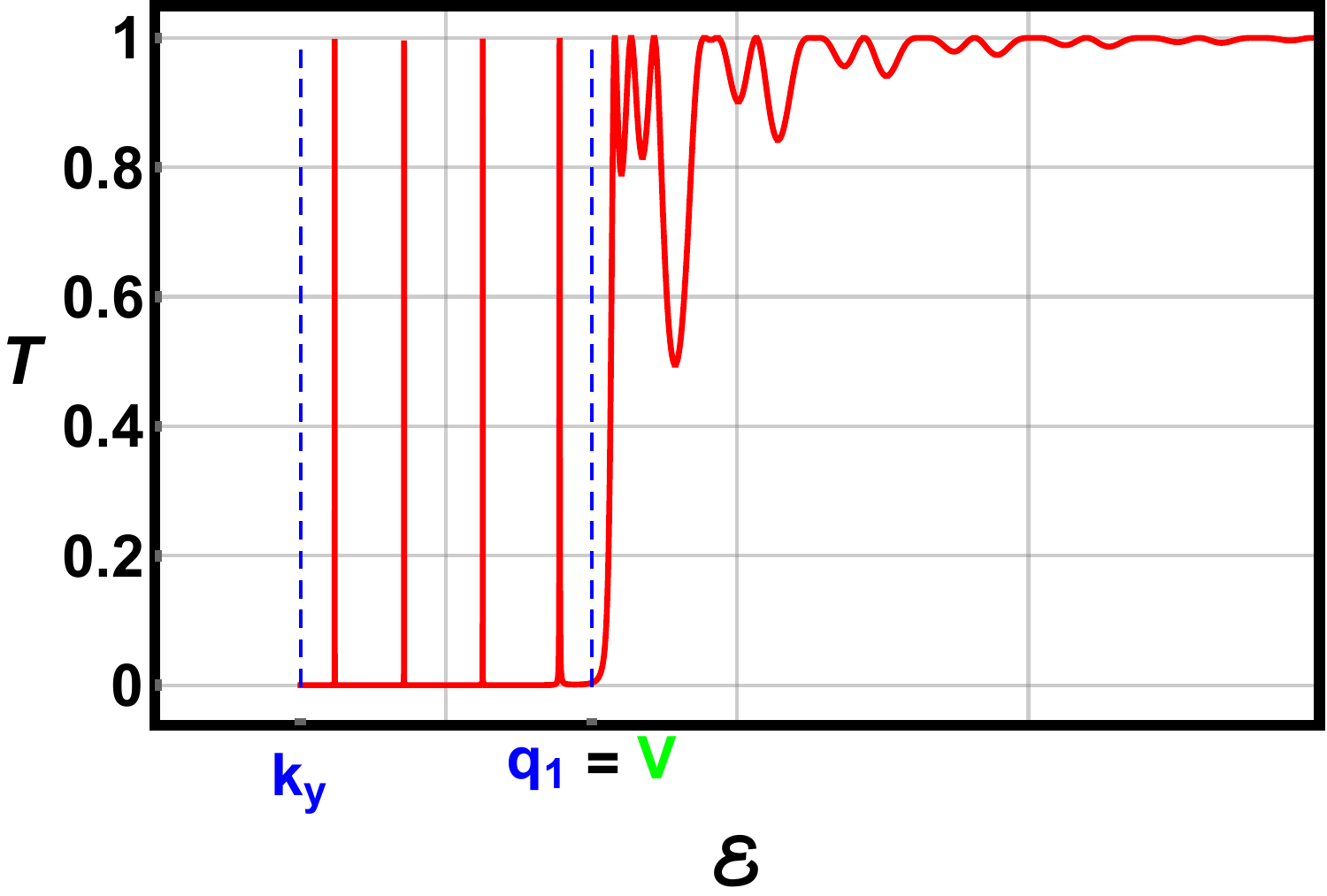}\label{fig05:SubFigG}}
     \subfloat[$\tau=2$,\quad $k_{y}=-1$]{
     \includegraphics[scale=0.17]{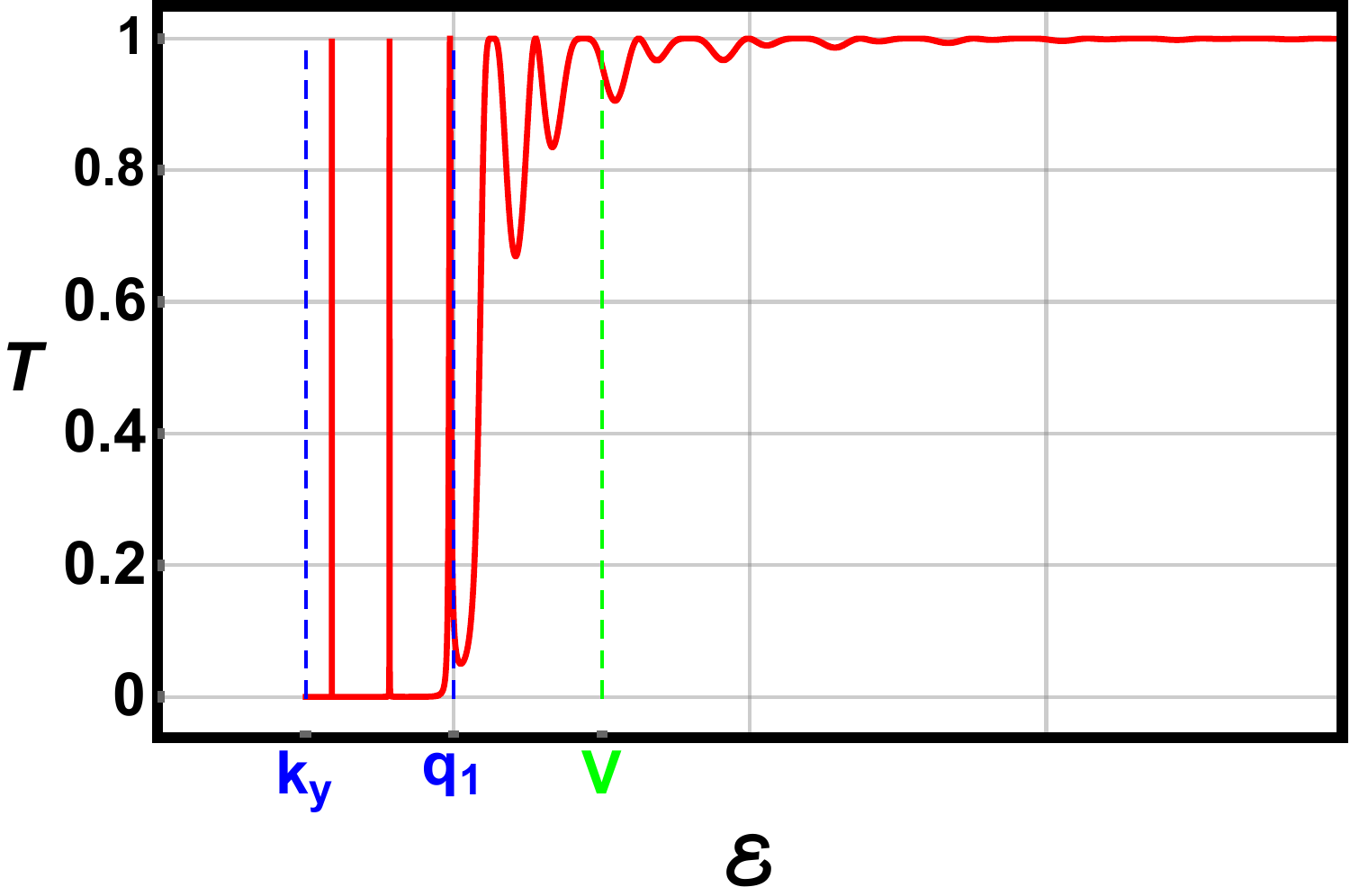}\label{fig05:SubFigH}}
    \caption{Transmission probability $T(\varepsilon)$ as a function of the incident
energy for fixed $k_{y}=\pm 1$ and different tilt parameters $\tau=0,0.5,1,2$. The
barriers have a height $V=3$, a width $d=4$, and are separated by the same distance $d$.
The upper panels correspond to $k_{y}=1$, while the lower panels are for $k_{y}=-1$. The
parameters $q_1$ and $q_2$ are defined as $q_{1}=k_{y}(\tau-1)+V$ and
$q_{2}=k_{y}(1+\tau)+V$.}
\label{fig05}
\end{figure}

Figure~\ref{fig05} shows transmission spectra at fixed transverse momenta $k_y=\pm 1$ for
different values of $\tau$. Four distinct regimes can be distinguished:
(i) a forbidden zone ($0\leq\varepsilon<|k_y|$) where no propagation is possible,
(ii) a tunneling region ($|k_y|\leq\varepsilon<q_1$) where constructive interference
between evanescent and propagating waves allows partial transmission,
(iii) a transmission gap ($q_1\leq\varepsilon<q_2$) characterized by sharp line-type
resonances associated with quasi-bound states inside the central well, and
(iv) a high-energy region ($\varepsilon\geq q_2$) where oscillatory Fabry–Pérot–like
transmission converges toward unity~\cite{Beenakker2008,RamezaniMasir2010,Shytov2008,Jellal2012}.

For $k_y=1$ [Figs.~\ref{fig05}(a–d)], increasing $\tau$ shifts the transmission gap toward
higher energies, while the number of line-type resonances inside the gap remains nearly
constant. In contrast, for $k_y=-1$ [Figs.~\ref{fig05}(e–h)], the gap shifts to lower
energies and the number of resonant peaks decreases once $\tau>1$, reflecting the
anisotropic nature of tilted Dirac cones.
In both cases, the tunneling window broadens for positive $k_y$ but shrinks for negative
$k_y$ as $\tau$ increases, confirming the strong directional dependence of transport.

Finally, Figures~\ref{fig04} and \ref{fig05} together demonstrate the robustness of the
Klein paradox in tilted Dirac systems: perfect transmission ($T=1$) occurs both at normal
incidence ($k_y=0$) and at the condition $k_y=-V/\tau$, where the effective refractive
indices of adjacent regions match ($n_1=n_2$). Under these circumstances, the barriers
become completely transparent, regardless of their height, width, or the incident
energy~\cite{Beenakker2008,Katsnelson2006,Allain2011}.
%========================================================
\subsection{Global View and Symmetry Breaking}
\begin{figure}[!h]\centering
    \subfloat[$k_{y}=1.5$]{
   \hspace{-0.4cm}\includegraphics[scale=0.365]{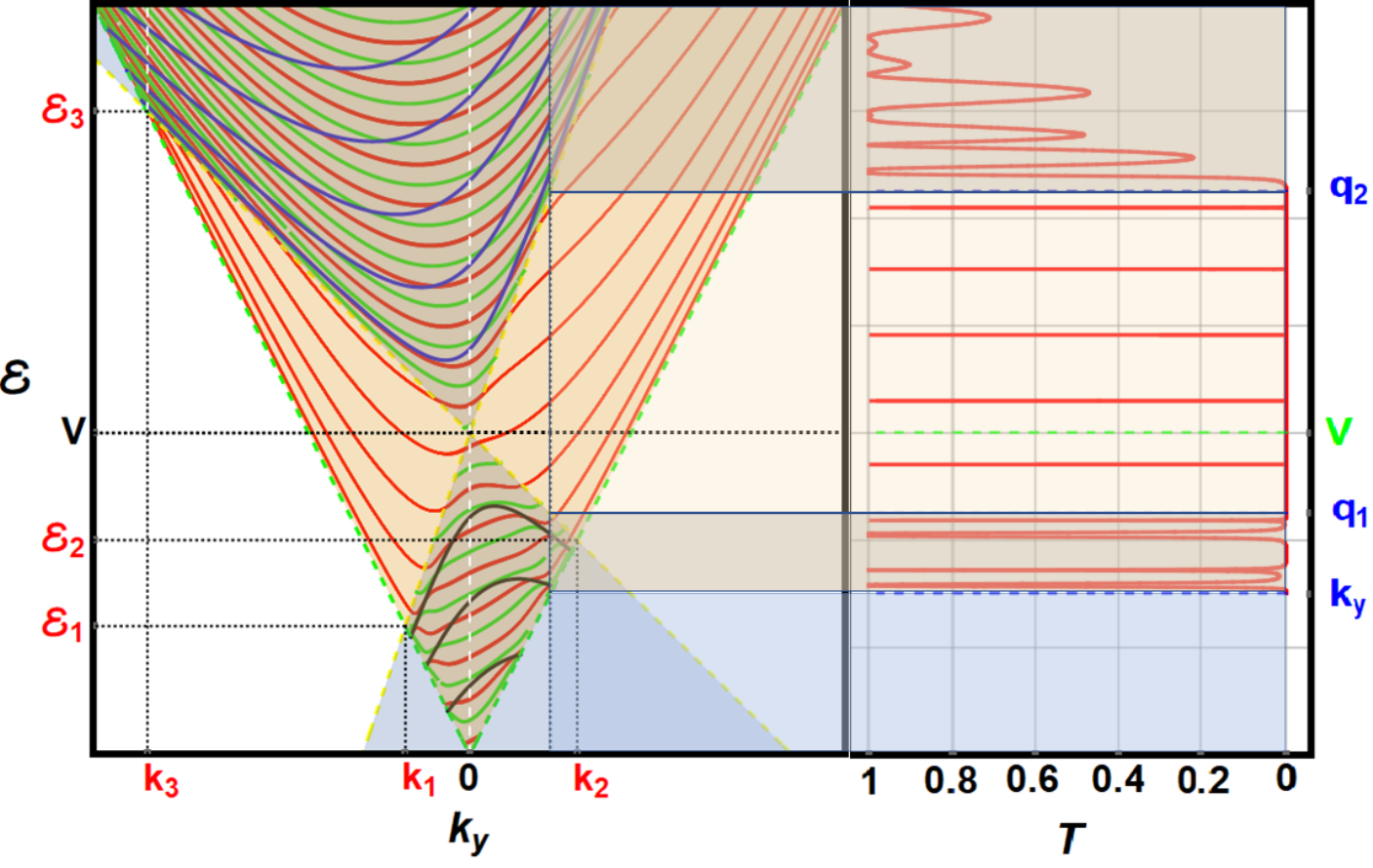}\label{fig5:SubFigA}
    }\subfloat[$k_{y}=-1.5$]{
   \includegraphics[scale=0.365]{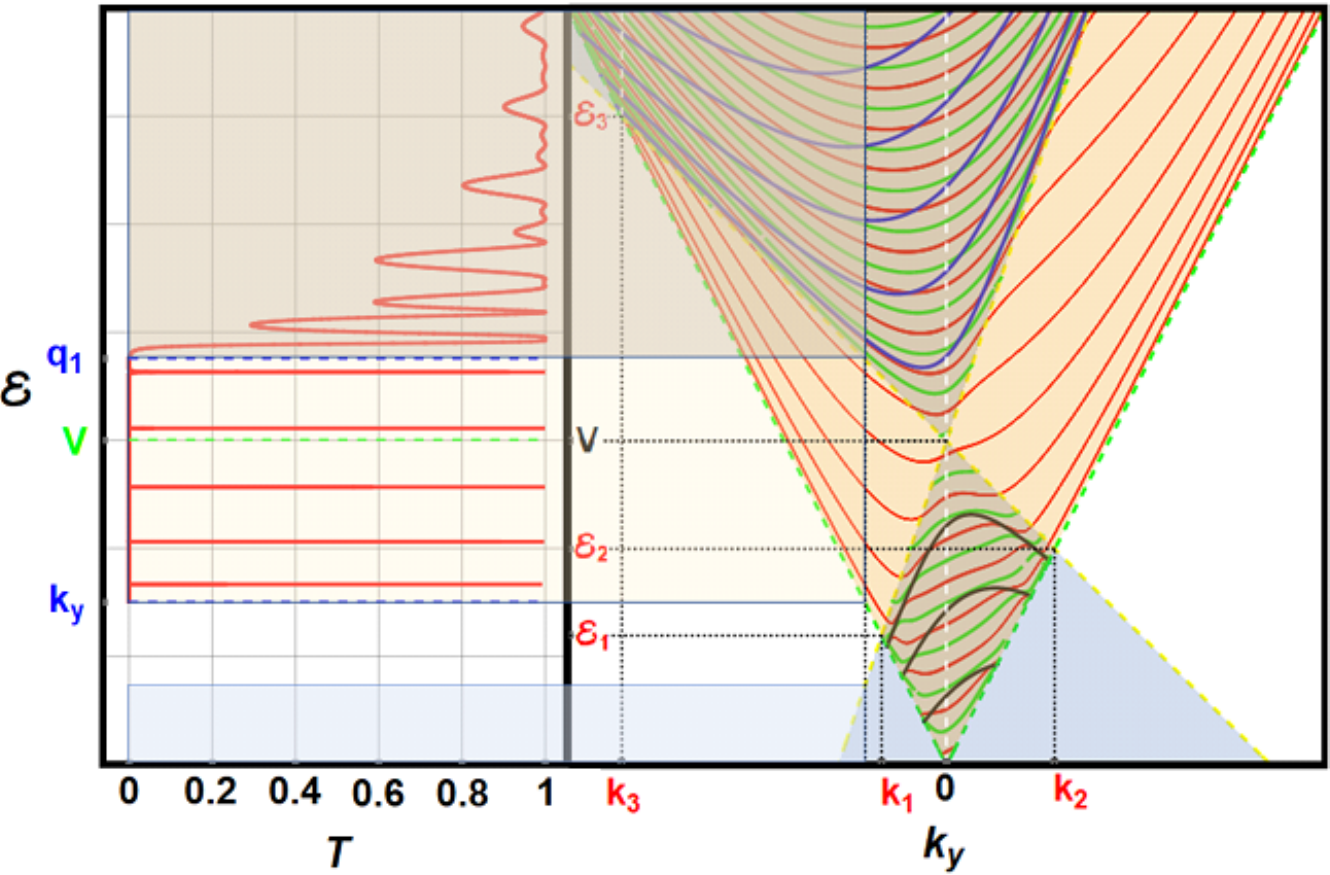}\label{fig5:SubFigB}
    }\\
   \subfloat[$k_{y}=1.5$]{
   \hspace{-0.2cm}\includegraphics[scale=0.25]{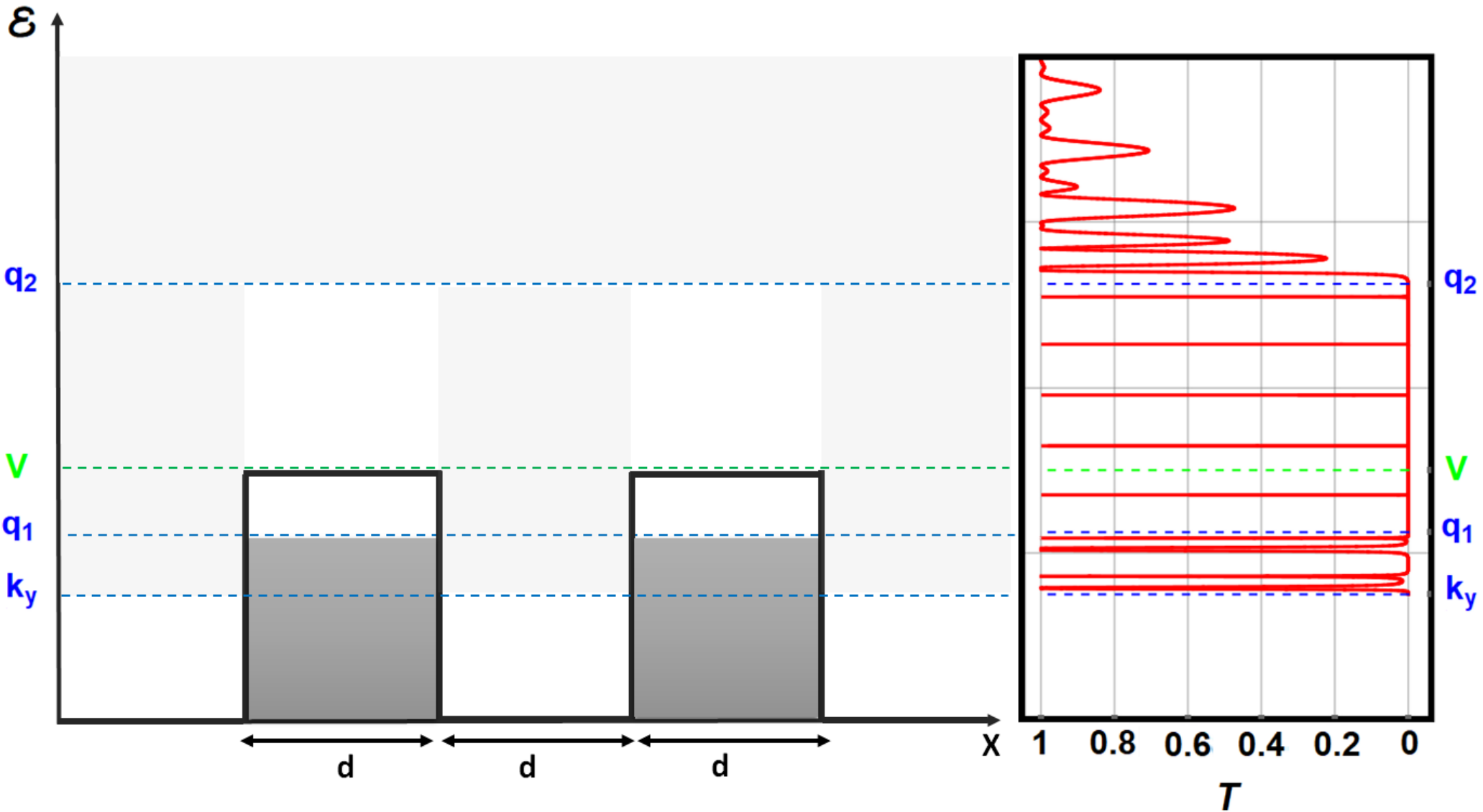}\label{fig5:SubFigC}
    }
    \subfloat[$k_{y}=-1.5$]{
   \hspace{-0.5cm}\includegraphics[scale=0.252]{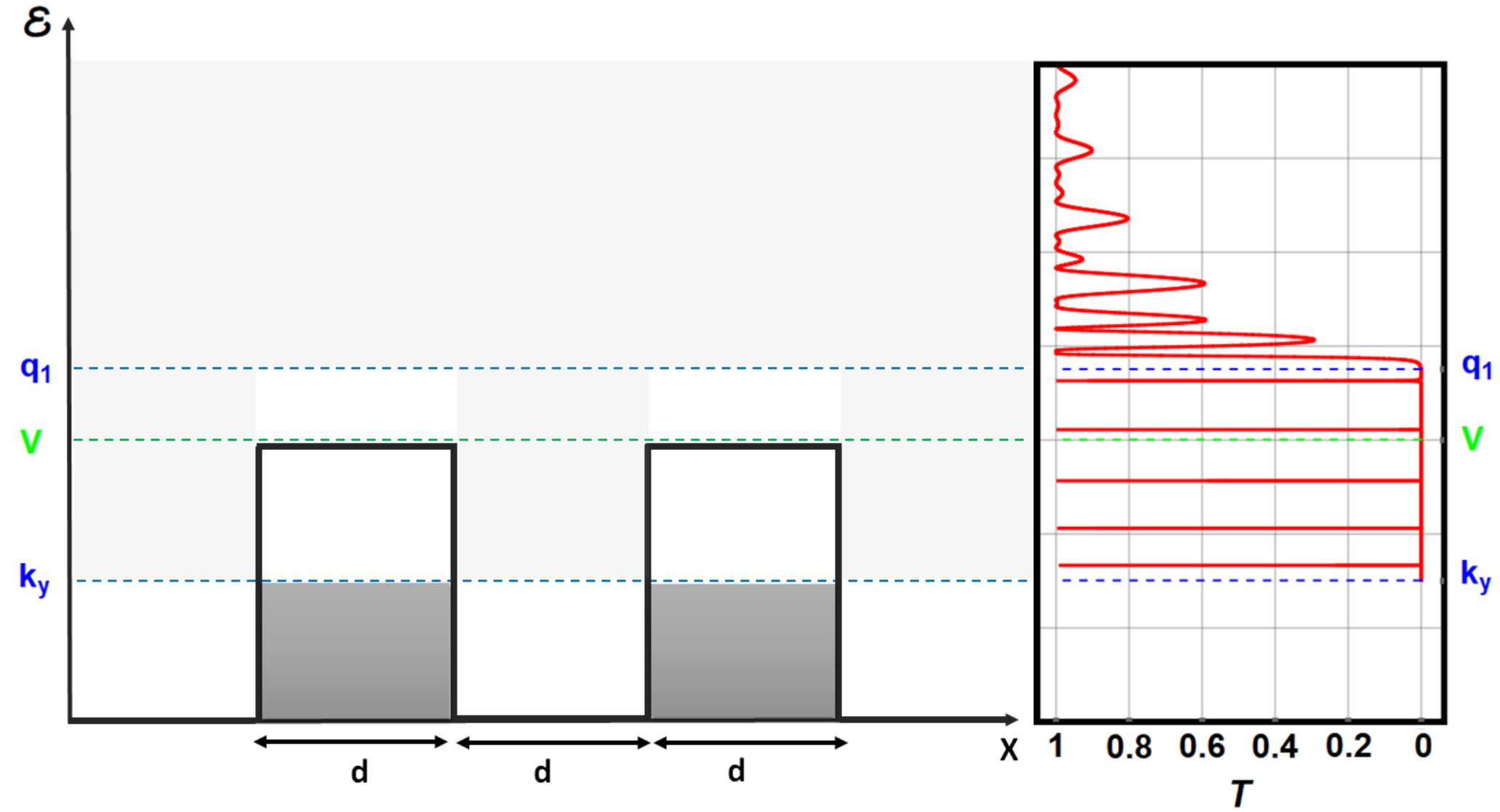}\label{fig5:SubFigD}
    }
     \caption{Global view of the transmission spectrum: interconnection between density
plots, symmetric cuts with respect to $k_y=0$, and the double-barrier profile.}
\label{fig06}
\end{figure}

%========================================================
Figure~\ref{fig06} provides a global view of the transmission through the double-barrier
structure by combining density plots of $T(\varepsilon,k_y)$ with two representative cuts
taken symmetrically with respect to $k_y=0$ (namely $k_y=\pm 1.5$), together with the
schematic profile of the potential barriers. This composite representation emphasizes the
interconnected nature of the transmission features across different visualizations.

The density maps [Figures~\ref{fig06}(a) and \ref{fig06}(b)] reveal the regions of allowed
and forbidden transmission, where the boundaries are determined by the critical energies
$\varepsilon_1=\tfrac{V}{2+\tau}$, $\varepsilon_2=\tfrac{V}{2-\tau}$, and
$\varepsilon_3=\tfrac{V}{\tau}$. These characteristic lines delimit the zones in which
propagating or evanescent modes dominate, thereby controlling the overall transmission
spectrum. The complementary cuts [Figures~\ref{fig06}(c) and
\ref{fig06}(d)] highlight the fine structure of the transmission probability for
$k_y=+1.5$ and $k_y=-1.5$, respectively, and allow for a direct comparison of the spectra
for opposite transverse momenta.

For positive incidence ($k_y=+1.5$), the transmission gap is shifted to higher energies
and contains several sharp line-type resonances associated with quasi-bound states in the
quantum well between the barriers. In contrast, for negative incidence ($k_y=-1.5$), the
transmission gap is displaced toward lower energies, and the number of such resonances is
significantly reduced. This asymmetry directly reflects the breaking of mirror symmetry
once $\tau \neq 0$, confirming that the tilt of the Dirac cone plays a decisive role in
determining whether resonant states are enhanced or
suppressed~\cite{Goerbig2008,Trescher2015}.

The inclusion of the barrier profile in Figures~\ref{fig06}(c) and \ref{fig06}(d)
clarifies the physical origin of these phenomena. The evanescent modes localized within
the barriers act as effective mirrors, confining Dirac fermions inside the central region
and giving rise to Fabry–Pérot-like oscillations. Depending on the sign of $k_y$, this
confinement either strengthens or weakens the coupling to the bound states of the well,
thereby modulating the number and intensity of line-type
resonances~\cite{Alhaidari2012,RamezaniMasir2010,Sun2011}.

Altogether, Figure~\ref{fig06} provides a comprehensive picture of the transmission
process: it shows how density maps, spectral cuts, and barrier geometry interconnect to
demonstrate the dual role of propagating and evanescent modes in shaping the resonance
spectrum, as well as the fundamental asymmetry introduced by the cone tilt parameter
$\tau$. This effect is of particular importance for the design of anisotropic transport
devices and energy filters in Dirac materials~\cite{Britnell2013,Sun2011,Trescher2015}.
%========================================================
\subsection{Angular Dependence of Transmission}
\begin{figure}[!h]\centering
    \subfloat[$\tau=0$]{
   \hspace{-0.8cm} \includegraphics[scale=0.17]{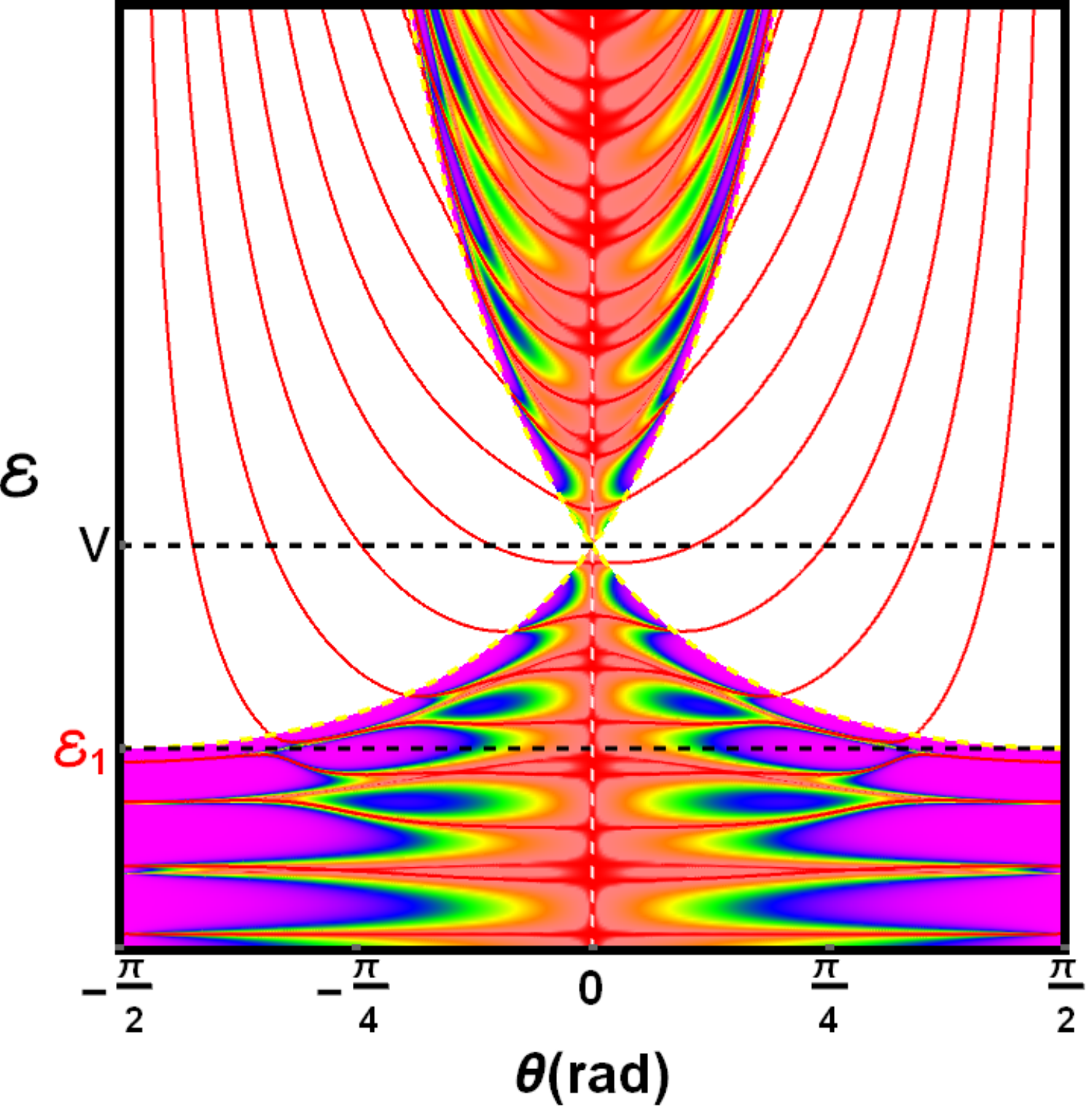}\label{fig07:SubFigA}}
    \subfloat[][$\tau=0.5$]{
      \hspace{-0.11cm}\includegraphics[scale=0.17]{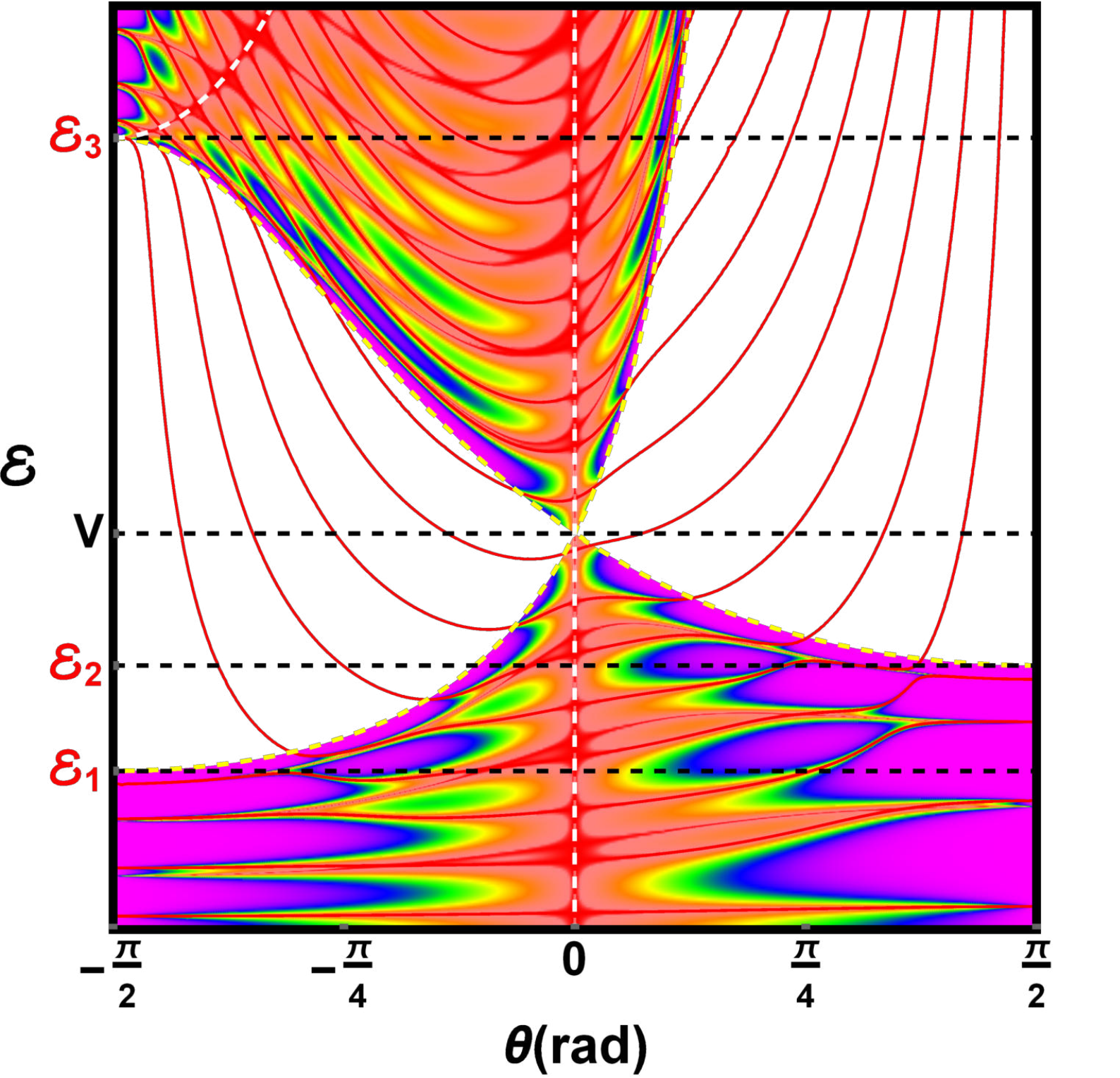}\label{fig07:SubFigB}}
    \subfloat[][$\tau= 1 $]{
      \hspace{-0.34cm}\includegraphics[scale=0.17]{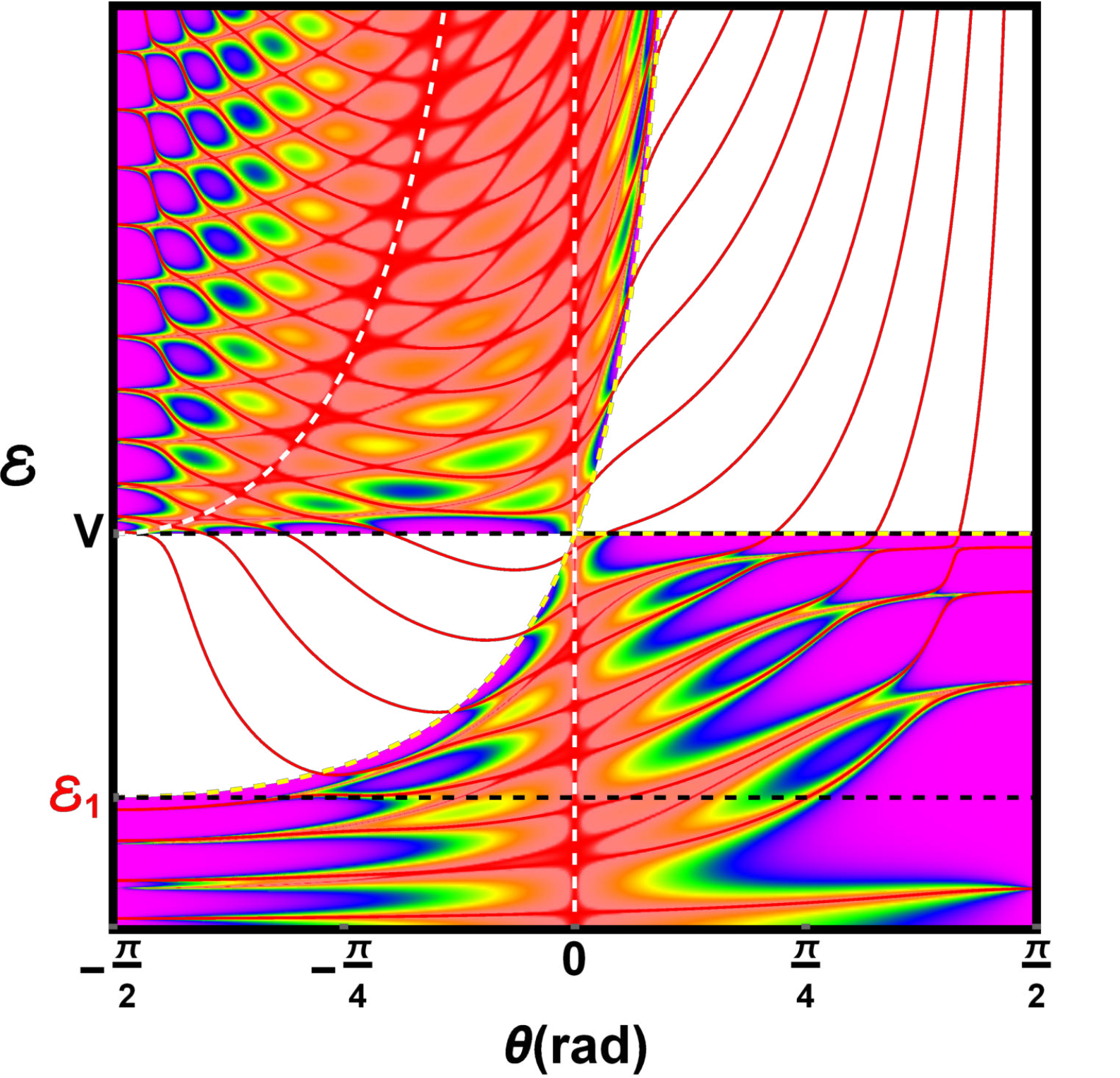}\label{fig07:SubFigC}}
    \subfloat[][$\tau=2$]{
   \hspace{-0.38cm}\includegraphics[scale=0.17]{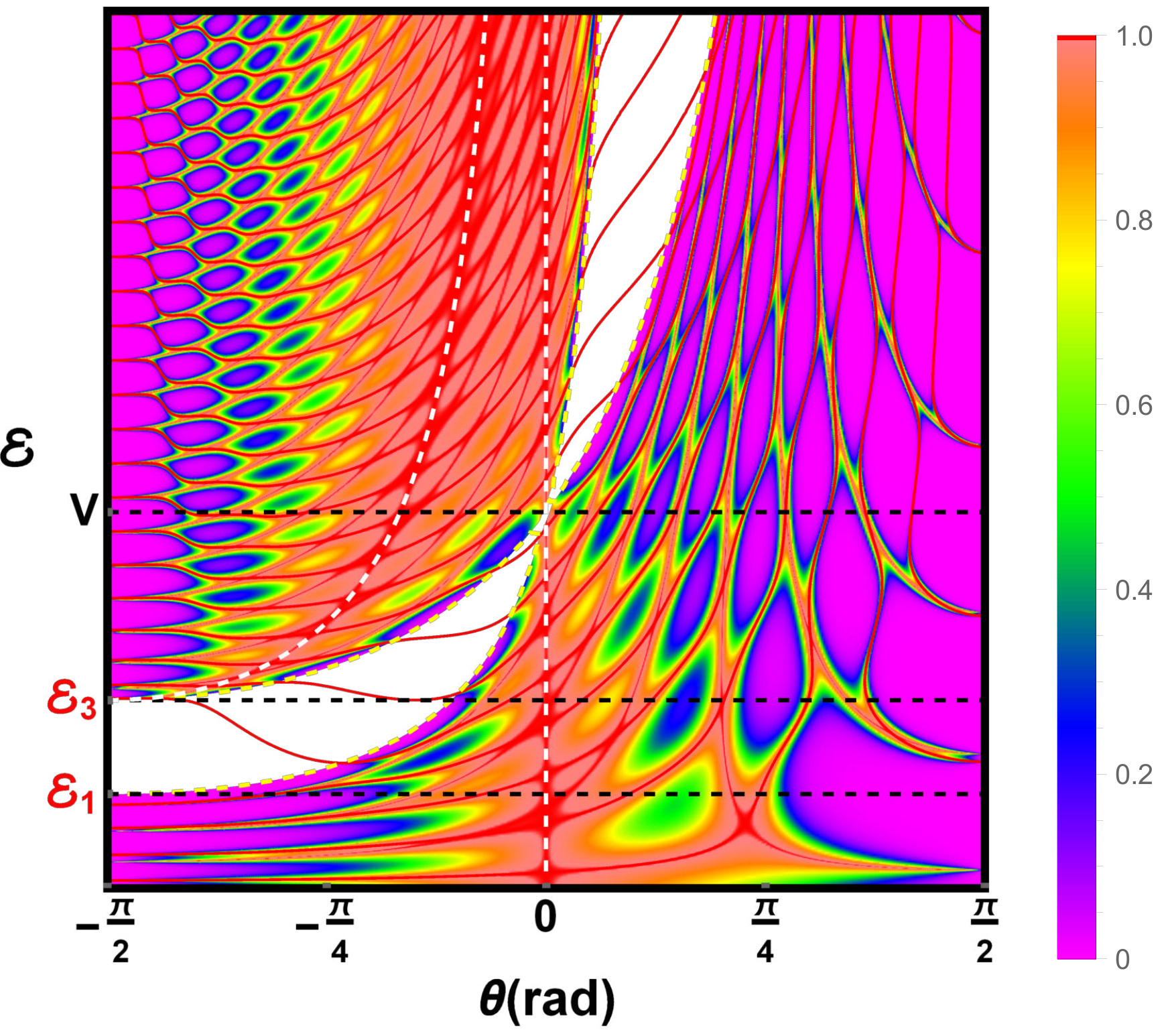}\label{fig07:SubFigD}}\\
\subfloat[][$\tau=0$ ]{
    \hspace{-0.6cm}\includegraphics[width=0.25\linewidth]{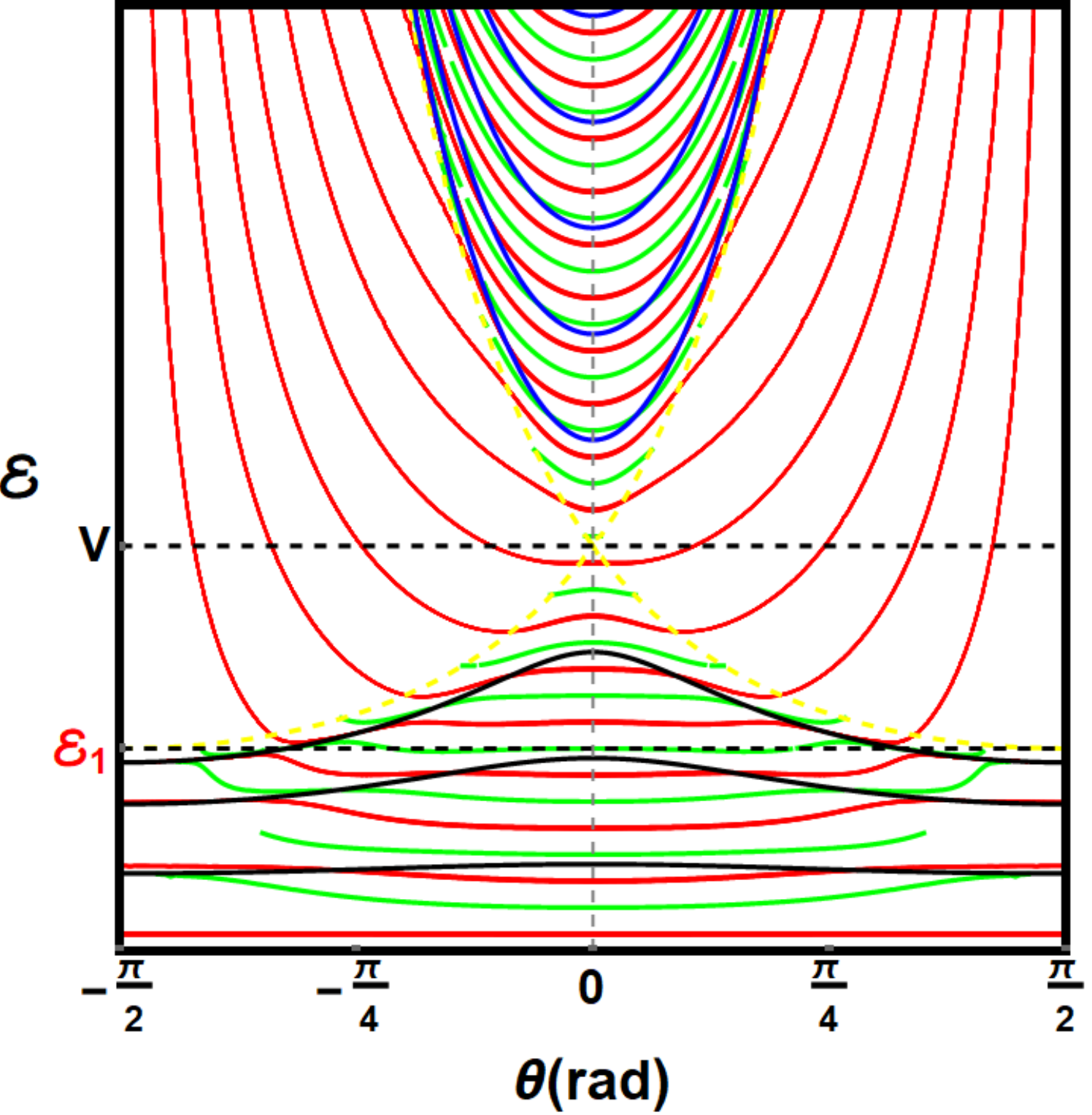}\label{fig07:SubFigE}}
 \subfloat[][$\tau=0.5$ ]{
    \includegraphics[width=0.25\linewidth]{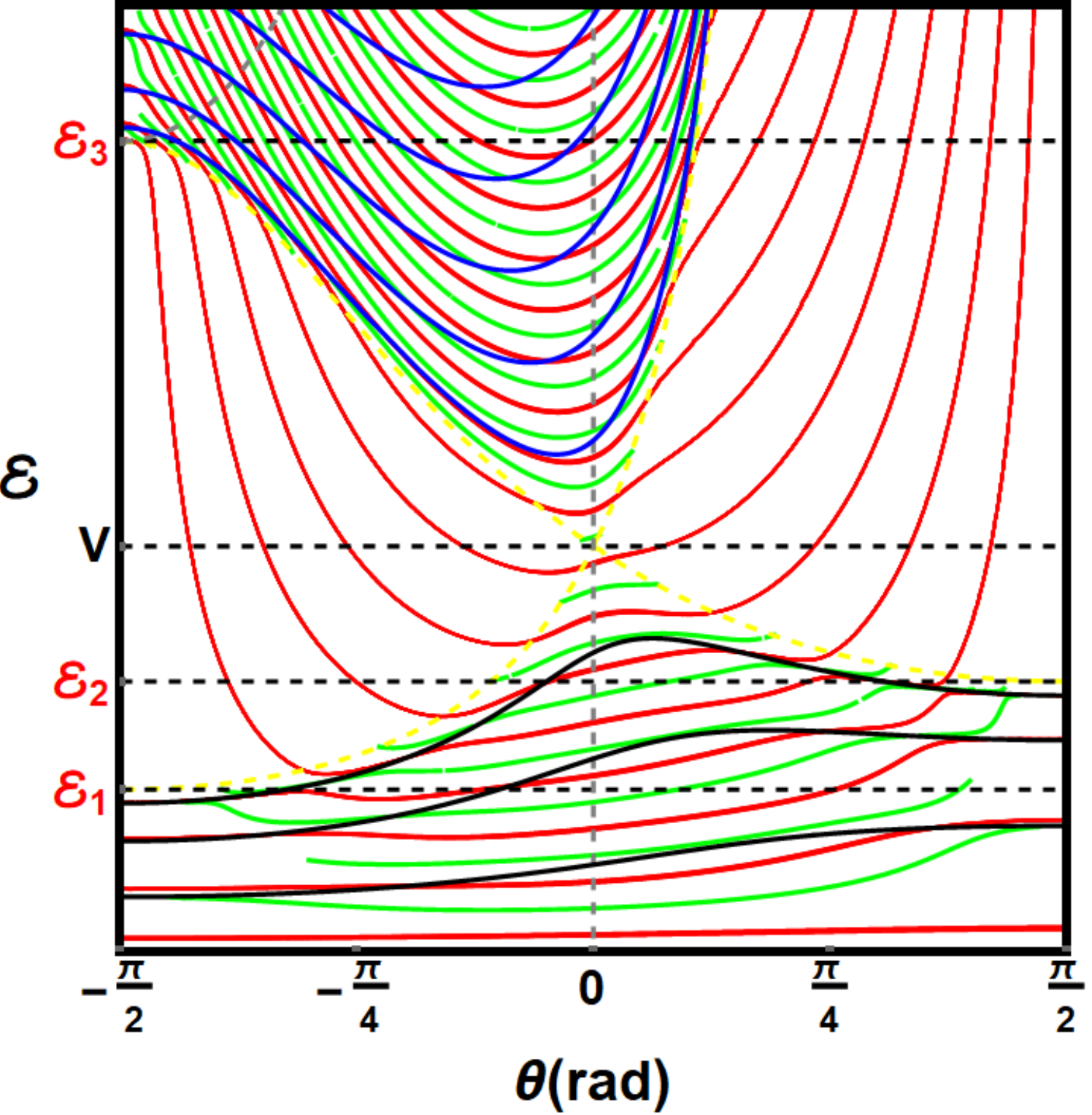}\label{fig07:SubFigF}}
\subfloat[][$\tau=1$ ]{
    \includegraphics[width=0.25\linewidth]{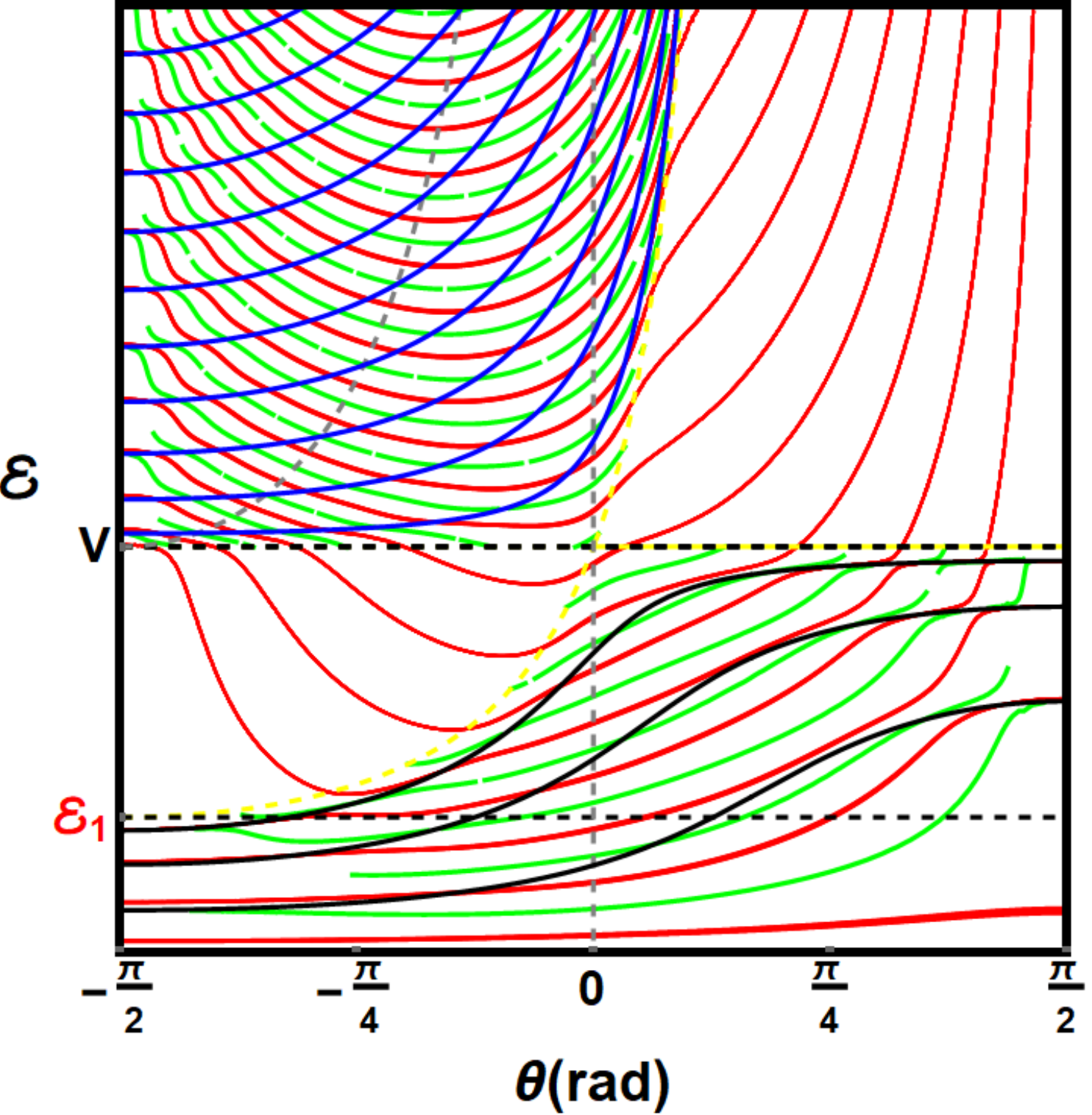}\label{fig07:SubFigG}}
    \subfloat[][$\tau=2$ ]{
    \includegraphics[width=0.25\linewidth]{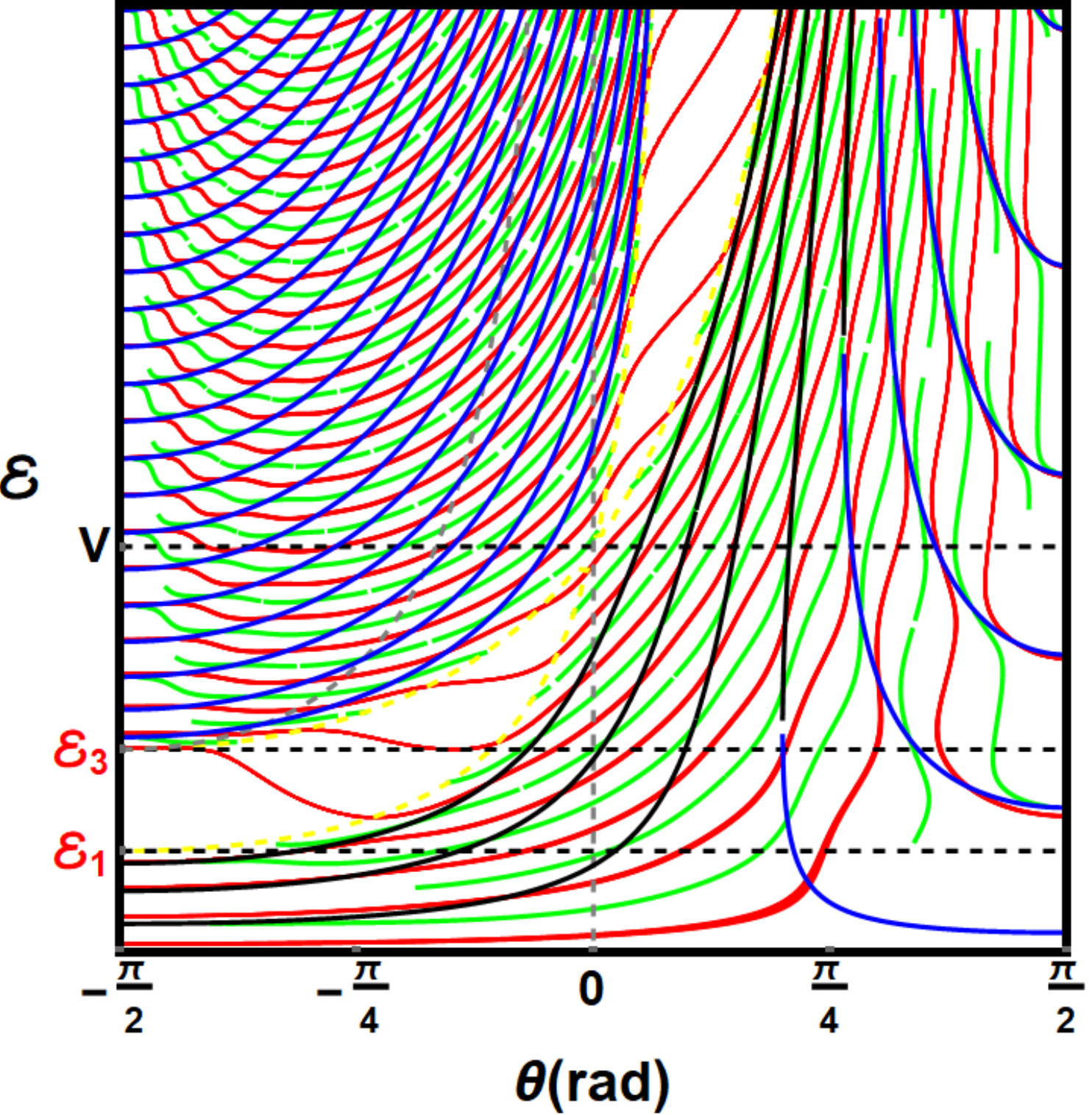}\label{fig07:SubFigH}}
    \caption{(Color online) Density plot of transmission probability $T$ versus incident
energy $\varepsilon$ and incident angle $\theta$ for a double-barrier structure. Both
barriers have the same height $V=3$, the same width $d$, and are separated by a distance
$d=4$. The plot is shown for four values of the parameter $\tau$. The incident energy
$\varepsilon$ is plotted on the vertical axis, while the incident angle $\theta$ is
plotted on the horizontal axis. The characteristic energies are defined as
$\varepsilon_{1}=\tfrac{V}{2+\tau}$, $\varepsilon_{2}=\tfrac{V}{2-\tau}$ and
$\varepsilon_{3}=\tfrac{V}{\tau}$.}
\label{fig07}
\end{figure}

%========================================================
The angular dependence of quantum transport in Dirac materials provides fundamental
insights into Klein tunneling, Fabry–Pérot resonances, and the emergence of anisotropic
effects when tilted cones are
considered~\cite{Katsnelson2006,Young2009,Allain2011,Goerbig2008,02}.
The density plot of the transmission probability $T(\theta,\varepsilon)$ is presented in
Figure~\ref{fig07} as a function of the incident angle $\theta$ and the incident energy
$\varepsilon$, using the relations $k_{y}=\varepsilon \sin\theta$ and $k_{x}=\varepsilon
\cos\theta$. The same parametric conditions as in Figure~\ref{fig04} are employed. While
Figure~\ref{fig04} displayed $T(k_y,\varepsilon)$ for different configurations of tilted
Dirac cones inside the barrier regions, the present figure highlights the angular
dependence of the transmission spectrum.

The allowed transmission zones, corresponding to the overlap of Dirac cones, are delimited
by the curves
\[
\varepsilon = \frac{V}{1+(1-\tau)\sin\theta}, \qquad
\varepsilon = \frac{V}{1-(1+\tau)\sin\theta},
\]
while the white regions correspond to forbidden transmission domains~\cite{Choubabi2024,Kamal2018,Shakouri2013,Choubabi2020}.
A key feature of these maps is the occurrence of perfect transmission, a hallmark of the
Klein paradox, which is centered around normal incidence ($\theta=0$) for a given value of
$\tau$. This phenomenon, first described in graphene by Katsnelson~\cite{Katsnelson2006},
is clearly illustrated by the dashed white lines in
Figures~\ref{fig07:SubFigA}–\ref{fig07:SubFigD} and by the grey dashed lines in the
resonance spectra of Figures~\ref{fig07:SubFigE}–\ref{fig07:SubFigH}.

Beyond normal incidence, Klein tunneling is also observed whenever
$\varepsilon=-\tfrac{V}{\tau\sin\theta}$, particularly for $\varepsilon_3 \geq V$
(Figures~\ref{fig07:SubFigB}, \ref{fig07:SubFigD}, \ref{fig07:SubFigF},
\ref{fig07:SubFigH}) and for $\varepsilon_3=V$ (Figures~\ref{fig07:SubFigC},
\ref{fig07:SubFigG}). This transparency arises from the matching of effective refractive
indices between adjacent regions, which makes the barriers invisible to Dirac
fermions~\cite{Allain2011,Choubabi2024,Katsnelson2006}.

When the tilt parameter is zero ($\tau=0$), the transmission spectrum is perfectly
symmetric with respect to normal incidence, as can be seen in Figure~\ref{fig07:SubFigA}.
In contrast, when $\tau \neq 0$, this symmetry is broken and the transmission properties
become anisotropic, with different behaviors for positive and negative incidence angles,
as illustrated in Figures~\ref{fig07:SubFigB}–\ref{fig07:SubFigD}. This anisotropy is a
direct manifestation of the tilted cone dispersion, as reported in
Refs.~\cite{Goerbig2008,Kapri2020,Nguyen2018,Pattrawutthiwong2021}.

The transmission maps also display two families of resonance peaks. The first corresponds
to Fabry–Pérot resonances, which appear within the allowed transmission zones and result
from constructive interference of electronic waves either inside the barriers or within
the central quantum well. Using equations~\ref{eq020} and \ref{eq011} in combination with
$k_{y}=\varepsilon\sin\theta$, one can determine the resonance energies in both regimes:
tunneling ($\varepsilon < V$) and propagating ($\varepsilon > V$). These resonances are
depicted by the black, blue, and red curves in
Figures~\ref{fig07:SubFigE}–\ref{fig07:SubFigH}. The second family corresponds to
line-type resonances associated with bound or quasi-bound states in the central well,
which extend into the forbidden transmission zones and are illustrated by the red curves.
The resonance energies associated with transmission minima, obtained from
equation~\ref{eq0111}, are shown as green curves.

The influence of the tilt parameter $\tau$ on the resonance spectrum is evident.
Increasing $\tau$ enhances the number of Fabry–Pérot resonances within the allowed zones,
while simultaneously reducing the number of line-type resonances originating from bound
states in the gaps. Overall, Figure~\ref{fig07} demonstrates that the resonance structure
depends explicitly not only on the tilt parameter of the Dirac cone but also on the sign
and magnitude of the incident angle, emphasizing the crucial role of anisotropy in
controlling quantum transport across tilted Dirac materials.%~\cite{Feuillet2019,Wu2023}.

\begin{figure}[!h]\centering
\subfloat[$\tau=0$]{
    \hspace{-0.7cm}\includegraphics[scale=0.17]{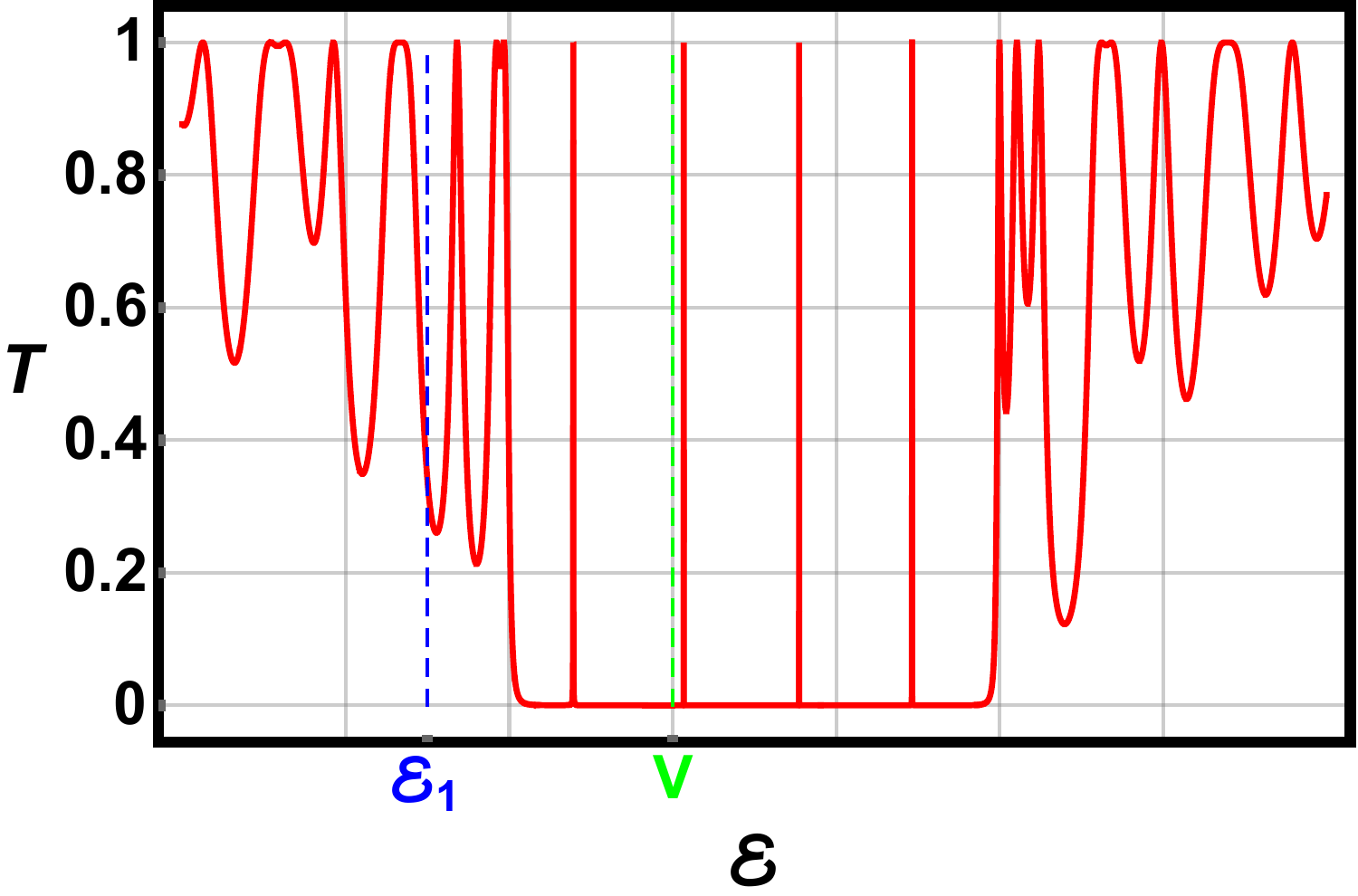}\label{fig08:SubFigA}}
    \subfloat[][$\tau=0.5$]{
   \includegraphics[scale=0.17]{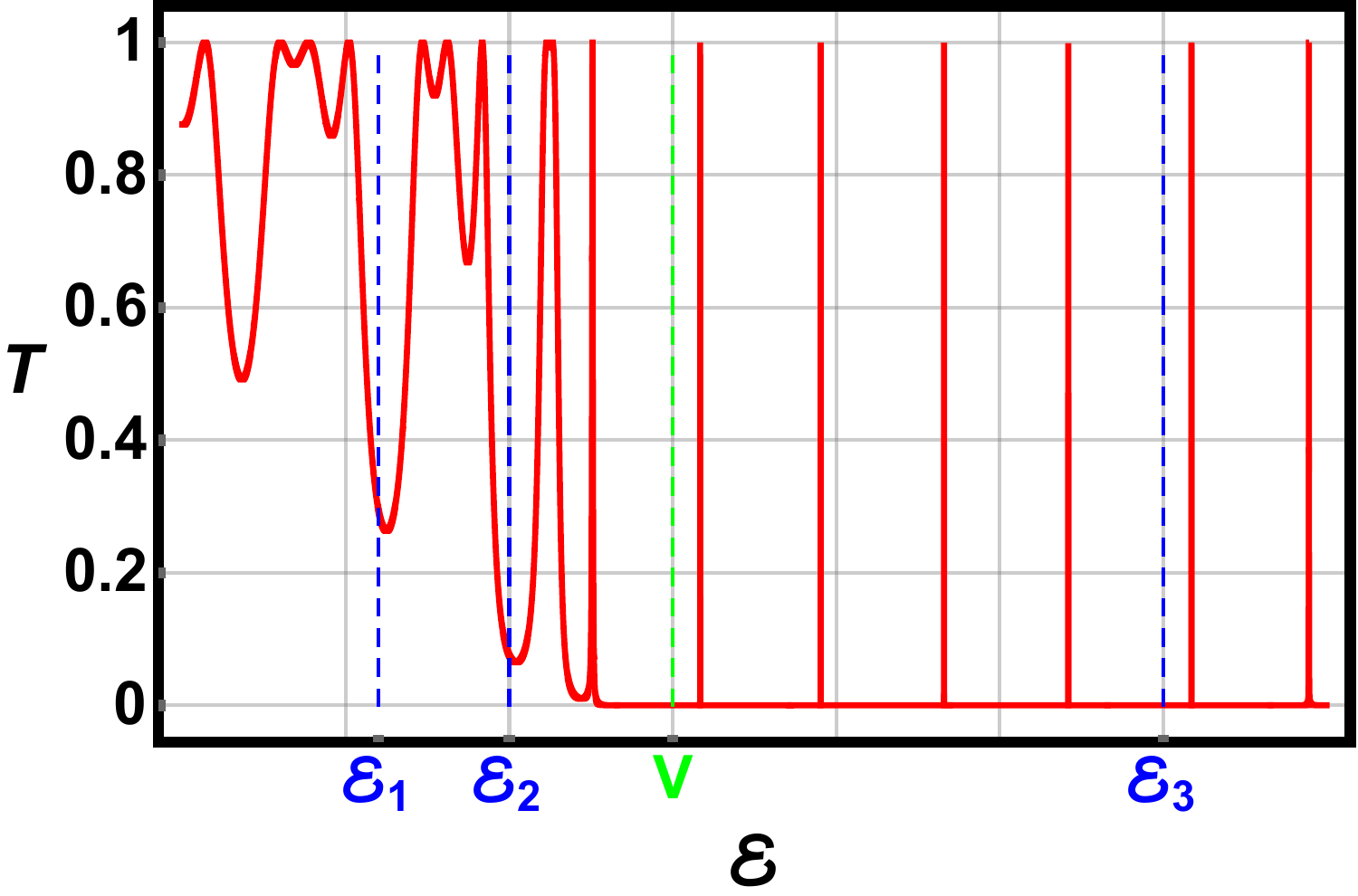}\label{fig08:SubFigB}}
    \subfloat[][$\tau=1$]{
   \includegraphics[scale=0.17]{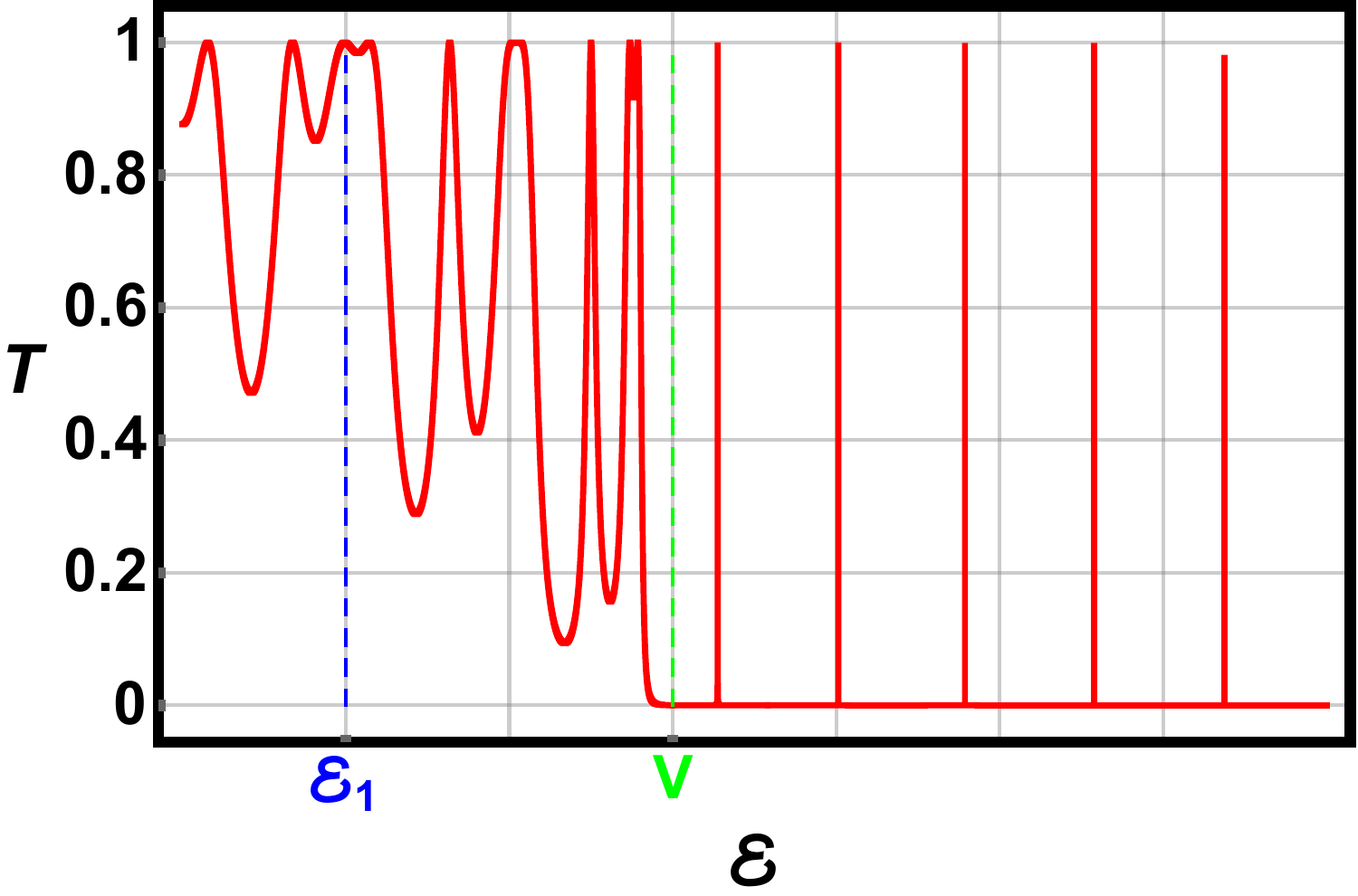}\label{fig08:SubFigC}}
\subfloat[][$\tau=2$]{
    \includegraphics[scale=0.17]{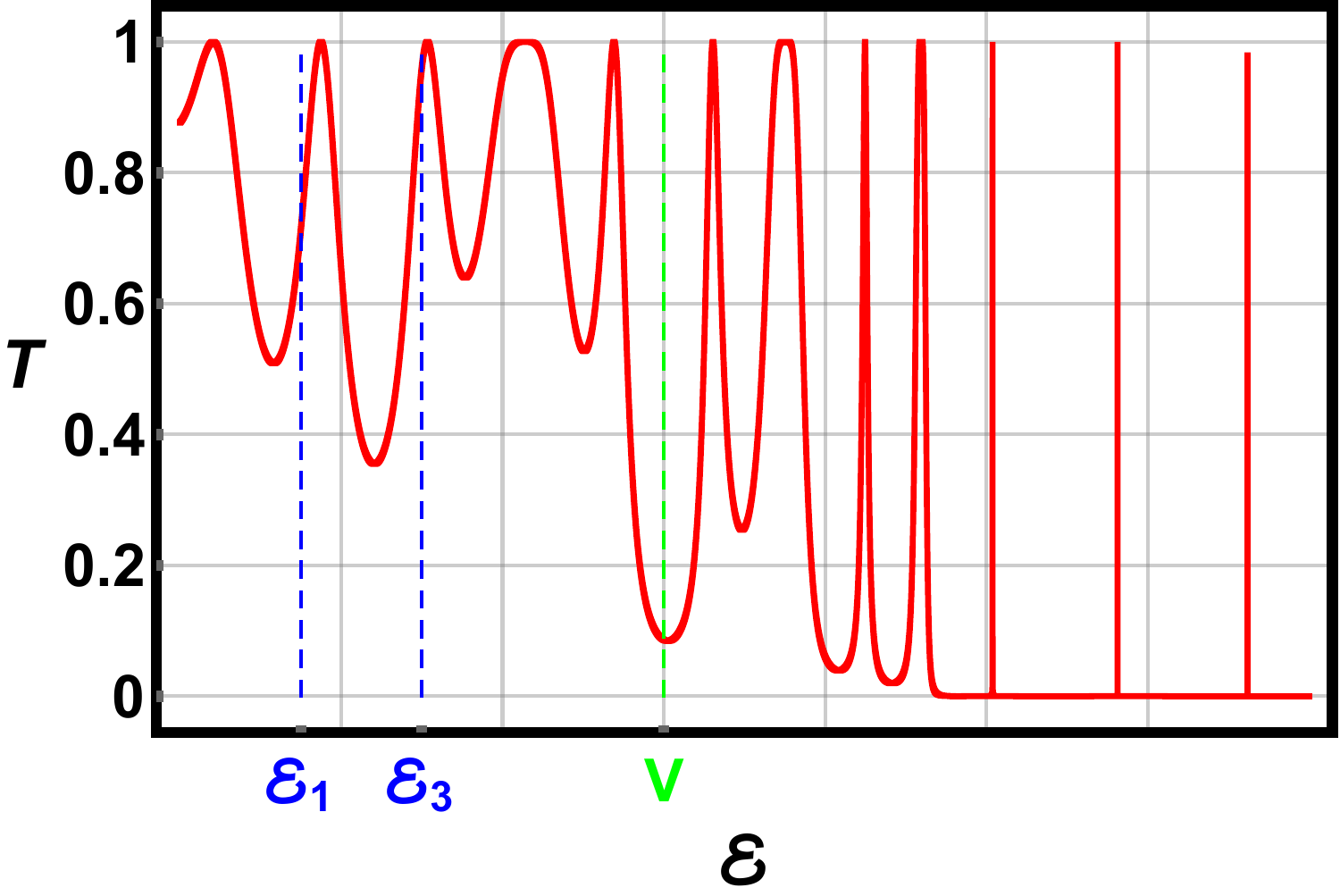}\label{fig08:SubFigD}}\\
    \subfloat[$\tau=0$]{
   \hspace{-0.7cm}\includegraphics[scale=0.17]{Fig32}\label{fig08:SubFigE}}
    \subfloat[][$\tau=0.5$]{
   \includegraphics[scale=0.17]{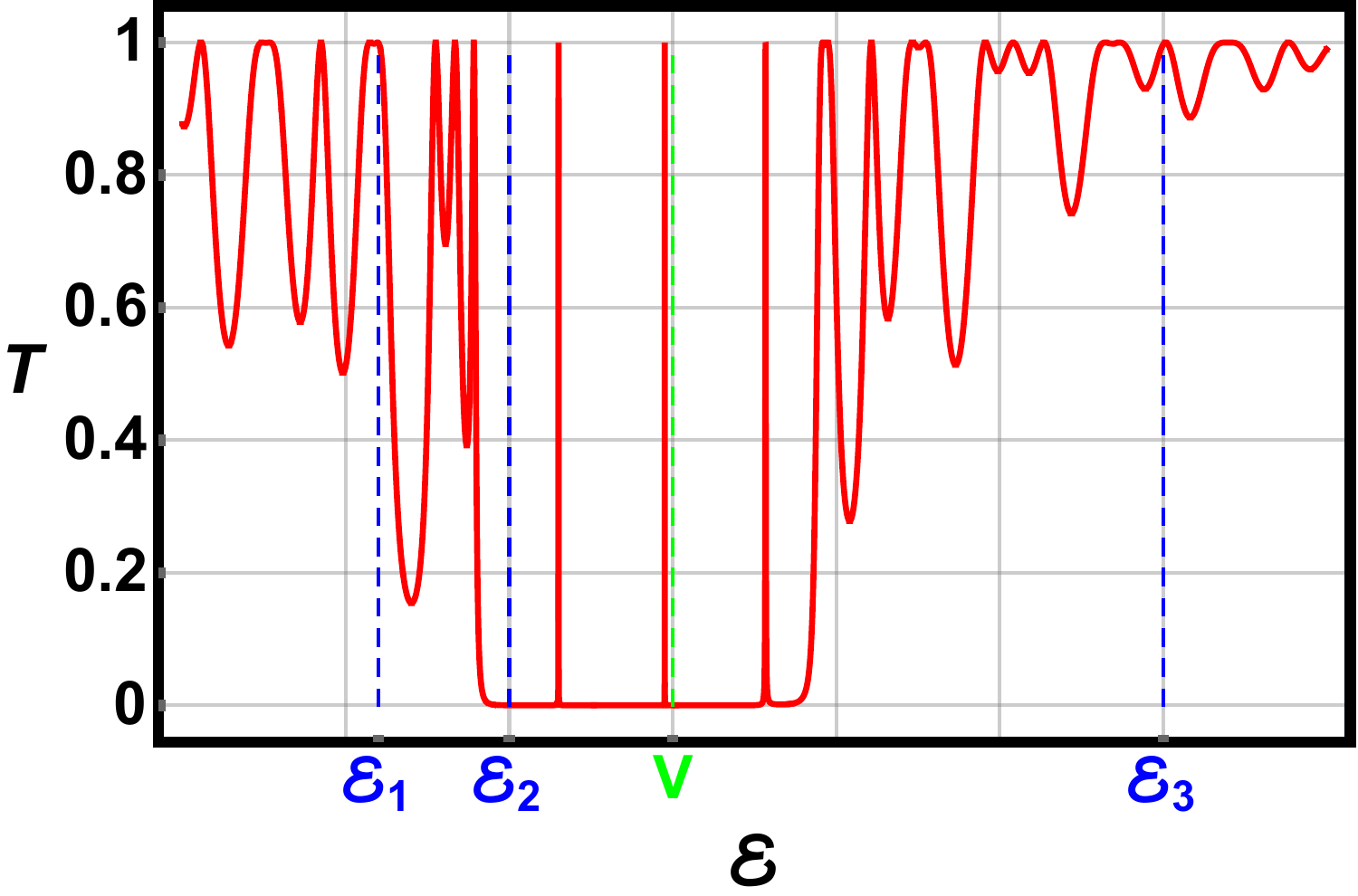}\label{fig08:SubFigF}}
    \subfloat[][$\tau=1$]{
   \includegraphics[scale=0.17]{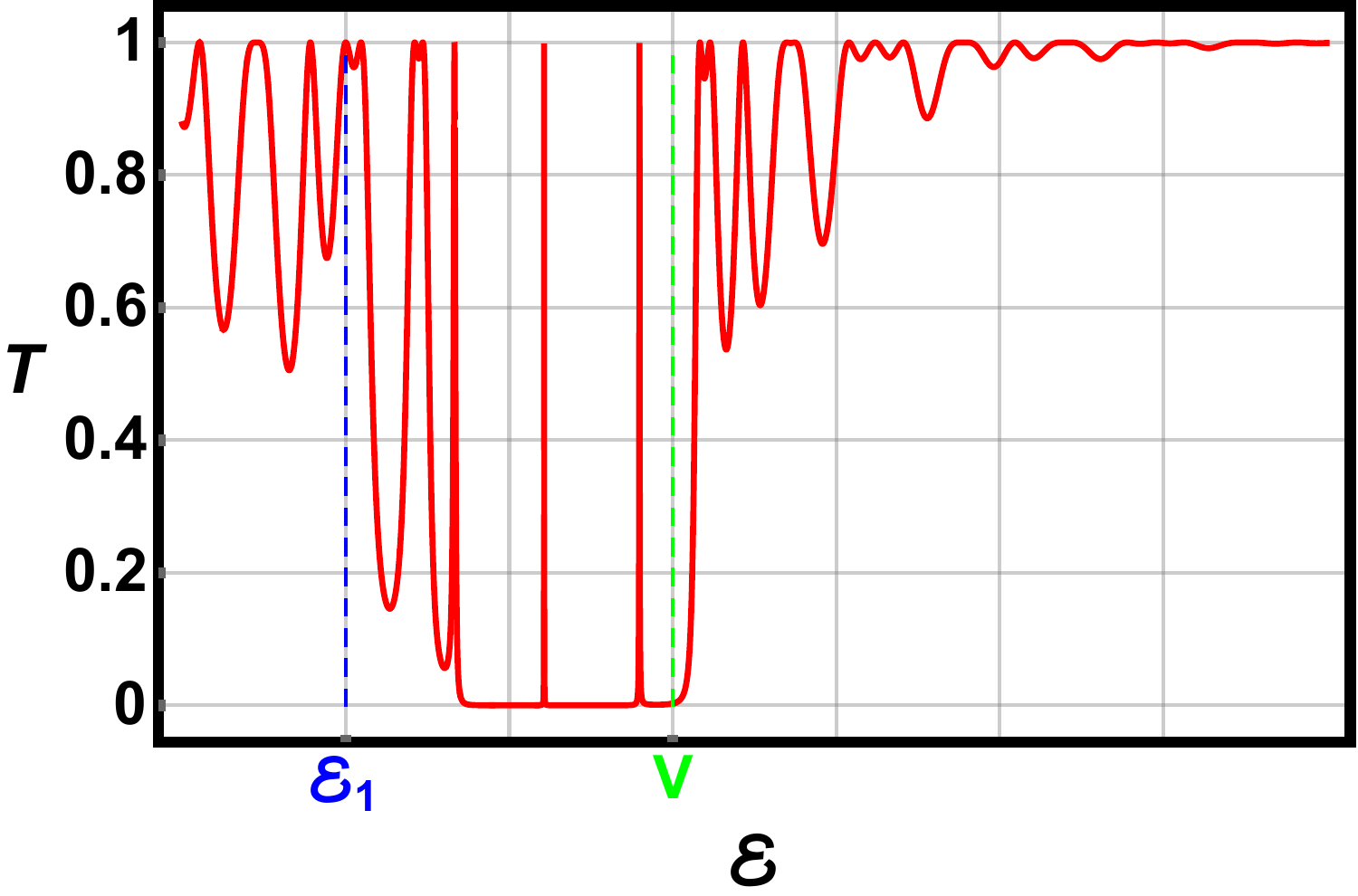}\label{fig08:SubFigG}}
\subfloat[][$\tau=2$ ]{
   \includegraphics[scale=0.17]{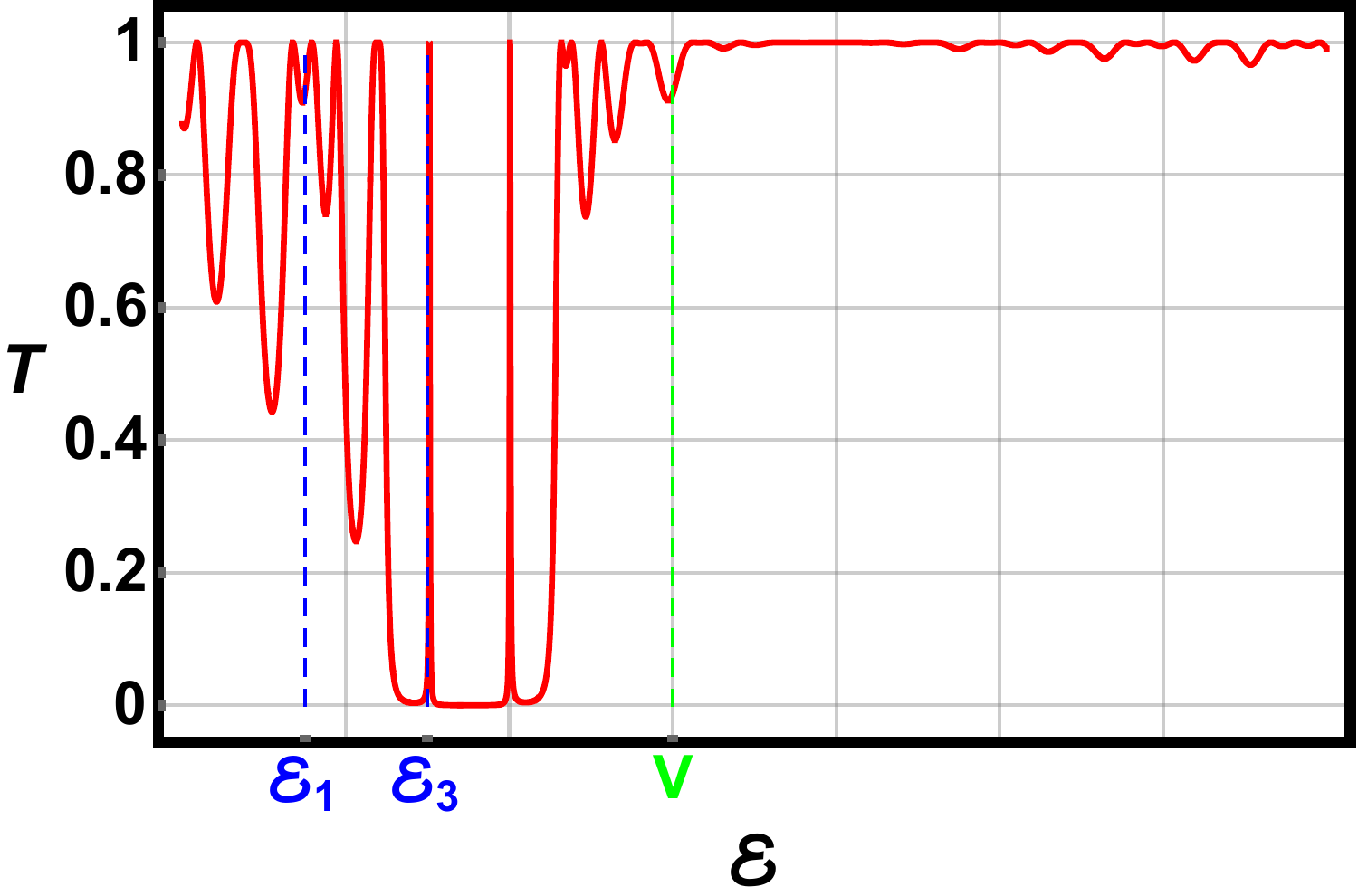}\label{fig08:SubFigH}}
    \caption{(Color online) Transmission probability $T(\varepsilon)$ as a function of
energy $\varepsilon$ for two distinct angles of incidence. The barriers have a height $V$,
a width $d$, and are separated by the same distance. The curves are plotted for four
values of the parameter $\tau=0, 0.5, 1$ and $2$. The upper graphs correspond to an angle
of incidence $\theta= \tfrac{\pi}{8}$, while the lower graphs are obtained for
$\theta=-\tfrac{\pi}{8}$.}
\label{fig08}
\end{figure}

To investigate in more detail the behavior of the different types of total transmission
peaks observed in both the allowed and forbidden regions, we performed vertical cuts of
the transmission density $T(\theta,\varepsilon)$. Two specific values of the incident
angle were selected, $\theta = \pm \tfrac{\pi}{8}$, for different values of the tilt
parameter $\tau$, as illustrated in Figure~\ref{fig08}.

For a positive incident angle ($\theta = \tfrac{\pi}{8}$), shown in
Figs.~\ref{fig08}(\ref{fig08:SubFigA}--\ref{fig08:SubFigD}), increasing the tilt parameter
$\tau$ leads to a multiplication of the total transmission peaks within the spectral gap.
These peaks are largely associated with \textit{line-type resonances}, which originate
from bound states in the well maintained by the back-and-forth motion of Dirac fermions
between the evanescent regions. At the same time, the transmission gap broadens and shifts
to higher energies, thereby enhancing resonant tunneling.

In contrast, for a negative incident angle ($\theta = -\tfrac{\pi}{8}$), as displayed in
Figs.~\ref{fig08}(\ref{fig08:SubFigE}--\ref{fig08:SubFigH}), increasing $\tau$ results in
a progressive reduction in the number of transmission peaks inside the gap. The
\textit{line-type resonances} become scarcer and less pronounced, while the transmission
gap shifts to lower energies and its width decreases, leading to a reduced contribution
from resonant tunneling.

This comparison highlights the crucial role played simultaneously by the tilt parameter
$\tau$ and by the sign of the incident angle. The combined modulation of these two
parameters allows not only the control of the density and location of the
\textit{line-type resonances}, but also the fine tuning of the transmission efficiency in
both allowed and forbidden regions. Such control offers a powerful tool for the
engineering of quantum nanoelectronic devices and opens new perspectives for the design of
directional and selective structures in tilted Dirac-cone
materials~\cite{Trescher2015,Sun2011,5}.
%========================================================
\subsection{Transmission in the $(\varepsilon, k_x)$ Plane}
\begin{figure}[!h]\centering
\subfloat[$\tau=0$]{
    \hspace{-0.8cm}\includegraphics[scale=0.17]{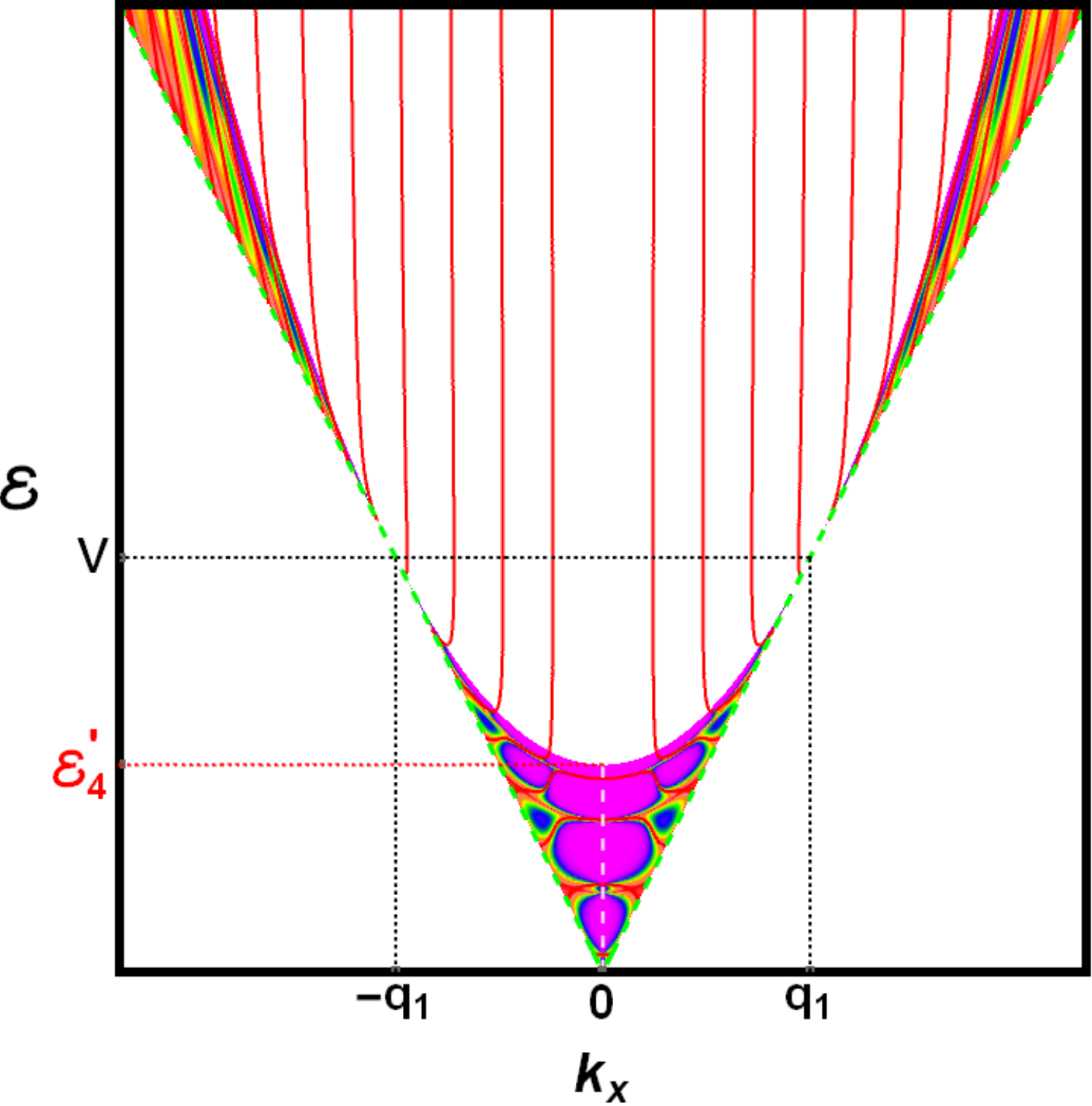}\label{fig09:SubFigA}}
    \subfloat[][$\tau=0.5$]{
   \includegraphics[scale=0.17]{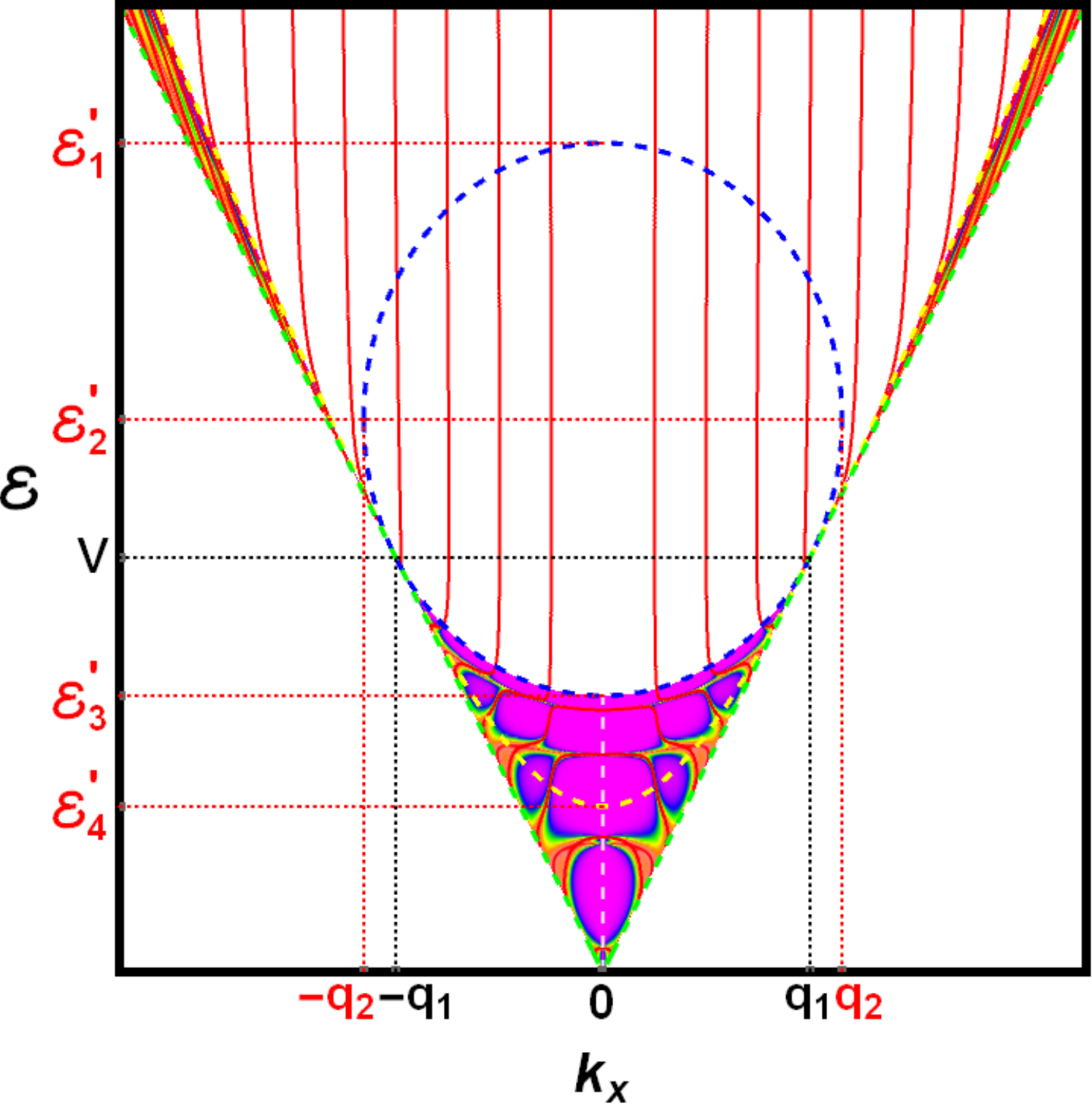}\label{fig09:SubFigB}}
    \subfloat[][$\tau=1$]{
   \includegraphics[scale=0.17]{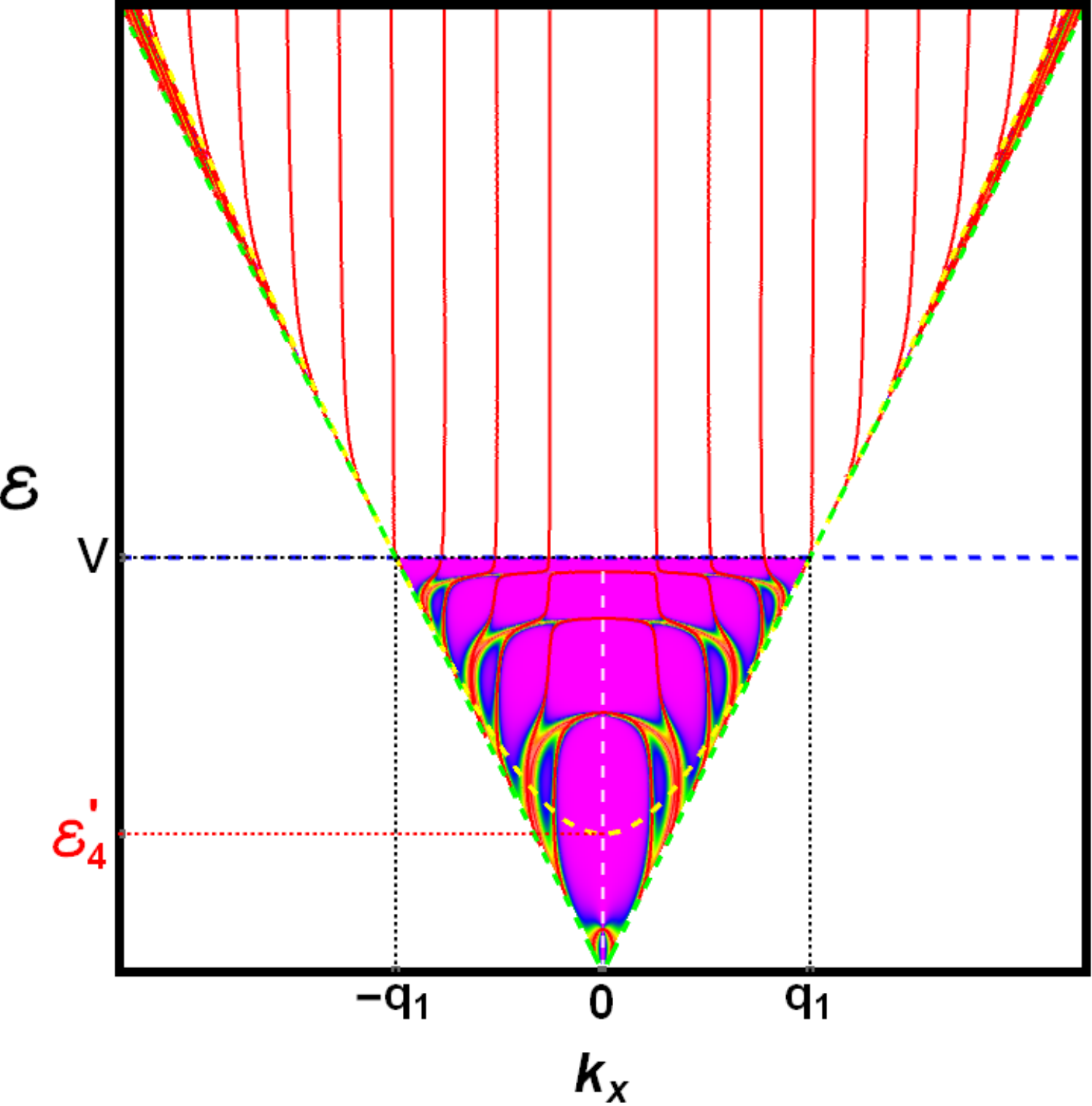}\label{fig09:SubFigC}}
\subfloat[][$\tau=1.5$]{
    \includegraphics[scale=0.17]{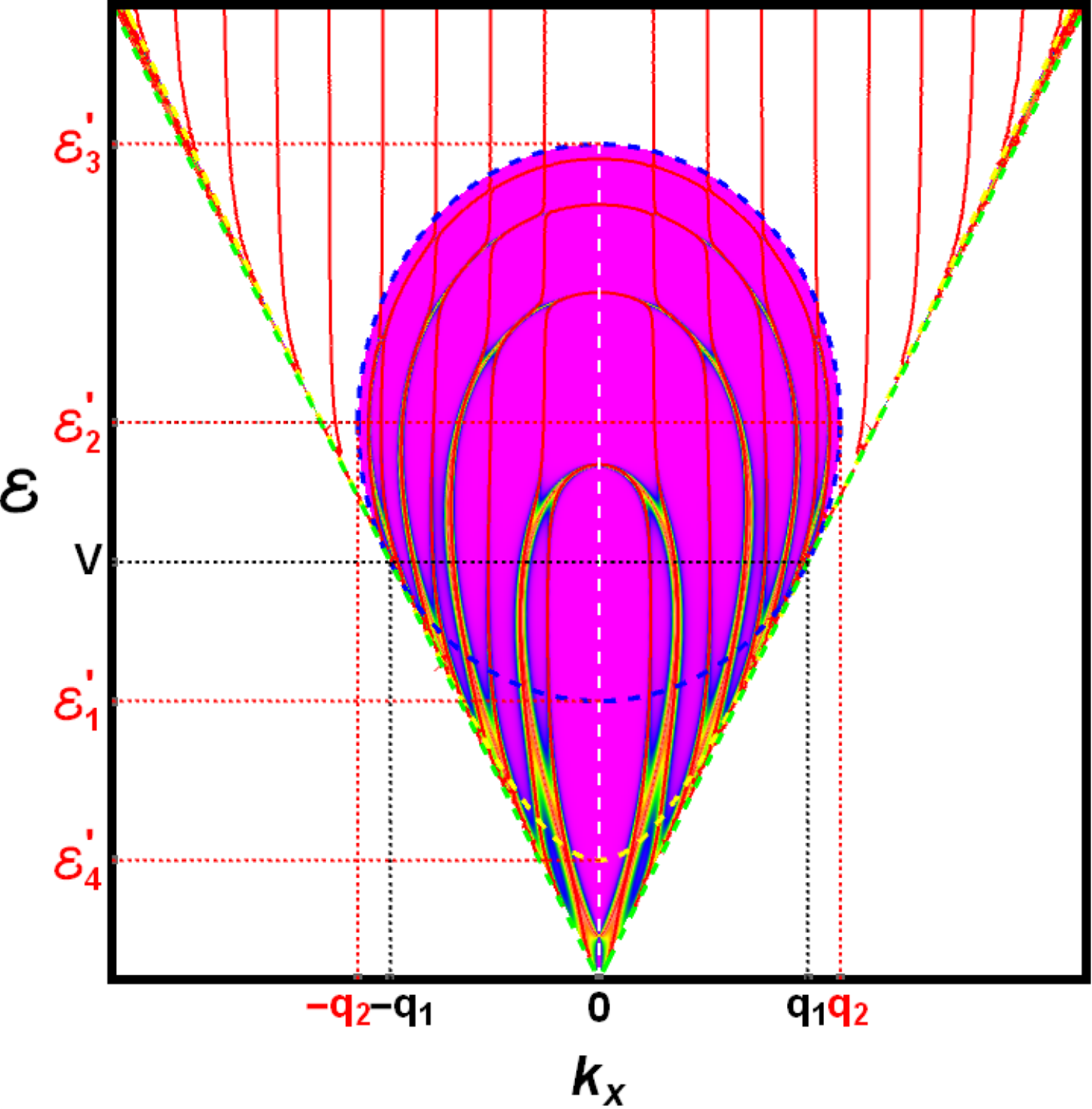}\label{fig09:SubFigD}}\\
    \subfloat[][$\tau=-0.5$]{
   \hspace{-0.8cm}\includegraphics[scale=0.17]{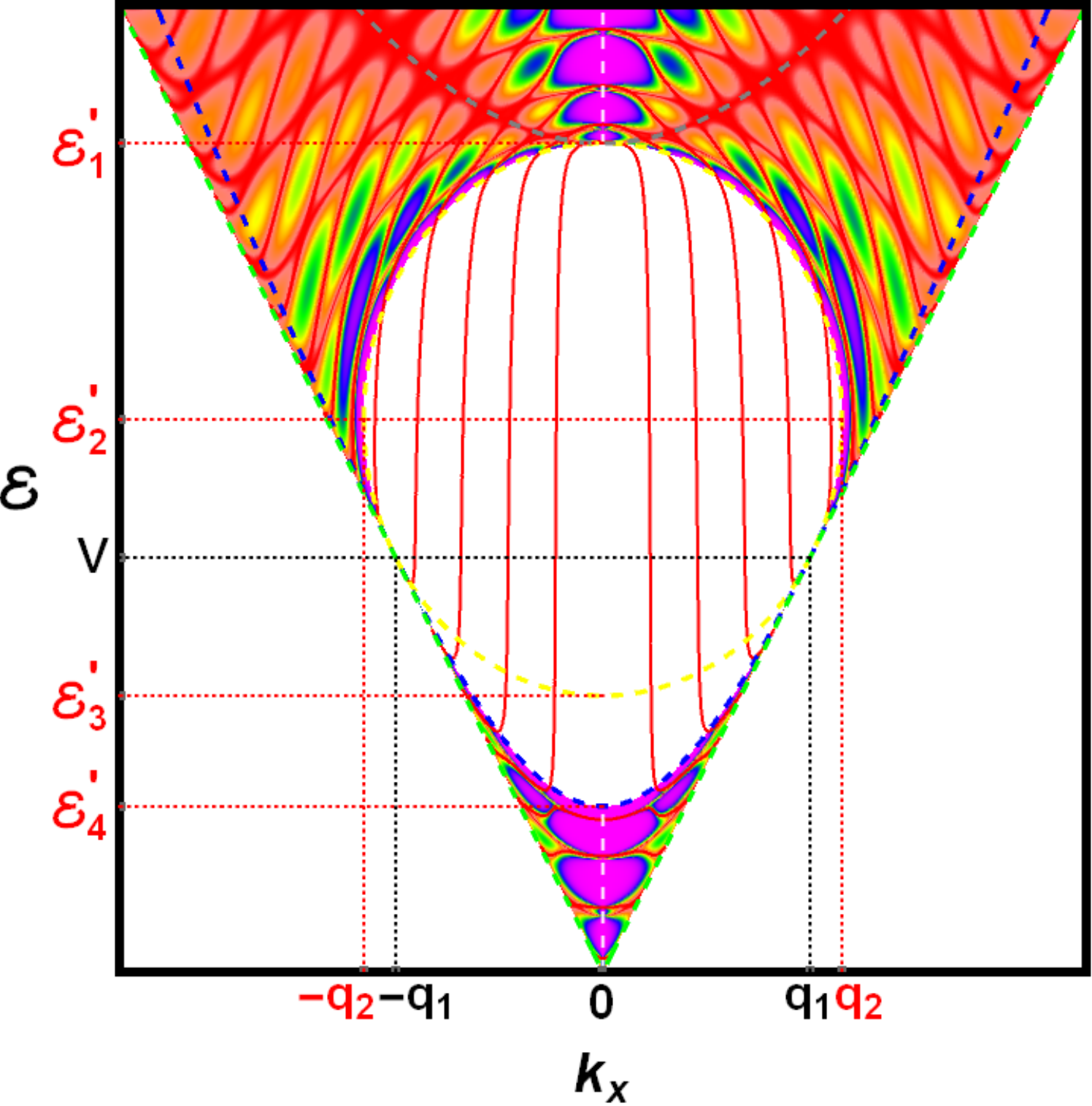}\label{fig09:SubFigE}}
    \subfloat[][$\tau=-1$]{
   \includegraphics[scale=0.17]{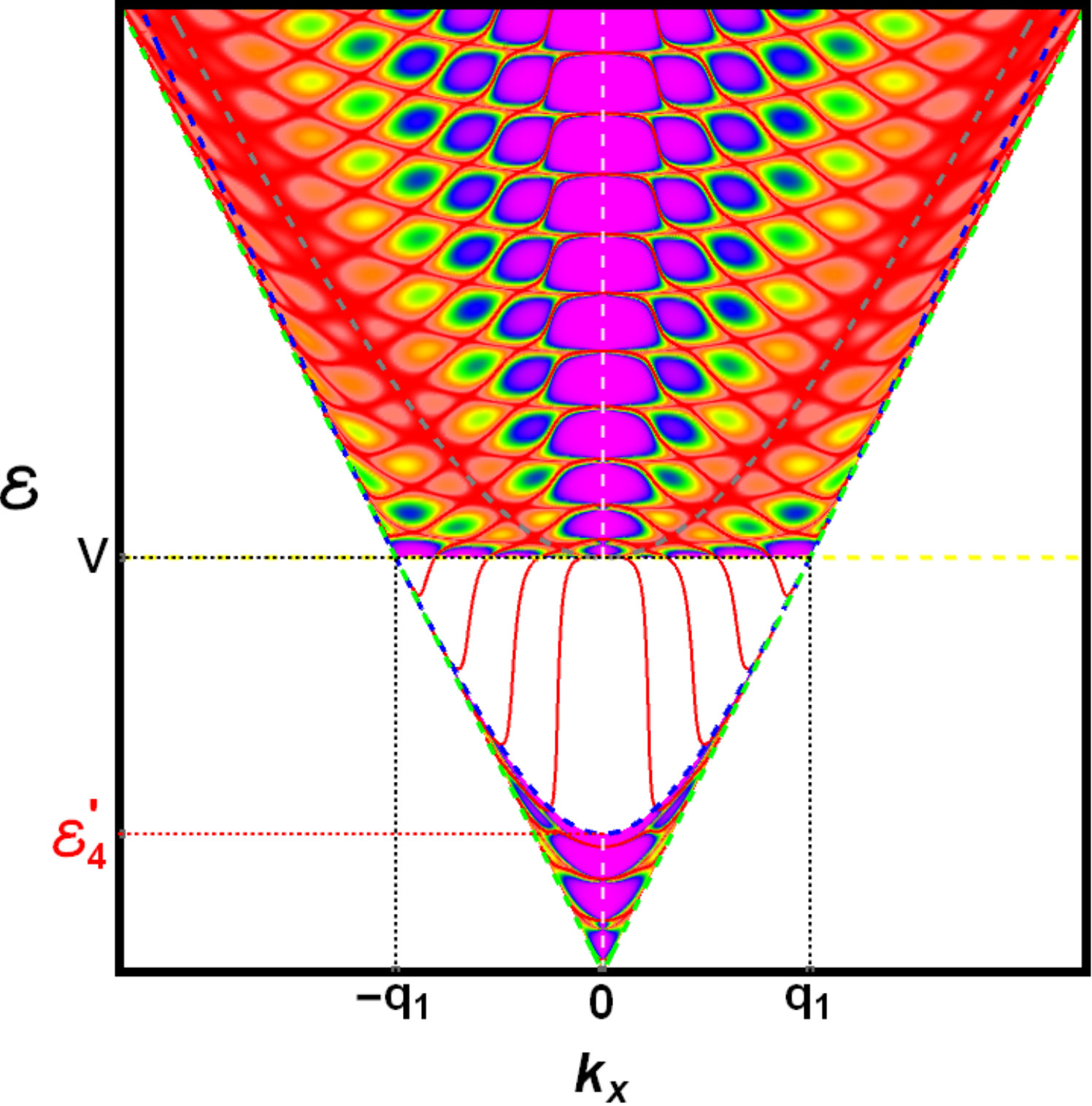}\label{fig09:SubFigF}}
\subfloat[][$\tau=-1.5$ ]{
   \includegraphics[scale=0.17]{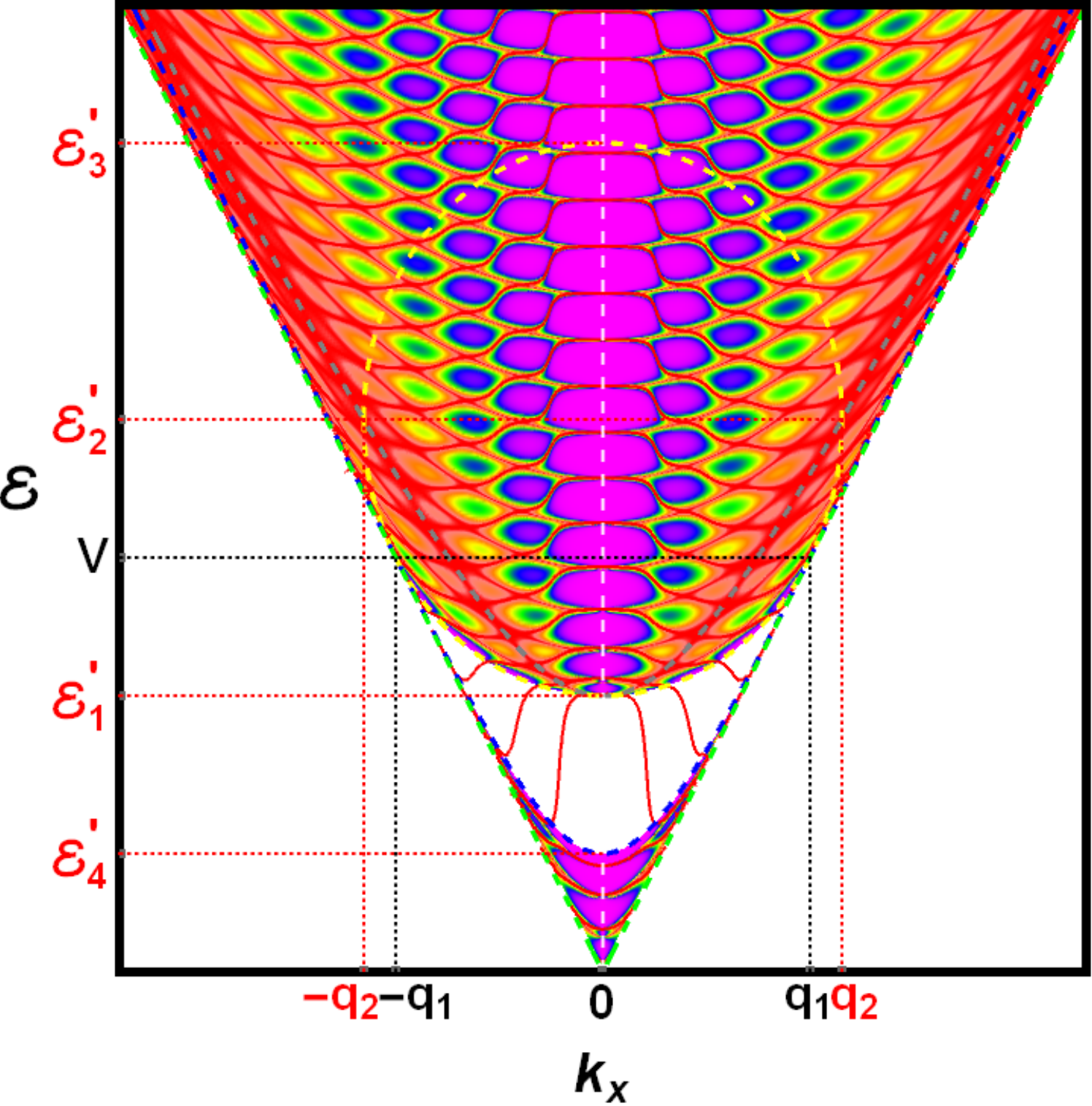}\label{fig09:SubFigG}
   \includegraphics[scale=0.17]{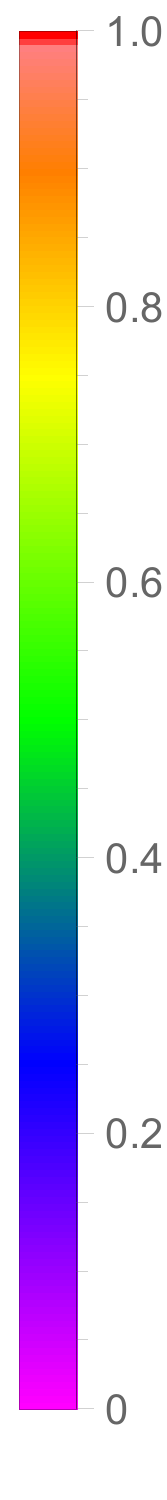}\label{fig09:SubFigG}}
    \caption{(Color online) Density plot of the transmission probability $T$ in the
$(\varepsilon,k_{x})$-plane for a double-barrier structure. Both barriers have the same
height $V=3$, the same width $d$, and are separated by a distance $d=4$. The plot is shown
for four values of the parameter $\tau$. The incident energy $\varepsilon$ is plotted on
the vertical axis, while the longitudinal momentum $k_{x}$ is plotted on the horizontal
axis.}
\label{fig09}
\end{figure}

%========================================================
The angular–momentum resolved analysis of quantum transport in Dirac systems provides
complementary information to the $(\varepsilon,k_y)$ representation, since the
longitudinal component $k_x$ directly governs the propagation along the barrier axis. The
density maps of the transmission probability $T(\varepsilon,k_x)$ for the double-barrier
structure are shown in Fig.~\ref{fig09}, as a function of the incident energy
$\varepsilon$ and the longitudinal wave vector component $k_x$, for different values of
the tilt parameter $\tau$. All panels exhibit a clear symmetry with respect to the axis
$k_x=0$. Moreover, the transmission strictly vanishes along this axis, reflecting the
absence of propagation when the longitudinal component of the wave vector is
zero~\cite{Choubabi2024,Katsnelson2006}.

Similar to the case of a single barrier~\cite{Katsnelson2006,Allain2011,Beenakker2008}, two distinct
families of resonance peaks emerge within the allowed transmission zones:
(i) those determined from Eq.~\ref{eq012}, represented by black and blue lines, and
(ii) those obtained from Eq.~\ref{eq011}, represented by red lines.

A key difference with respect to the single-barrier case is the multiplication of these
peaks in the double-barrier system, as well as the appearance of \textit{line-type
resonances}. These resonances extend far beyond the boundaries of the allowed regions and
penetrate the forbidden transmission zones, while remaining confined between the limits
defined by the allowed transmission regions. This phenomenon is unique to the
double-barrier structure and does not occur in the single-barrier
case~\cite{Britnell2013,Baltateanu2019,Sun2011,Jellal2012}.

For $\varepsilon < V$, the allowed transmission regions are bounded by the straight lines
$\varepsilon = \pm k_x$ (green dashed lines) and by elliptic or hyperbolic branches
depending on $\tau$. Inside these regions,
perfect transmission resonances appear at discrete energy values in agreement with the
analytical solutions of Eqs.~\ref{eq011} and \ref{eq012}.

For $\varepsilon > V$, the allowed zones are similarly delimited by $\varepsilon = \pm
k_x$ and the upper branches of hyperbolas (yellow or blue dashed lines). Again, resonances
with perfect transmission occur within these regions, while additional peaks extend
outside the allowed zones into the forbidden transmission regions, indicating strong
coupling between the bound states in the well and the barriers.

A remarkable feature is the persistence of the Klein paradox, clearly identified along the
straight lines $\varepsilon = \pm k_x$, which correspond to normal incidence ($k_y=0$).
Under these conditions, the barriers become completely transparent to Dirac fermions,
regardless of their width or height~\cite{Katsnelson2006,Allain2011,Beenakker2008}.

\begin{figure}[!h]\centering
\subfloat[ $\tau$=0.5]{
    \hspace{-0.01cm}\includegraphics[width=0.3\linewidth]{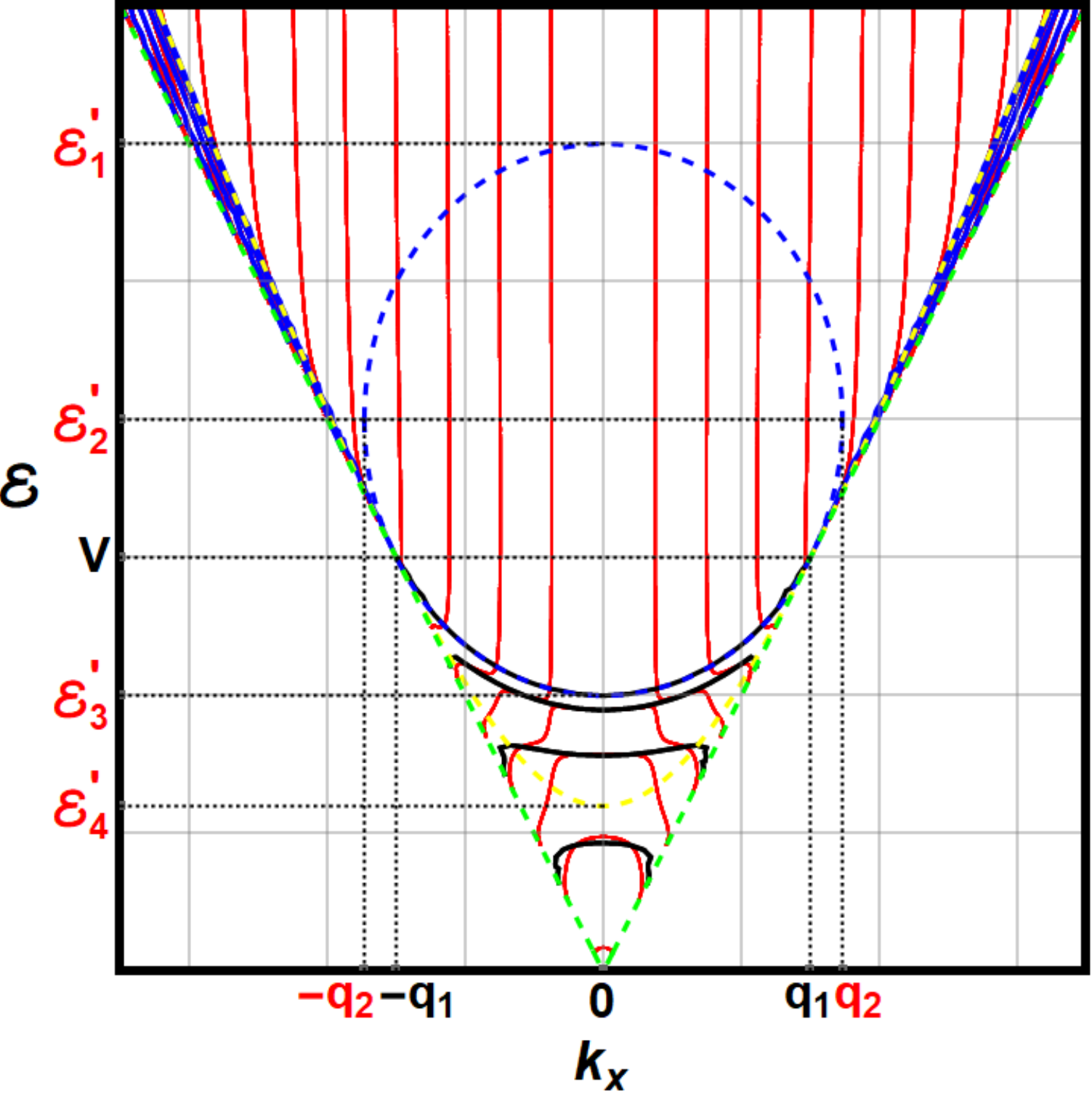}\label{fig016:SubFigA}}
    \subfloat[][ $\tau$=-0.5]{
   \hspace{-0.01cm}\includegraphics[width=0.3\linewidth]{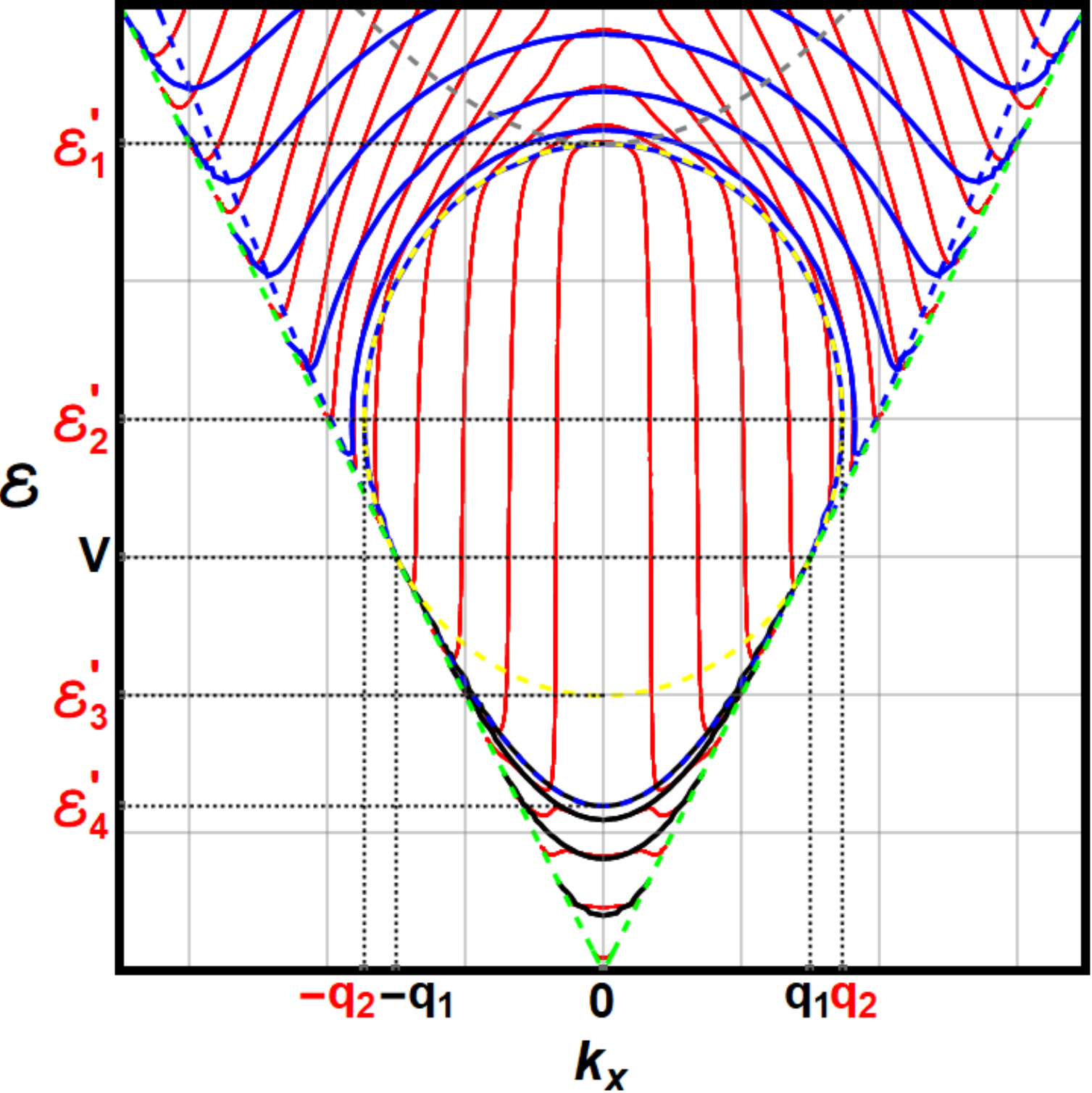}\label{fig016:SubFigB}}
    \caption{(Color online) Spectral curves of bound states and resonant scattering states
with perfect transmission ($T=1$), obtained from Eqs.~\ref{eq012}, \ref{eq011}, and
\ref{eq0111}. The results highlight the influence of the tilt parameter sign ($\tau = \pm
0.5$) on the formation and symmetry of resonances in a tilted Dirac cone double-barrier
structure.}
\label{fig016}
\end{figure}

Figure~\ref{fig016} further illustrates these features by explicitly showing the spectral
curves of bound states and the curves of resonant scattering states corresponding to
perfect transmission ($T=1$). For $\tau=0.5$ [Fig.~\ref{fig016:SubFigA}], where the tilted
Dirac cones of the two barriers are oriented in the negative $k_y$ direction, the spectral
structure reveals families of resonances inside the allowed transmission regions, some of
which also extend into forbidden zones. In contrast, for $\tau=-0.5$
[Fig.~\ref{fig016:SubFigB}], the cones tilt in the opposite direction, modifying the
resonance structure while maintaining the overall symmetry with respect to $k_x=0$.

Taken together, Figs.~\ref{fig09} and \ref{fig016} highlight the crucial role of the tilt
parameter $\tau$ in shaping not only the position and density of resonance peaks but also
the emergence of \textit{line-type resonances} that propagate into forbidden transmission
zones. This dependence, absent in single-barrier systems, constitutes a distinctive
signature of resonant tunneling in double-barrier structures with tilted Dirac
cones~\cite{RamezaniMasir2010,Sun2011,Barbier2010,Alhaidari2012}.
%========================================================
\subsection{Comparison Between Single and Double Barriers}
\begin{figure}[!h]
\centering
    \subfloat[][]{
   \hspace{-0.8cm}\includegraphics[scale=0.17]{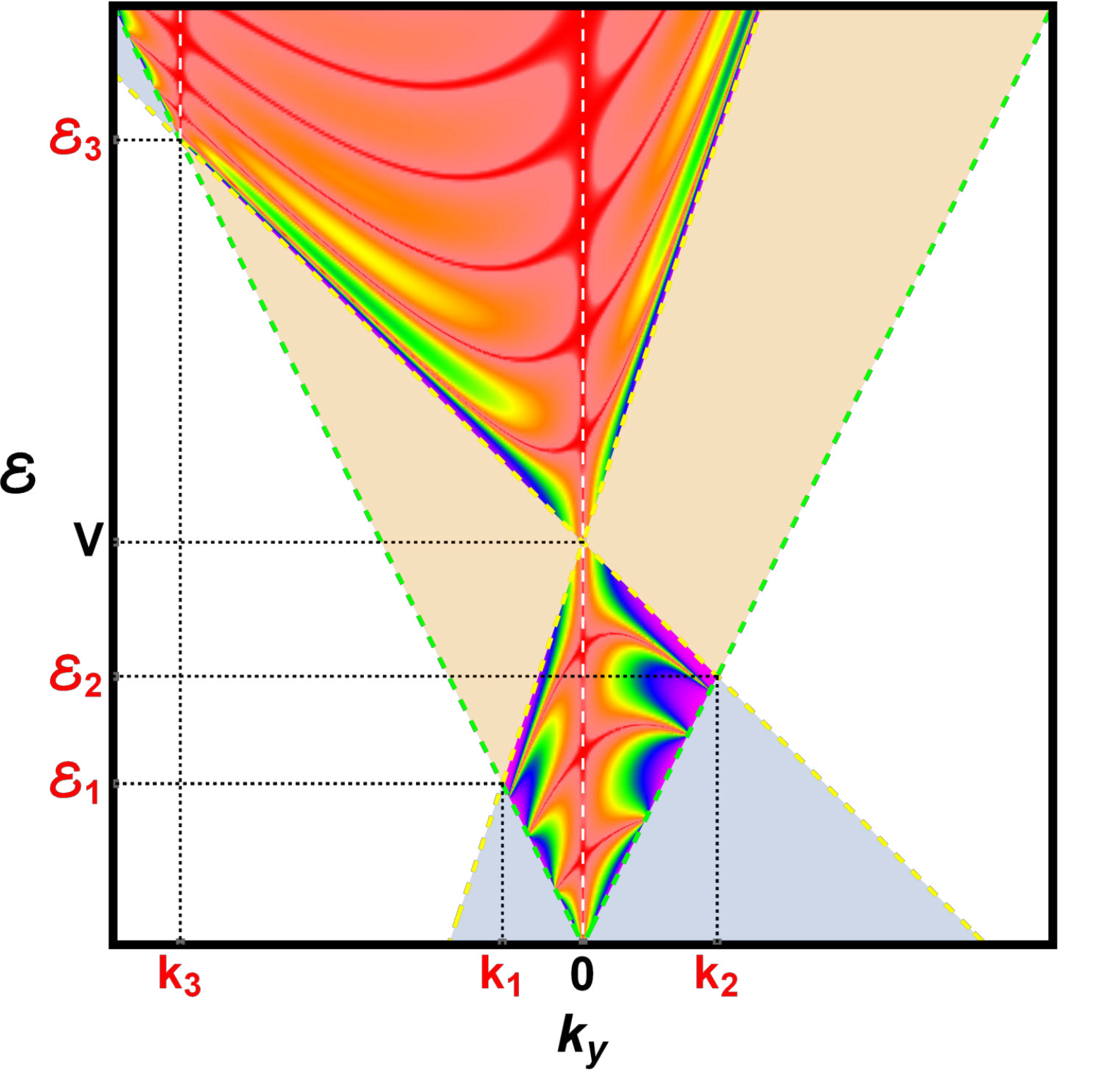}\label{fig010:SubFigA}}
    \subfloat[][]{
   \includegraphics[scale=0.17]{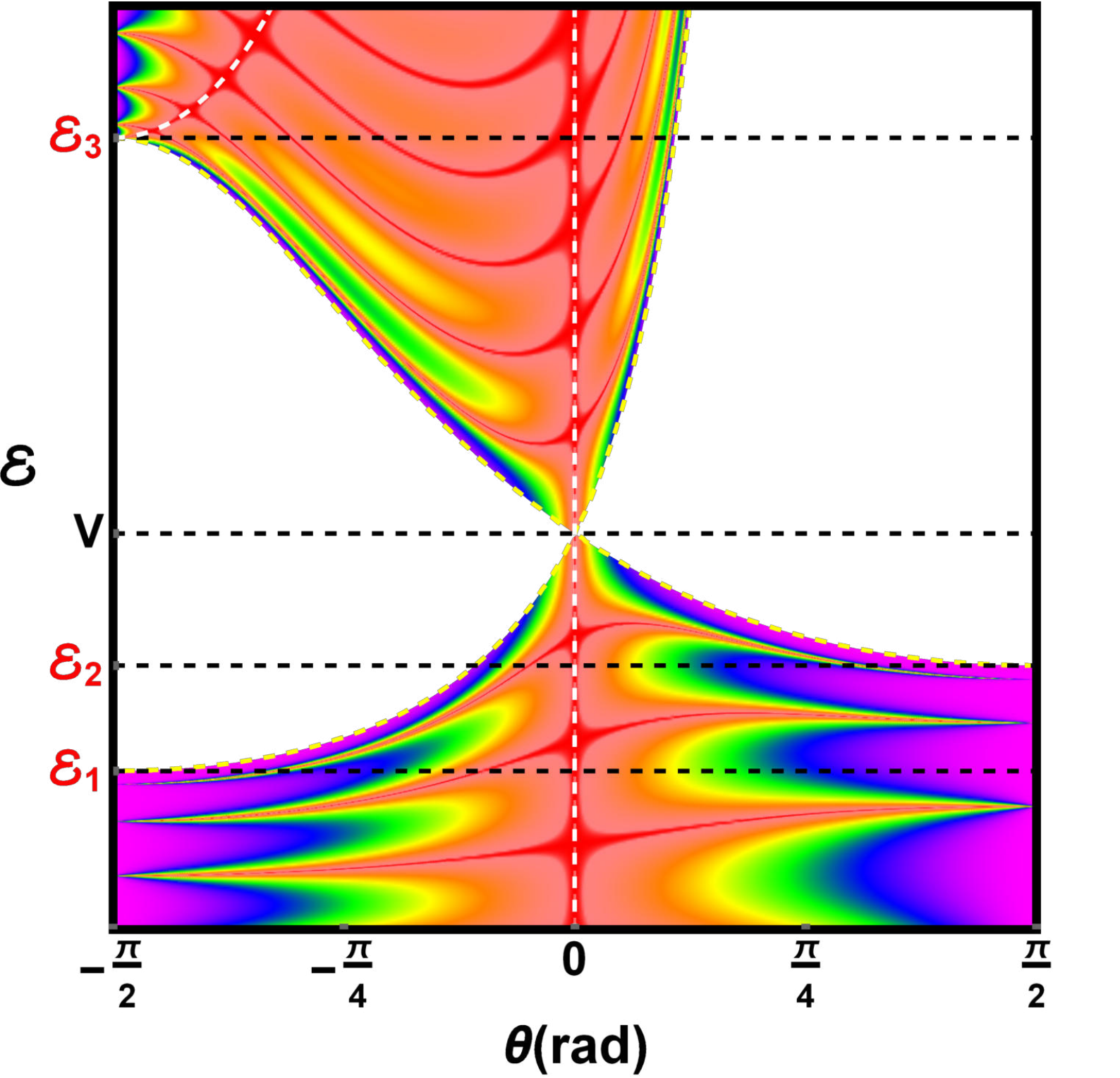}\label{fig010:SubFigB}}
\subfloat[][]{
    \hspace{-0.25cm}\includegraphics[scale=0.17]{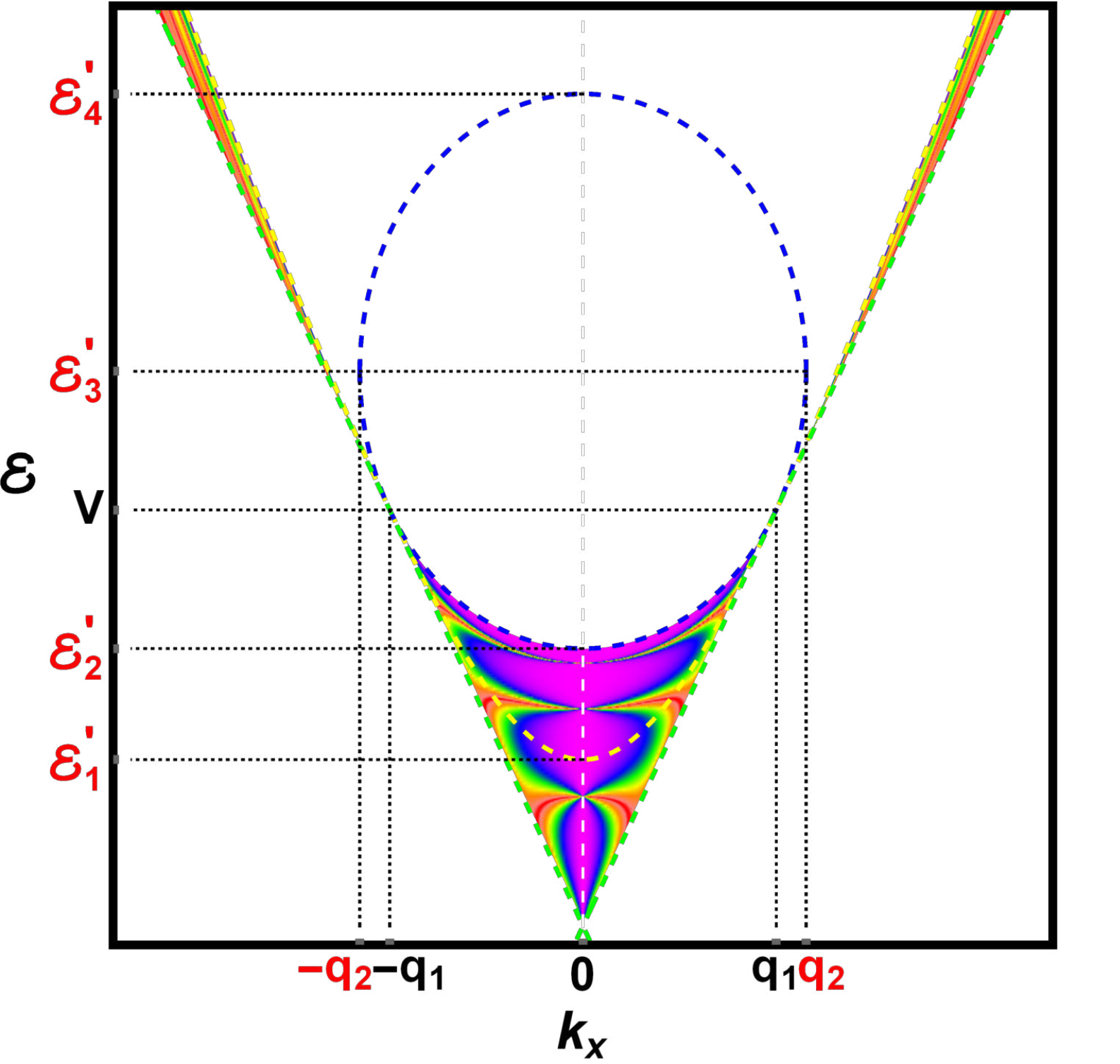}\label{fig010:SubFigC}}
    \subfloat[][]{
    \hspace{-0.3cm}\includegraphics[scale=0.17]{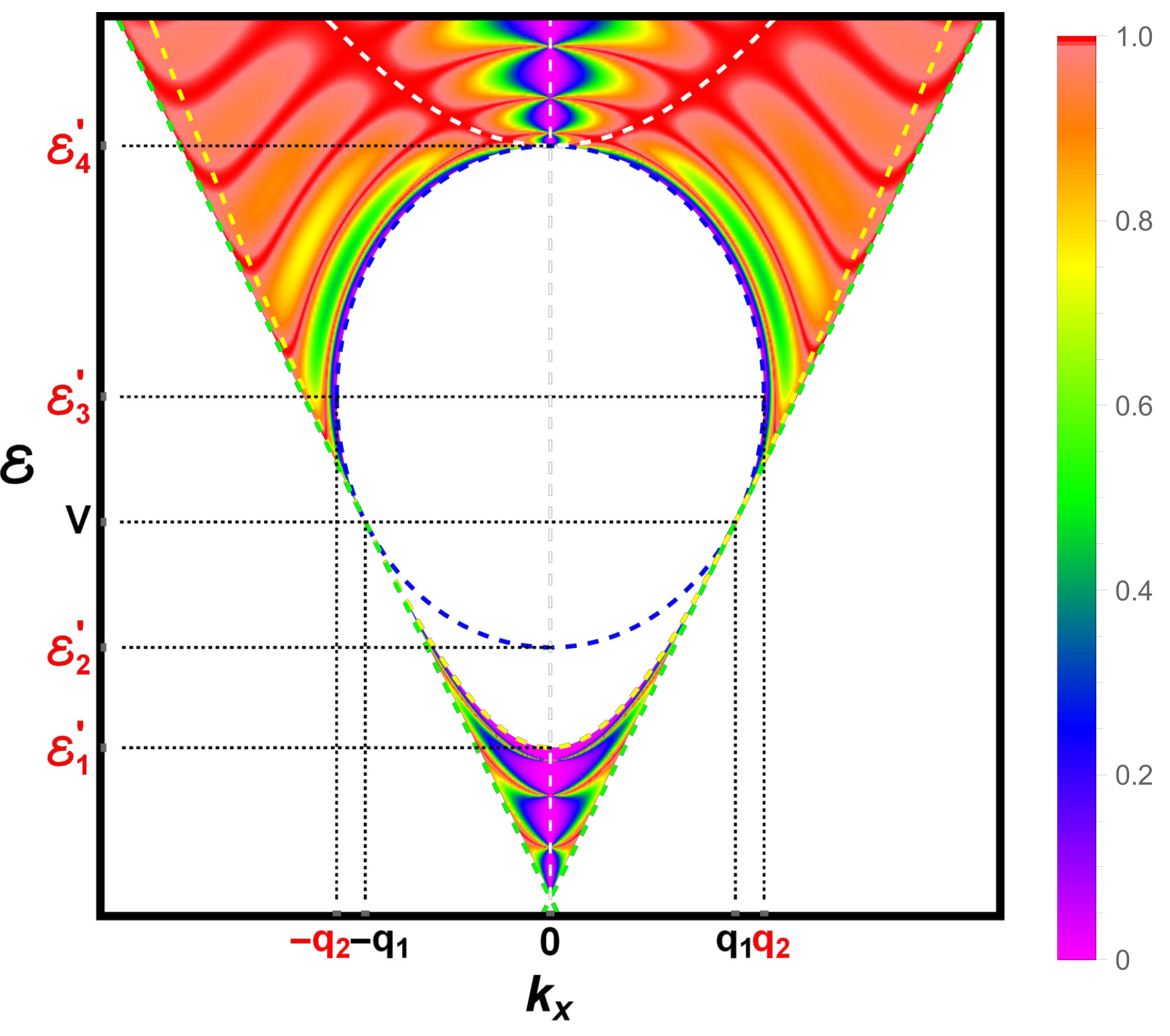}\label{fig010:SubFigD}}\\
    \subfloat[][]{
   \hspace{-0.8cm}\includegraphics[scale=0.17]{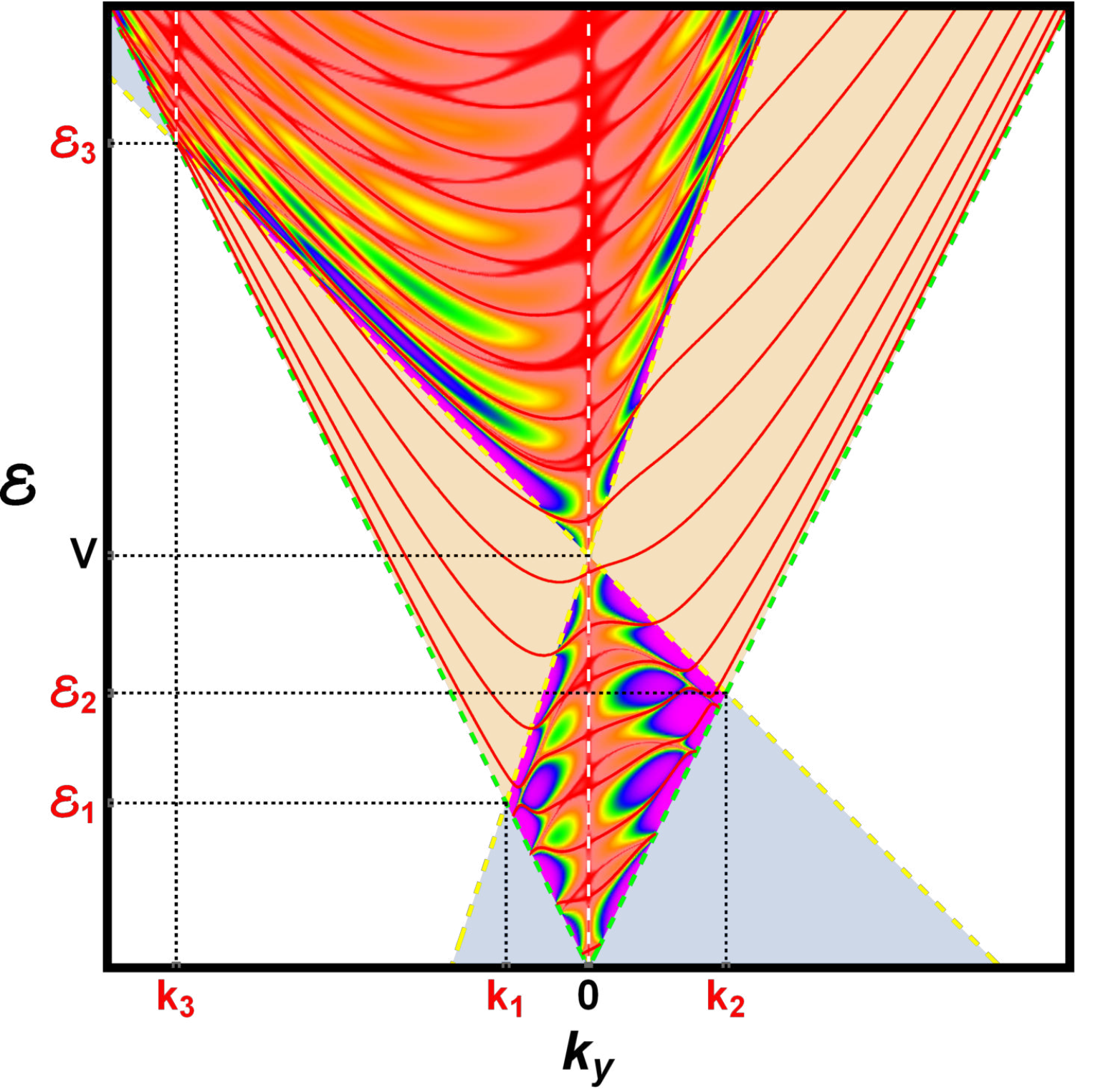}\label{fig010:SubFigE}}
    \subfloat[][]{
   \includegraphics[scale=0.17]{fig22}\label{fig010:SubFigF}}
   \subfloat[][]{
  \hspace{-0.25cm}\includegraphics[scale=0.17]{Fig40}\label{fig010:SubFigG}}
\subfloat[][]{
   \hspace{-0.1cm}\includegraphics[scale=0.17]{Fig43}\label{fig010:SubFigH}}
    \caption{(Color online) Comparative transmission profiles for single- and
double-barrier tilted Dirac cone structures, illustrated through transmission probability
density maps. The results highlight the coexistence of Fabry–Pérot and line-type
resonances, for a tilt parameter fixed at $\tau = 0.5$, $d = 4$ and $V = 3$.}
\label{fig010}
\end{figure}

%========================================================
The comparison between single- and double-barrier configurations in tilted Dirac materials
provides valuable insight into the role of quantum interference in transport.
Figure~\ref{fig010} presents a comparative analysis of the electronic transmission
properties for both geometries under the influence of a tilted Dirac cone with the tilt
parameter fixed at $\tau = 0.5$.

The upper panels (\ref{fig010:SubFigA}--\ref{fig010:SubFigD}) correspond to the
single-barrier configuration, while the lower panels
(\ref{fig010:SubFigE}--\ref{fig010:SubFigH}) depict the double-barrier case. In both
systems, the allowed and forbidden transmission regions (white areas) exhibit similar
boundaries, indicating that the incidence conditions and accessible wave vectors remain
unchanged when duplicating the barriers. However, significant differences emerge in the
fine structure of the transmission spectra: in the double-barrier case, the number of
resonance peaks increases markedly due to multiple reflections between the interfaces,
giving rise to enhanced Fabry–Pérot–type oscillations~\cite{RamezaniMasir2010,Lu2013,Pereira2010}.
Moreover, \textit{line-type resonances} appear and extend into the forbidden regions, a
phenomenon absent in the single-barrier
case~\cite{RamezaniMasir2010,Sun2011,Barbier2010,Alhaidari2012,Jellal2012}.

The Klein paradox remains preserved in all configurations. At normal incidence ($k_y=0$,
or $\theta=0$), perfect transmission occurs regardless of the barrier height, width, or
multiplicity~\cite{Katsnelson2006,Allain2011,Beenakker2008}. This phenomenon also reappears under
specific conditions, such as $k_y=-V/\tau$ for $\varepsilon > \varepsilon_3$, or
$\varepsilon=-V/(\tau\sin\theta)$ for $\varepsilon > V$, which correspond to situations
where the effective refractive indices of adjacent regions are
equal~\cite{Choubabi2024,Allain2011,Katsnelson2006}.

Finally, the spectra plotted as a function of $k_x$ [panels (\ref{fig010:SubFigC},
\ref{fig010:SubFigD}, \ref{fig010:SubFigG}, \ref{fig010:SubFigH})] confirm this behavior.
The Klein paradox is associated with the parabolic branch
$\varepsilon=\sqrt{k_x^2+V^2/\tau^2}$ (gray dashed line), while the transmission strictly
vanishes along the axis $k_x=0$, reflecting the absence of propagation when the
longitudinal wave-vector component is zero.
%========================================================
\subsection{Analytical Validation of Resonances}
\begin{figure}[h!]\centering
\subfloat[$k_{y}=0.6$, $\tau=0.5$]{
\hspace{-0.01cm}\includegraphics[width=0.4\linewidth]{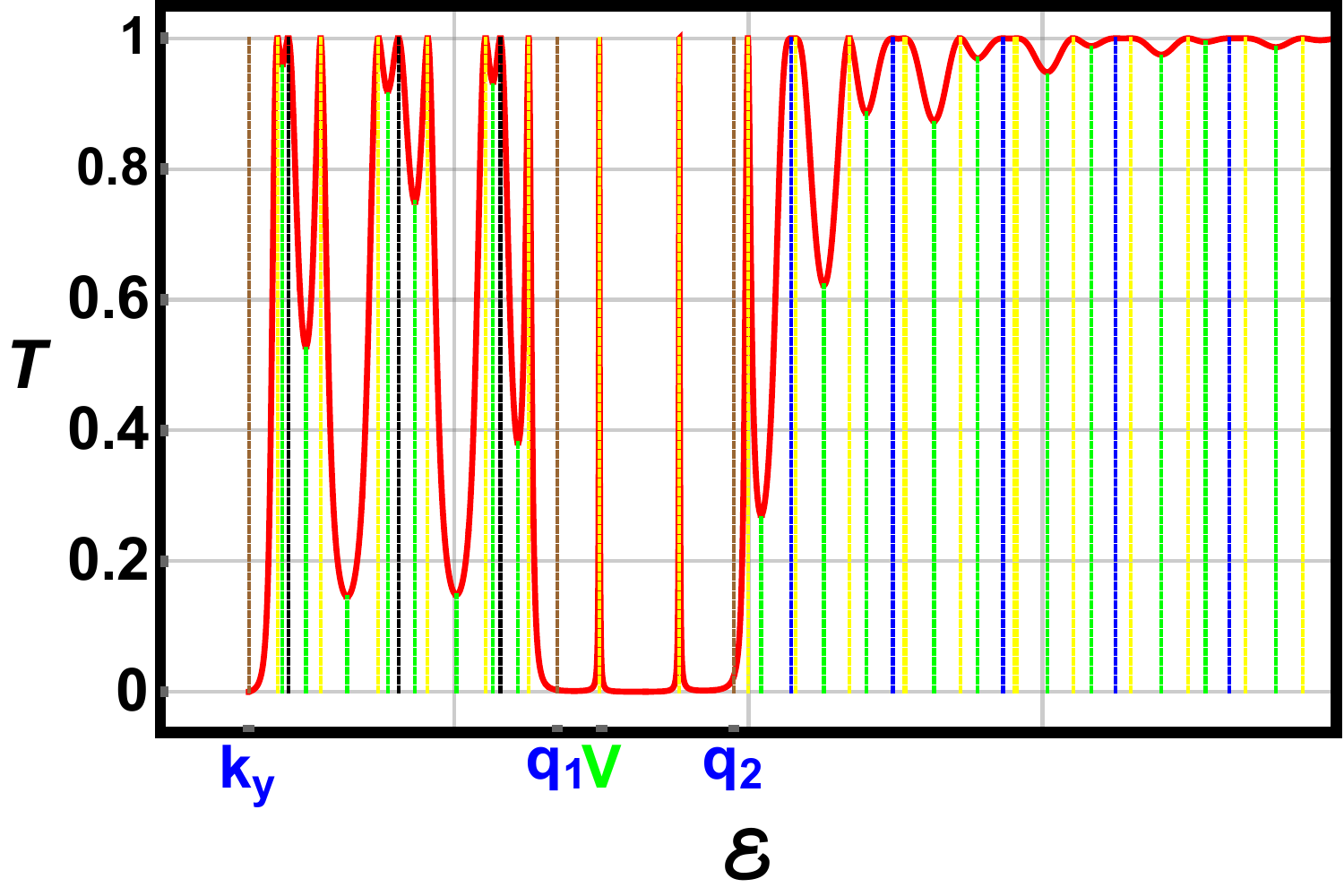}\label{fig017:SubFigA}}
\subfloat[$\theta=\frac{\pi}{8}$, $\tau=0.5$]{
\hspace{-0.01cm}\includegraphics[width=0.4\linewidth]{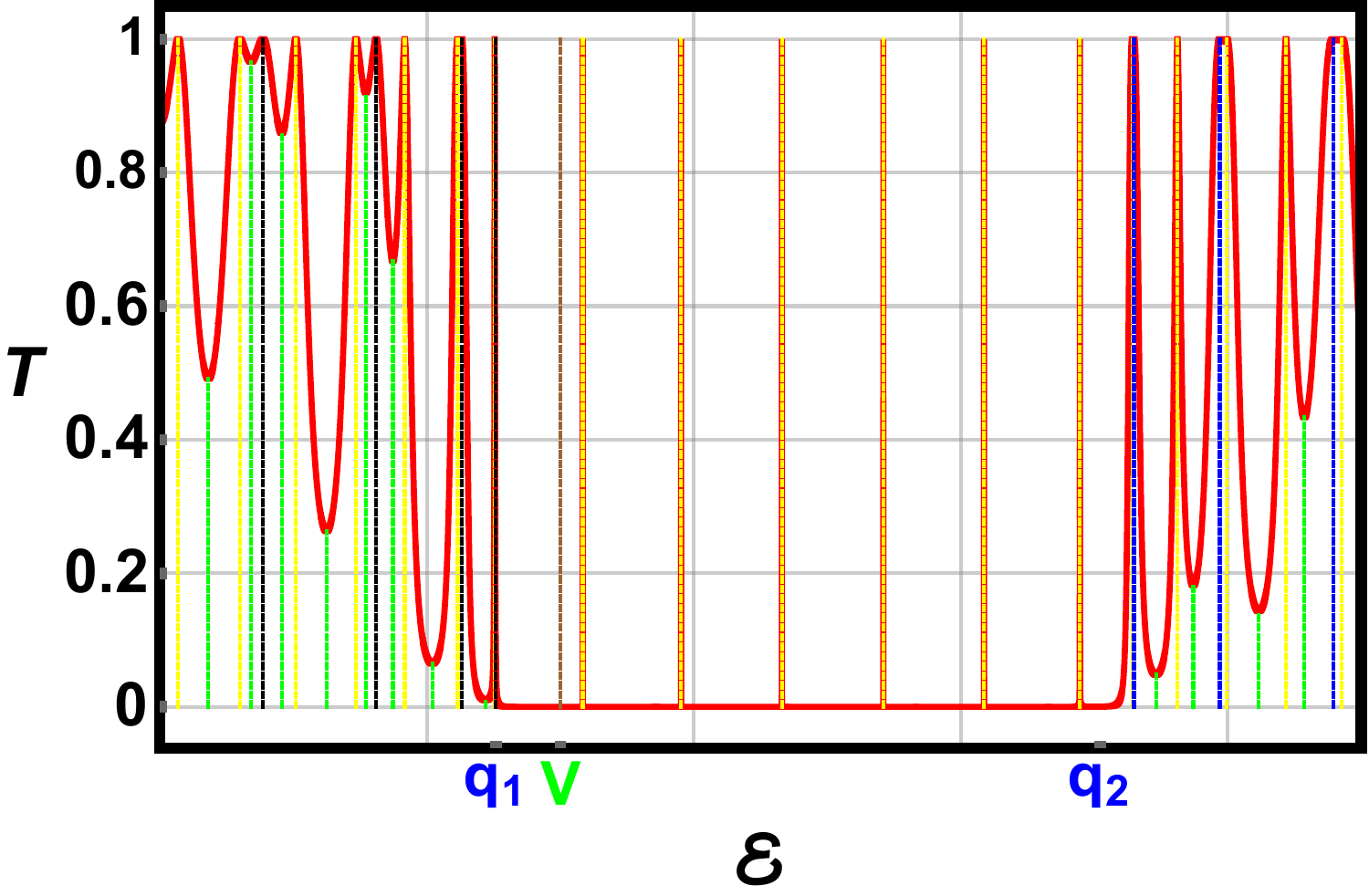}\label{fig017:SubFigB}}\\
\subfloat[$\varepsilon=1.8$, $\tau$=0.5]{
\hspace{-0.01cm}\includegraphics[width=0.4\linewidth]{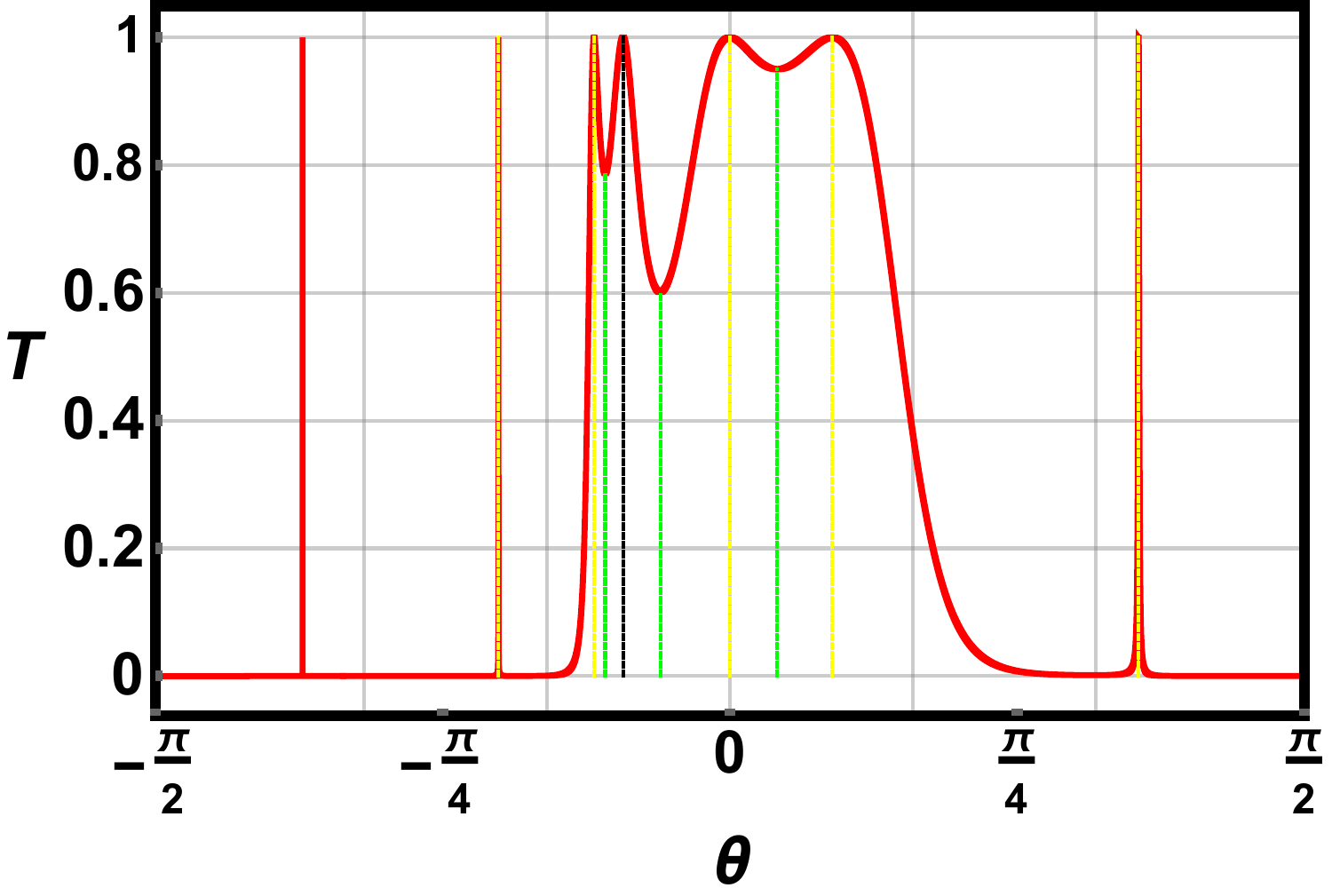}\label{fig017:SubFigE}}
\subfloat[][$k_{x}=1$, $\tau=-0.5$]{
\hspace{-0.01cm}\includegraphics[width=0.4\linewidth]{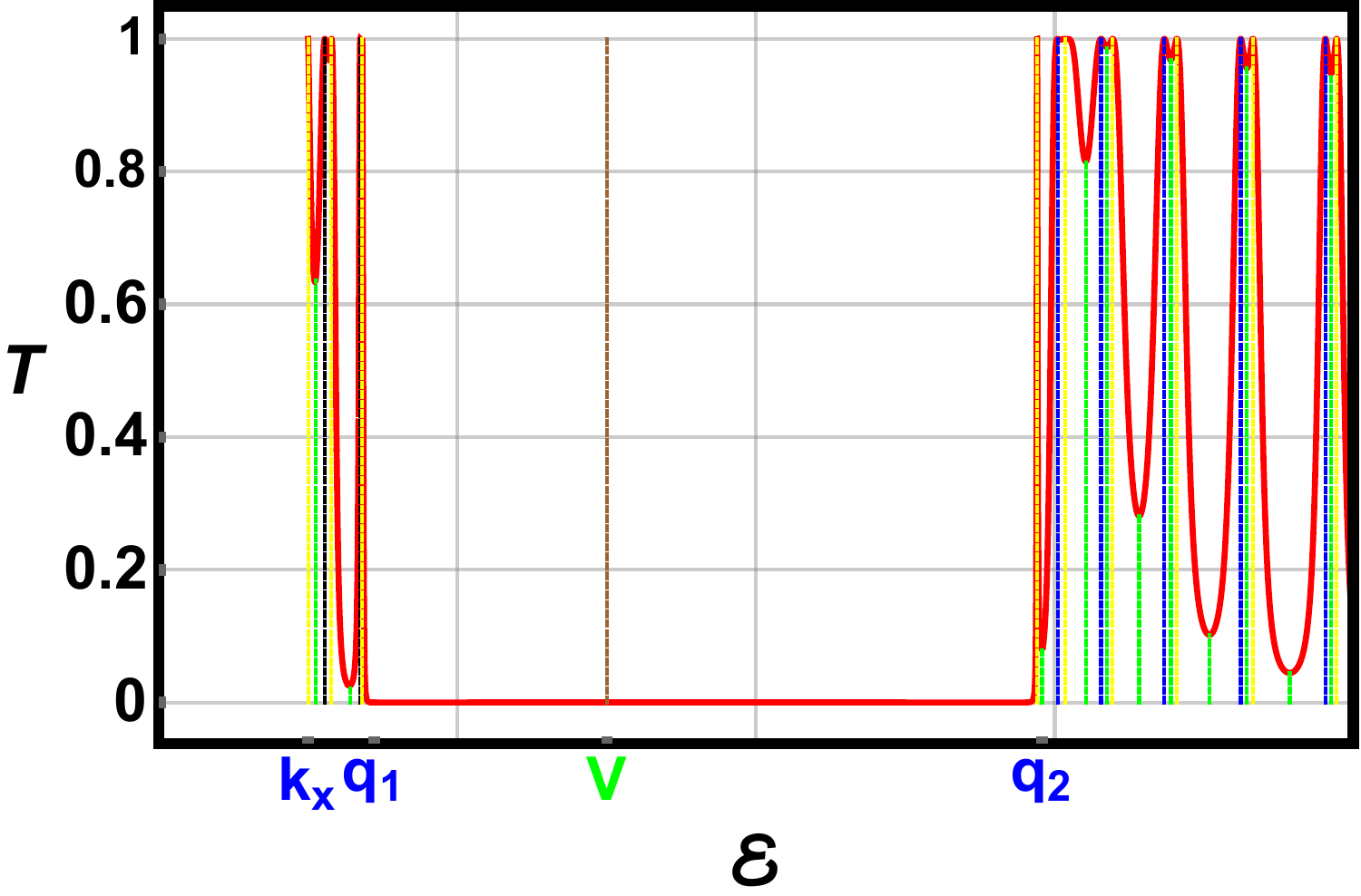}\label{fig017:SubFigF}}
\caption{(Color online) Transmission spectra $T(\varepsilon)$ and $T(\theta)$ showing
Fabry–Pérot and line-type resonances in a tilted Dirac cone double-barrier structure. The
results are obtained for $k_{y}=0.6$, $\theta=\pi/8$, $\varepsilon=1.8$, and $k_{x}=1$,
consistent with the analytical solutions of Eqs.~\ref{eq011}, \ref{eq012}, and
\ref{eq0111}. The tilt parameter is fixed at $\tau=0.5$, with $d=4$ and $V=3$.}
\label{fig017}
\end{figure}

%========================================================
The validation of the resonance conditions through analytical predictions is crucial for
assessing the robustness of the transport model. Figure~\ref{fig017} illustrates the
transmission resonances $T(\varepsilon)$ and $T(\theta)$ obtained for different fixed
values of wave and energy parameters.

Subfigures (a) and (b) show the dependence of the transmission on the energy
$\varepsilon$, respectively for $k_y = 0.6$ and $\theta = \pi/8$, with the tilt parameter
fixed at $\tau = 0.5$. In these cases, the transmission exhibits a series of resonance
peaks perfectly aligned with the vertical dashed lines (black, blue, and yellow), which
correspond to the analytical solutions of Eqs.~\ref{eq011} and~\ref{eq012}. These
solutions characterize the conditions of resonant tunneling, where the transmission
reaches unity ($T = 1$). In parallel, transmission
minima, indicated by the green dashed lines, also appear and correspond to the energies
obtained from Eq.~\ref{eq0111}, confirming the predictive power of the analytical
framework.

Subfigure (c) shows the evolution of the transmission $T(\theta)$ for a fixed energy
$\varepsilon = 1.8$ and $\tau = 0.5$. One can clearly distinguish discrete angular
resonances that coincide with the analytical solutions, separated by regions of strongly
suppressed transmission. This behavior confirms the selective nature of resonances with
respect to the angle of incidence, in line with earlier studies of angle-resolved
Fabry–Pérot interference in graphene-based systems~\cite{Katsnelson2006,Young2009,Benlakhouy2025}.

Finally, subfigure (d), plotted for $k_x = 1$ and $\tau = 0.5$, highlights the excellent
agreement between numerical results and analytical predictions, with total transmission
peaks located precisely at the positions predicted by the resonance equations. This
correspondence provides strong evidence that the analytical model accurately captures the
essential physics of resonant tunneling and interference in tilted Dirac
systems~\cite{Katsnelson2006,Goerbig2008,Nguyen2018,Zhang2023b}.

\medskip
In conclusion, the results demonstrate that the interplay between barrier geometry, cone
tilt, and incidence conditions gives rise to a rich variety of transmission features,
including Fabry–Pérot oscillations, line-type resonances, and Klein
tunneling~\cite{Britnell2013,Pattrawutthiwong2021,Zhang2023b,Sun2011,Jellal2012}. The strong dependence of
the transmission spectrum on the tilt parameter highlights its importance as a tuning knob
for controlling quantum transport. These findings provide useful guidelines for designing
resonant tunneling devices and anisotropic filters based on tilted Dirac materials.
%========================================================
\section{Conclusion}
%========================================================
In this work, we investigated the transport properties of massless Dirac fermions in a
system composed of multiple junctions of the type graphene–tilted Dirac cone
material–graphene, with a particular emphasis on the role of the tilt parameter $\tau$.
Starting from the eigenstates of the tilted Dirac Hamiltonian and applying continuity
conditions at the interfaces through the transfer matrix method, we derived the
transmission probabilities across single- and double-barrier
configurations.

Our results highlight that the tilt parameter $\tau$ has a profound influence on the
transmission spectra. Increasing $\tau$ not only modifies the boundaries of the allowed
and forbidden regions but also induces strong anisotropy with respect to the incidence
angle~\cite{Pattrawutthiwong2021,Goerbig2008,Nguyen2018,Zhang2023b}. Within the double-barrier
configuration, sharp \textit{line-type resonances} emerge inside the transmission gaps due
to the confinement of bound states between the two barriers. These resonances, absent in
the single-barrier case, can be understood as a Fabry–Pérot-like interference mechanism
where evanescent states act as effective mirrors~\cite{RamezaniMasir2010,Sun2011,Barbier2010,Allain2011,Young2009,Jellal2012}. Their
number, position, and intensity are highly sensitive to $\tau$, the incidence angle, and
the structural parameters of the barriers (height, width, and separation).

A systematic comparison between single- and double-barrier systems confirmed that, while
both share similar shapes of allowed and forbidden regions, the double-barrier structure
enhances Fabry–Pérot resonances and introduces additional line-type resonances extending
into forbidden zones. Furthermore, the Klein paradox persists in all configurations:
perfect transmission is maintained at normal incidence ($k_y=0$) and at $k_y=-V/\tau$,
where the effective refractive indices of adjacent regions
match~\cite{Katsnelson2006,Choubabi2024}. This remarkable transparency occurs
independently of the barrier width, height, or incident energy.

These findings provide a deeper physical understanding of quantum transport in tilted
Dirac cone materials and open new perspectives for device engineering. By exploiting the
tunability of $\tau$ and structural parameters, one can design resonant tunneling diodes,
energy-selective filters, or anisotropic electron
waveguides~\cite{Britnell2013,Valagiannopoulos2019,Sun2011}. Beyond the present
analysis, future work could address more complex architectures and explore the influence
of external fields (electric, magnetic, or strain), further broadening the technological
potential of tilted Dirac systems.
%========================================================

%========================================================
\end{document}